\documentclass[aps,prc,preprint,showpacs,showkeys,floatfix,nofootinbib,%
superscriptaddress]{revtex4}
\usepackage{graphicx}

\usepackage{multirow}
\usepackage{bm}
\usepackage{color}

\def\be{\begin{equation}} \def\ee{\end{equation}} \def\bea{\begin{eqnarray}}
\def\eea{\end{eqnarray}} 

\newcommand{\eqn}[1]{\label{eq:#1}}
\newcommand{\refeq}[1]{(\ref{eq:#1})}

\newcommand{\Equation}{Equation~\refeq}
\newcommand{\Eq}{Eq.~\refeq}
\newcommand{\Eqs}[2]{Eqs.~(\ref{eq:#1}) and (\ref{eq:#2})}
\newcommand{\Eqss}[2]{Eqs.~(\ref{eq:#1})--(\ref{eq:#2})}

\def\ket#1{\vert#1\rangle}
\def\bra#1{\langle#1\vert}

\date{\today}

\def\gtap{\ \raise.3ex\hbox{$>$\kern-.75em\lower1ex\hbox{$\sim$}}\ }
\def\ltap{\ \raise.3ex\hbox{$<$\kern-.75em\lower1ex\hbox{$\sim$}}\ }
\begin{document}
\title{Dynamical Model of  Coherent Pion Production in 
 Neutrino-Nucleus Scattering}
\author{S. X. Nakamura}\email{satoshi@jlab.org}
\thanks{Current affiliation: 
Excited Baryon Analysis Center (EBAC),
Thomas Jefferson National Accelerator Facility, Newport News, VA 23606, USA}
\affiliation{Instituto de F\'isica, Universidade de S\~ao Paulo,
C.P. 66318, 05315-970, S\~ao Paulo, SP, Brazil}
%\affiliation{Theory Group, TRIUMF,
%4004 Wesbrook Mall, Vancouver, BC V6T 2A3, Canada}
\author{T. Sato}
\affiliation{Department of Physics, Osaka University, Toyonaka, Osaka,
 560-0043, Japan}
\author{T.-S. H. Lee}
\affiliation{Physics Division, Argonne National Laboratory, Argonne, Illinois 60439, USA}
\author{B. Szczerbinska}
\affiliation{Dakota State University, College of Arts \& Sciences,
Madison, SD, 57042-1799, USA}
\author{K. Kubodera}
\affiliation{Department of Physics and Astronomy, University of South
 Carolina, Columbia, SC, 29208, USA}

\begin{abstract}
We study coherent pion production in neutrino-nucleus scattering in
the energy region relevant to neutrino oscillation experiments
of current interest.
Our approach is based on a combined use of the Sato-Lee model 
of  electroweak pion production on a nucleon
and the $\Delta$-hole model of pion-nucleus reactions.
Thus we develop a model which describes pion-nucleus scattering and
electroweak coherent pion production in a unified manner.
Numerical calculations are carried out for the case of the $^{12}$C
target.
All the free parameters in our model are fixed by fitting 
to both total and elastic
differential cross sections for $\pi-^{12}$C scattering.
Then we demonstrate the reliability of our approach by confronting our
prediction for the coherent pion photo-productions with data.
Finally, we calculate total and differential cross sections for neutrino-induced
coherent pion production, and some of the results are (will be) compared with 
the recent (forthcoming) data from K2K, SciBooNE and MiniBooNE.
We also study effect of the non-locality of the $\Delta$-propagation in
the nucleus, and compare the elementary amplitudes used in different
 microscopic calculations.

\end{abstract}
\maketitle

\section{Introduction}
\label{intro}

The detailed theoretical study 
of neutrino-nucleus reactions is of great current  importance 
due to the ever increasing precision of neutrino oscillation experiments  
(recently carried out, on-going and forthcoming).
Since most of these experiments measure the neutrino flux 
through neutrino-nucleus scattering, 
reliable theoretical estimates of the relevant cross sections
are prerequisite for the accurate interpretation of the data.
Some of these experiments (T2K, MiniBooNE, etc.) use neutrinos 
in an energy range within which the dominant processes are
the quasi-elastic nucleon knockout  and the 
quasi-free single-pion production through the excitation
of the $\Delta$ (1232) resonance.
Meanwhile, coherent single-pion production 
in this energy region (albeit not a dominant process) is 
also of considerable interest,
since it allows us to study,
with no ambiguity concerning the final nuclear state,
the details of  the $\Delta$-excitation mechanism 
and medium effects on the pion; 
the knowledge of these details is essential 
for  predicting the dominant quasi-free pion production processes.
In this paper we focus on the coherent single-pion production process. 

There have indeed been quite active experimental efforts
to investigate neutrino-induced coherent single-pion production
in the $\Delta$-excitation region.
K2K\cite{hasegawa} and SciBooNE\cite{hiraide} investigated
charged-current (CC) coherent pion production,
while MiniBooNE\cite{miniboone} studied neutral-current (NC) 
coherent pion production. 
Furthermore, results for the anti-neutrino-induced
coherent pion-production processes are expected 
to become available soon from MiniBooNE
for the NC process~\cite{anti_nc},
and from SciBooNE
for the CC process~\cite{anti_cc}.
It is to be remarked, however, 
that the recent experimental results offer
a rather puzzling situation.
The experiments at K2K\cite{hasegawa} and SciBooNE\cite{hiraide} 
report that the CC process is not observed, whereas
the MiniBooNE experiment~\cite{miniboone} concludes
that the NC process is observed.
Now, from the isospin factors,
we expect an approximate relation
$\sigma_{CC}\sim 2\sigma_{NC}$.
Although the muon mass can reduce the phase space for
the CC process at low energies, 
we still expect that $\sigma_{CC}$ should 
be of a significant size compared with $\sigma_{NC}$, and hence 
the above experimental results seem quite puzzling.
In this connection it is to be noted that 
the MiniBooNE's use of the Rein-Sehgal (RS) model~\cite{RS} 
in analyzing the NC data has recently be questioned~\cite{amaro}:
for a critical review of the RS model, see Refs.~\cite{amaro,hernandez}.
The CC data analyses in Refs.\cite{hasegawa,hiraide}
did not rely on a particular theoretical model 
for coherent pion production itself but,
in dealing with some other neutrino-nucleus reactions
that entered into the analyses,
certain models whose reliability was open to debate
needed to be invoked.

The theoretical treatment of coherent pion production
can be categorized into two types: 
a PCAC-based model and a microscopic model.
In the former approach, the hadronic matrix element 
for neutrino-induced pion production is related to the
pion-nucleus (or pion-nucleon) scattering amplitude through 
the PCAC relation.
Meanwhile, in the microscopic approach,
the hadronic matrix element is calculated by summing
the elementary amplitude for weak pion-production 
off a single nucleon embedded in a nuclear environment.

A prominent example of the PCAC-based approach
is the model due to Rein and Sehgal (RS model)~\cite{RS}. 
Because of its success in the high energy neutrino process~\cite{rs_success}
($E_\nu$\gtap 2~GeV,  where $E_\nu$ is the incident neutrino energy)
and its simplicity, the RS model has been extensively used
in analyzing data in neutrino-oscillation experiments. 
Several authors, however, have recently pointed out that
the RS model does not give a reasonable description for
relatively low-energy neutrino processes 
($E_\nu$\ltap 2~GeV)\cite{amaro,hernandez}, and that 
the use of the RS model may have led to 
the puzzling experimental situation currently facing us.
There have been several proposals~\cite{hernandez,RS2,paschos,bs_pcac}
to remedy some of the 
possible insufficiencies
in the original RS model.

Meanwhile, 
in order to build a quantitatively reliable microscopic approach, 
it is obviously of primary importance to start with a model
that can describe with sufficient accuracy
electroweak pion production off a free single nucleon. 
Furthermore, for pion production off a nuclear target, 
we need to consider medium effects
such as the final-state interactions (FSI) between 
the outgoing pion and nucleus, etc.
Recently there have been several microscopic 
calculations\cite{amaro,singh,valencia1,valencia2},
the most elaborate one being that by Amaro et al.~\cite{amaro}.
These calculations differ in the way the elementary process 
($\nu_\mu N\to \mu^+ N \pi$) is modeled and/or in the way
the medium effects are taken into account.
For example, only the resonant $\Delta$-excitation mechanism is considered
in Refs.~\cite{singh,valencia1}, while the non-resonant mechanism is additionally
considered in Refs.~\cite{amaro,valencia2}.
It was shown in Refs.~\cite{amaro,valencia2} that the inclusion of the
non-resonant mechanism leads to a reduction of the cross section by a
factor of $\sim 2$, even though both models are constructed 
in such a manner that the data for the elementary process 
are reproduced fairly well.
\footnote{
Unfortunately, conclusive data for the elementary neutrino process are
still lacking, which leads to theoretical uncertainty.
}
This result indicates the importance of modelling the elementary
process with a sound and systematic approach which has been extensively
tested by available data.

The purpose of the present article is to develop an alternative
microscopic model for coherent pion production.
An important ingredient of our formalism 
is a reliable dynamical model for the elementary process,
and for that we shall employ the Sato-Lee (SL) model~\cite{SL,SUL}.
The SL model was first developed as a systematic framework
for studying the resonance properties 
by analyzing data on pion production in photon (electron)-nucleon 
scattering in the $\Delta$-resonance region~\cite{SL,SL2}.
The SL model treats the resonant and non-resonant mechanisms 
on the same footing, and is known to provide
a reasonably accurate description 
of an extensive set of pion production data.
The SL model was further extended to the weak sector
in Ref.~\cite{SUL}, and was
shown to be able to reproduce data for neutrino-induced pion production
off a nucleon.
As has been done in the previous microscopic calculations, 
we also need to incorporate the nuclear medium effects.
In the energy region of our interest, the $\Delta$-hole approach has
proved to be successful in describing various processes 
involving pion-nucleus dynamics.
These situations motivate us to develop a model for coherent pion production by 
combining the SL model and the $\Delta$-hole model,
and this is what we attempt in this article.
We shall limit ourselves here to a case
where the target nucleus (and hence the final nucleus also)
has spin 0,
and employ a simplified $\Delta$-hole model
proposed in Ref.~\cite{karaoglu}.
As for concrete numerical calculations,
we concentrate on the $^{12}$C target, 
which has been and will continue to be an important nuclear target
in many of neutrino-oscillation experiments.
To test the reliability of our approach, 
we first calculate observables for coherent photo-pion production 
on $^{12}$C using the same theoretical framework
and show that the calculated results agree well with data.
We then proceed to calculate observables for coherent neutrino-pion
production on $^{12}$C
and present numerical results that can be compared with 
the recent data from K2K and SciBooNE.
We shall also present theoretical predictions
for those quantities for which experimental data will soon become available.

The fact that the previous microscopic calculations
exhibit rather large model-dependence
makes it particularly interesting to use the SL model,
which has been highly successful in the single nucleon sector. 
The SL model provides a consistent set of amplitudes 
for pion production and pion-nucleon scattering 
on a single nucleon;
all these amplitudes are obtained in a systematic manner
from the same Lagrangian.
In our approach this consistency can be further taken over
to the description of the FSI between the final pion and the nucleus.
Thus, based on the SL amplitudes,
we can construct a pion-nucleus optical potential 
that is consistent with the transition operators
for electroweak pion production off a nucleus.
To the best of our knowledge, our approach
is the first to provide a consistent framework 
for treating the medium effect on the pion 
and electroweak pion production on the same footing.
This point is worth emphasizing because it is this consistency that
enables us to {\it predict} cross sections 
for electroweak coherent pion production 
{\it with no adjustable parameters},
once we fix certain parameters (see below) 
relevant to medium effects by fitting to
the pion-nucleus scattering data.

Another point to be noted is 
that our model takes into account the non-local effect
for in-medium $\Delta$-propagation.
For neutrino-induced coherent pion production, 
neither the RS-based nor previous microscopic models
have included this effect.
As pointed out in Ref.~\cite{non-local},
the non-local effect could reduce the cross section by a factor of $\sim$ 2
($\sim$ 1.7) for $E_\nu$ = 0.5 (1) GeV.
We consider it important to take due account
of the possibly large non-local effect.

Our calculation adopts the following procedure.
We first construct a pion-nucleus optical potential,
employing the SL $\pi N$ scattering (on- and off-shell) amplitudes 
as basic ingredient.
The medium modification of the $\Delta$-propagation in a nucleus 
is considered with the use of the $\Delta$-hole model~\cite{karaoglu}.
All the free parameters in our model (spreading potential, phenomenological
terms in the optical potential) are fixed by
fitting to pion-nucleus scattering data.
After these parameters are determined,
we are in a position to make prediction 
on the coherent pion production process.
Before calculating the neutrino-induced process, we test the reliability 
of our model by comparing our predictions for the photo-induced process with data.
After finding satisfactory results for the photo-process, 
we proceed to calculate neutrino-induced coherent pion production.

The organization of this paper is as follows.
Sec.~\ref{sec_model} is dedicated 
to the explanation of our approach.
We first introduce the elementary amplitudes of the SL model.
We then give expressions for calculating the
electroweak coherent pion production amplitudes in
terms of the SL amplitudes and derive the cross section formulae.
The expression for the constructed optical potential and its
relation with the scattering amplitude are also given there.
We present numerical results in Sec.~\ref{sec_result}, and give
a conclusion in Sec.~\ref{sec_conclusion}.
Appendix \ref{app_mutlipole}
provides the definition of the multipole amplitudes,
while Appendix \ref{app_2cm} explains 
the Lorentz transformation used in our calculation.
In Appendix \ref{app_misc} 
we give expressions for quantities 
that appear in the $\Delta$-hole model.

\section{Formulation}
\label{sec_model}

The kinematics of the reactions
under consideration is as follows.
We consider coherent pion production in 
neutrino($\nu_\ell$)-nucleus($t$) scattering: 
$\nu_\ell (p_\nu)+ t (p_t) \to \ell^- (p_\ell^\prime) +
\pi^+ (k) + t (p_t^\prime)$ for the CC process, and
$\nu_\ell (p_\nu)+ t (p_t) \to \nu_\ell (p_\ell^\prime) +
\pi^0 (k) + t (p_t^\prime)$ for the NC process;
we also consider the antineutrino-counterparts.
The four-momentum for each particle 
in the laboratory frame (LAB)
is given in the parentheses.
The four-momentum transfer from the leptons is denoted by
$q^\mu\equiv p_\nu^\mu-p_\ell^{\prime\,\mu}$.
We choose a right-handed coordinate system in which the $z$-axis lies
along the incident neutrino momentum $\bm{p}_\nu$, 
and the $y$-axis is taken along 
$\bm{p}_\nu\times\bm{p}_\ell^\prime$.
In evaluating a nuclear matrix element, it is convenient to work in 
the pion-nucleus center-of-mass frame (ACM).
The kinematical variables in ACM are denoted by
$\bm{q}_A$, $\bm{k}_A$, etc. 
We also work in the pion-{\it nucleon} CM frame (2CM),
when calculating the elementary SL amplitudes.
The kinematical variables in 2CM are denoted by
$\bm{q}_2$, $\bm{k}_2$, etc. 
When working in ACM (2CM), we choose a coordinate system in
which the $z$-axis lies along $\bm{q}_A$ ($\bm{q}_2$) and
the $y$-axis is along $\bm{p}_{\nu,A}\times\bm{p}_{\ell,A}^\prime$
($\bm{p}_{\nu,2}\times\bm{p}_{\ell,2}^\prime$).

\subsection{The SL Model}

We express nuclear transition amplitudes for coherent pion production
in terms of the elementary amplitudes derived from the SL model~\cite{SUL}. 
In this section, therefore, we introduce the SL amplitudes.
The differential cross section in the LAB frame
for pion production in the
neutrino-{\it nucleon} CC reaction, 
$\nu_\ell (p_\nu)+ N (p_N) \to \ell^- (p_\ell^\prime) +
\pi^+ (k) + N (p_N^\prime)$,
is given by (cf. Eq.~(10) of Ref.~\cite{SUL})
\begin{eqnarray}
\eqn{xs1}
\frac{d^5\sigma}{d E_\ell^\prime d\Omega_\ell^\prime d\Omega_\pi}
=\frac{G_F^2\cos^2\theta_c}{2}
\left({ {|\bm{k}| \over \omega_\pi} +
 {|\bm{k}| - {\hat{\bm{k}}\cdot (\bm{p}_\nu -
 \bm{p}_\ell^\prime) }\over E_N^\prime}}\right)^{-1}
\frac{|\bm{p}_\ell^\prime|}{|\bm{p}_\nu|}
{|\bm{k}|^2 m_N^2 \over \omega_\pi E_N E_N^\prime}
 \frac{L^{\mu\nu}W_{\mu\nu}}{(2\pi)^5} \ ,
\end{eqnarray}
where $G_F=1.16637 \times 10^{-5}$ GeV$^{-2}$ is the Fermi constant, and
$\theta_c$ is the Cabbibo angle ($\cos\theta_c=0.974$).
$E_\ell^\prime$ and $\omega_\pi$ are the energies of the final lepton
and pion, respectively, 
$m_N$ is the nucleon mass, and 
$E_N$ ($E_N^\prime$) is the initial (final) nucleon energy.
$W_{\mu\nu}$ and $L^{\mu\nu}$ represent the hadron and
the lepton tensors, respectively, and their definitions are found in
Ref.~\cite{SUL} [Eqs.~(11), (12)]. 
The above cross section can be written as
\begin{eqnarray}
\eqn{xs1-2}
\frac{d^5\sigma}{d E_\ell^\prime d\Omega_\ell^\prime d\Omega_\pi}
&=&\frac{G_F^2\cos^2\theta_c}{2}
\left({ {|\bm{k}| \over \omega_\pi} +
 {|\bm{k}| - {\hat{\bm{k}}\cdot (\bm{p}_\nu -
 \bm{p}_\ell^\prime) }\over E_N^\prime}}\right)^{-1}
\frac{|\bm{p}_\ell^\prime||\bm{k}|^2}{(2\pi)^5|\bm{p}_\nu|}
{E_{\ell,2}^\prime p_{\nu,2}}
\\\nonumber
&\times& {1\over 2}\sum_{s_N
 s_N^\prime}\sum_{s_\ell^\prime} |\Gamma_{2L} (F^V-F^A)|^2 \ ,
\end{eqnarray}
where $F^V$ and $F^A$ are the transition amplitudes in which 
the hadronic vector and the axial-vector currents are respectively
contracted with the leptonic current.
The symbol $s_N$ ($s_N^\prime$) is the z-component of the initial
(final) nucleon spin, while $s_\ell^\prime$ denotes the final lepton
spin.
The energies of the final lepton and the initial neutrino in 2CM are
denoted by $E_{\ell,2}^\prime$ and $p_{\nu,2}$, respectively.
$F^V$ and $F^A$, including both hadronic and lepton currents,
are calculated in 2CM, and then
embedded in the cross section expression given in LAB.
The factor $\Gamma_{2L}$ arises from the relevant Lorentz transformation
(see Appendix B):
\begin{eqnarray}
\Gamma_{2L} = \sqrt{\omega_{\pi,2}E_{N,2}E^\prime_{N,2}\over
\omega_{\pi}E_{N}E^\prime_{N}} \ ,
\end{eqnarray}
where $\omega_{\pi,2}$, $E_{N,2}$ and $E^\prime_{N,2}$ are the
energies of the pion, the incident nucleon and the final nucleon in
2CM. 
The spin structure of $F^V$ and $F^A$ can be parametrized as
\begin{eqnarray}
F^V  =   - i\vec{\sigma}\cdot \vec{\epsilon}_\perp F_1^V
 -  \vec{\sigma}\cdot \hat{k}_2 \vec{\sigma}\cdot\hat{q}_2\times\vec{\epsilon}_\perp
                                                   F_2^V
 - i\vec{\sigma}\cdot\hat{q}_2\hat{k}_2\cdot\vec{\epsilon}_\perp F_3^V
 - i\vec{\sigma}\cdot\hat{k}_2\hat{k}_2\cdot\vec{\epsilon}_\perp F_4^V
 \nonumber \\
 - i\vec{\sigma}\cdot\hat{q}_2\hat{q}_2\cdot\vec{\epsilon} F_5^V
 - i\vec{\sigma}\cdot\hat{k}_2\hat{q}_2\cdot\vec{\epsilon} F_6^V
 + i \vec{\sigma}\cdot\hat{k}_2 \epsilon_0 F_7^V 
 + i \vec{\sigma}\cdot\hat{q}_2\epsilon_0 F_8^V\,,
  \eqn{fvec}
\end{eqnarray}
where $\vec{\epsilon}_\perp = \hat{q}_2\times(\vec{\epsilon} \times \hat{q}_2)$
and
\begin{eqnarray}
F^A  =   - i\vec{\sigma}\cdot\hat{k}_2\vec{\sigma}\cdot \vec{\epsilon}_\perp F_1^A
 -  \vec{\sigma}\cdot\hat{q}_2\times\vec{\epsilon}_\perp
                                                   F_2^A
 - i\vec{\sigma}\cdot\hat{k}_2
    \vec{\sigma}\cdot\hat{q}_2\hat{k}_2\cdot\vec{\epsilon}_\perp F_3^A
 - i\hat{k}_2\cdot\vec{\epsilon}_\perp F_4^A
 \nonumber \\ 
 - i\vec{\sigma}\cdot\hat{k}_2
    \vec{\sigma}\cdot\hat{q}_2\hat{q}_2\cdot\vec{\epsilon} F_5^A
 - i\hat{q}_2\cdot\vec{\epsilon} F_6^A
 + i \epsilon_0  F_7^A
 + i \vec{\sigma}\cdot\hat{k}_2\vec{\sigma}\cdot\hat{q}_2 \epsilon_0 F_8^A\,.
  \eqn{faxi}
\end{eqnarray}
The lepton-current matrix element $\epsilon^\mu$ is 
given by 
$\epsilon^\mu=\bra{\ell}\bar{\psi}_l\gamma^\mu(1-\gamma_5)\psi_\nu\ket{\nu_\ell}$.
We have introduced parametrization for $F^A$ simply via
$F^A=\vec{\sigma}\cdot\hat{k}_2F^V$.
The amplitudes, $F_i^V$ and $F_i^A$, are expressed
in terms of the multipole amplitudes $E_{l\pm}^{V,A},M_{l\pm}^{V,A},
S_{l\pm}^{V,A}$ and $L_{l\pm}^A$, which are
functions of $q^2$ and $W$ (the $\pi N$ invariant mass) 
and computed in 2CM.
Their explicit expressions are presented in Appendix \ref{app_mutlipole}.

In a coherent process on a spin-zero target
under consideration, only the spin non-flip terms of the transition
amplitudes contribute. 
We therefore can work with $\bar{F}^{V(A)}$ defined by
\begin{eqnarray}
\eqn{non-flip}
\bar{F}^{V(A)} = {1\over 2} {\rm Tr} [F^{V(A)}] \ ,
\end{eqnarray}
where the trace is taken for nucleon spin space.
Their explicit forms are
\begin{eqnarray}
\eqn{bar_fv}
\bar{F}^V  =   
 -  \hat{k}_2 \cdot\hat{q}_2\times\vec{\epsilon}_\perp  F_2^V \ ,
\end{eqnarray}
and
\begin{eqnarray}
\eqn{bar_fa}
\bar{F}^A  &=&   - i\hat{k}_2\cdot \vec{\epsilon}_\perp F_1^A
 - i\hat{k}_2\cdot\hat{q}_2
    \hat{k}_2\cdot\vec{\epsilon}_\perp F_3^A
 - i\hat{k}_2\cdot\vec{\epsilon}_\perp F_4^A \nonumber \\
&& - i\hat{k}_2\cdot\hat{q}_2
    \hat{q}_2\cdot\vec{\epsilon} F_5^A
 - i\hat{q}_2\cdot\vec{\epsilon} F_6^A
 + i \epsilon_0  F_7^A
 + i \hat{k}_2\cdot\hat{q}_2 \epsilon_0 F_8^A\ .
\end{eqnarray}
In particular, the resonant parts of the elementary amplitudes 
are given by
\begin{eqnarray}
\bar{F}^V_R  - \bar{F}^A_R 
 &=&   \left( -2 \hat{k}_2\cdot\hat{q}_2\times\vec{\epsilon}_\perp M_{R\ 1+}^{(3/2),V}
-2 i \hat{k}_2\cdot \vec{\epsilon}_\perp E_{R\ 1+}^{(3/2),A}\right.\\\nonumber
&&\left. -4 i \hat{k}_2\cdot\hat{q}_2 \epsilon_0 S_{R\ 1+}^{(3/2),A}
+4 i \hat{k}_2\cdot\hat{q}_2 \hat{q}_2\cdot\vec{\epsilon} L_{R\ 1+}^{(3/2),A}
\right)\Lambda^{3/2}_{ij} \ ,
\end{eqnarray}
where the suffix ``$R$'' stands for the resonant parts
of the corresponding multipole amplitudes associated with the excitation
of the $\Delta$ resonance.
From the resonant amplitude 
we can factor out
the $\Delta$-propagator, $D(W)$, as 
\begin{eqnarray}
\eqn{amp-res}
\bar{F}^V_R  - \bar{F}^A_R = {N(k_2,q_2)\over D(W)} \ ,
\end{eqnarray}
and
\begin{eqnarray}
\eqn{delta-prop}
D(W)=W-m_\Delta-\Sigma_\Delta(W) \ ,
\end{eqnarray}
where 
$m_\Delta$ and $\Sigma_\Delta$ are the bare mass and
self energy of the $\Delta$-resonance, respectively.

We next discuss the T-matrix element for $\pi N$ scattering, 
which serves as an input 
for constructing an optical potential for pion-nucleus scattering.
A calculational procedure for the $\pi N$ T-matrix within the SL model
can be found in Ref.~\cite{SL}.
A distorted wave obtained with this optical potential 
will be used to take account of the final-state interaction 
in coherent pion production.
The T-matrix is decomposed into the resonant ($t_R$) and 
non-resonant ($t_{nr}$) parts as
\begin{eqnarray}
\eqn{t-mat}
t^{(c)}_{\pi N} = t^{(c)}_R + t^{(c)}_{nr} \ ,
\end{eqnarray}
where the superfix $c$ specifies a channel;
in our model the resonance amplitude exists
only for the $P_{33}$ channel.
The on-shell component of the T-matrix given in \Eq{t-mat} is related to
the phase shift by
\begin{eqnarray}
\eqn{phase}
t^{(c)}_{\pi N} = - {W\over \pi \omega_{\pi,2} E_{N,2}}
{e^{2i\delta^{(c)}}-1\over 2i k_2^o} \ ,
\end{eqnarray}
where $W$ is the invariant mass of the $\pi N$ system, and
$\omega_{\pi,2}$ =$\sqrt{k_2^{o\,2}+m_\pi^2}$ and 
$E_{N,2}=\sqrt{k_2^{o\,2}+m_N^2}$ are the on-shell energies of the pion
and the nucleon in 2CM, respectively. 
The resonant amplitude is expressed as
\begin{eqnarray}
\eqn{t-res}
t^{(P_{33})}_R (k_2^\prime,k_2; W) = - {F_{\pi N\Delta}(k_2^\prime)
F_{\pi N\Delta}(k_2)\over D(W)} \ ,
\end{eqnarray}
where $F_{\pi N\Delta}(k_2)$ is the dressed $\pi N\Delta$ vertex, and
$D(W)$ is the $\Delta$ propagator introduced in \Eq{delta-prop}.
We note that the four-momenta, $k_2$ and $k_2^\prime$,
are in general off-energy-shell.

\subsection{Coherent pion production in neutrino-nucleus scattering}

Similarly to \Eq{t-mat} for the $\pi N$ scattering amplitude, the
weak amplitudes $\bar{F}^{V(A)}$, defined in \Eqss{non-flip}{bar_fa}, also have a
resonant $\bar{F}^{V(A)}_{R}$ and non-resonant $\bar{F}^{V(A)}_{nr}$ parts.
 Accordingly, the transition amplitudes
of coherent pion production on nuclei have the resonant and non-resonant
parts. We now describe how these two components are calculated in our
approach. 

\subsubsection{transition matrix element: resonant part}
\label{sec_res}

The main task in calculating the resonant part of
coherent pion production on nuclei is to account for 
the medium effects on $\Delta$ propagation in the elementary resonant
amplitudes
$F^{V(A)}_R$. Here we follow the
procedure of the $\Delta$-hole model of  
pion-nucleus reactions by modifying
the $\Delta$ propagator in \Eq{delta-prop}.
Thus it is useful to first briefly
 explain how the $\Delta$-hole model is formulated
by considering the elastic pion-nucleus scattering;
for a full account of the formulation see
Refs.~\cite{annal99,annal108,annal120,taniguchi}.

The $\Delta$-hole model is formulated within the projection operator
formalism\cite{annal99}. 
The nuclear Fock space is divided into four spaces; 
$P_0$, $P_1$, $D$ and $Q$.
The $P_0$-space is spanned by the pion and the nuclear ground state,
the $P_1$-space  by the pion and one-particle one-hole states,
the $D$-space by the one-$\Delta$ one-hole configurations,
and $Q=1-P_0-P_1-D$ contains the reminder of the full space.
A projected Hamiltonian is written as, e.g., $H_{P_0D}= P_0 H D$.
Starting with the Schr\"odinger equation in the full space 
($H\ket{\Psi} = E\ket{\Psi}$), 
we can apply the standard projection operator techniques\cite{annal99}
to obtain an equation, defined only in the $P_0$-space, to
describe 
 the pion-nucleus elastic scattering T-matrix. In the $\Delta$-hole model, one 
further imposes the condition that the  $D$-space is the doorway
of the transitions between $P=P_0+P_1$  and  $Q$ spaces; namely
$H_{PQ}=H_{QP}=0$. The  pion-nucleus scattering amplitude due to the
$\Delta$ excitation can then be written as
\begin{eqnarray}
\eqn{elastic_amp}
 T_{P_0P_0}(E) = H_{P_0D} G_{\Delta h}(E) H_{DP_0} \ ,
\end{eqnarray} 
where the total energy defined in ACM ($E + A m_N$) is
given by
\begin{eqnarray}
E + A m_N = q_A^0 + \sqrt{\bm{q}_A^2 + (Am_N)^2}
= \sqrt{\bm{k}_A^2 + m_\pi^2} + \sqrt{\bm{k}_A^2 + (Am_N)^2} \ ,
\end{eqnarray}
where $A$ is the mass number.
The $\Delta$-hole propagator $G_{\Delta h}$ in \Eq{elastic_amp}
is defined by
\begin{eqnarray}
\eqn{nuclear-delta-prop}
G_{\Delta h}^{-1} = 
D(E-H_\Delta)-W_{el} - \Sigma_{\rm pauli} - \Sigma_{\rm spr} \ .
\end{eqnarray}
Here $D(E-H_\Delta)$ can be calculated from \Eq{delta-prop} with
$H_\Delta$ being the Hamiltonian for the $\Delta$-particle in the
nuclear many-body system. 
The effects due to the $Q$-space are included in
the so-called spreading potential,
$\Sigma_{\rm spr}$. 
A microscopic calculation of the spreading potential is very complicated
since it involves the calculation of pion absorption by two or more nucleons.
It is therefore a common practice 
to determine $\Sigma_{\rm spr}$ phenomenologically
by fitting to the pion-nucleus scattering data.
Excitations to the $P_1$-space are included in the $\Delta$ self energy
$\Sigma_\Delta(W)$ of $D(E-H_\Delta)$ [see \Eq{delta-prop}]
with a correction due to  the Pauli blocking ($\Sigma_{\rm pauli}$).
De-excitation to the $P_0$-space
is the rescattering in the elastic mode,
and is denoted by $W_{el}$.
In our actual calculation, we expand $G_{\Delta h}$ in term of
$W_{el}$, and the expansion series is resummed by solving the
Lippmann-Schwinger equation.

The calculations of the pion-nucleus
scattering amplitude in \Eq{elastic_amp} require a diagonalization of the
$\Delta$-hole propagator $G_{\Delta h}$ of \Eq{nuclear-delta-prop}.
For the diagonalization,
it is practically convenient to work with the oscillator basis for 
the $\Delta$ state, defined by the Hamiltonian $H_\Delta$,
and the nucleon hole state. 
This diagonalization is a difficult numerical task.
Although an efficient method using  the doorway state expansion has
been developed~\cite{annal120}, the diagonalization of $G_{\Delta h}$ is
still difficult, particularly for heavier nuclei.
In Ref.~\cite{karaoglu}, Karaoglu and Moniz (KM) proposed a simplified
calculation with the $\Delta$-hole model in which 
$G_{\Delta h}$ is calculated
with a local density approximation rather than a
diagonalization.
In their simplified treatment, $\Sigma_{\rm pauli}$ is calculated by a
nuclear matter calculation\cite{pauli}, and their result is
given in Appendix \ref{app_misc}.
Their parametrization of the spreading potential $\Sigma_{\rm spr}$
in terms of a central and a spin-orbit terms
 are also given in Appendix \ref{app_misc}.
Each term of the spreading potential has a complex strength, 
which are determined by fitting to the pion-nucleus scattering data.
KM applied their approach to $\pi$-$^{16}$O scattering, and
found a good agreement between their calculation with data,
and also with the full $\Delta$-hole calculation
\cite{annal108,annal120} except for the most central partial waves.
Encouraged by this success,
we follow this simplified version of the $\Delta$-hole model to include the
medium effects on the $\Delta$ propagation in defining the
electroweak pion production matrix elements.

Schematically, the resonant part of the
transition matrix element, ${\cal M}^A_R$, of 
weak coherent pion production
on nuclei induced by the charged current can be obtained by replacing the 
initial $H_{DP_0}$ of \Eq{elastic_amp}
by $H_{DP_0^\prime}$ where $P_0^\prime$ is the space spanned by the
(axial-)vector current and the nucleus in the ground state.
In terms of the single particle wave functions $\psi_j(\bm{p}_N)$
of the nucleons in the initial and final nuclear states, 
we thus have\footnote{
In Ref.~\cite{koch_moniz}, the authors carried out a 
calculation for photon-induced coherent pion production
by diagonalizing $G_{\Delta h}$.
}
\begin{eqnarray}
\eqn{mr2}
{\cal M}^A_{R} &=& \sum_{j} 
\int {d^3p_N \over (2\pi)^3}{d^3p_N^\prime \over (2\pi)^3}
\psi_j^*(\bm{p}_N^\prime) 
{\Gamma_{2A}
N(k_2,q_2) 
(2\pi)^3 \delta(\bm{p}_N+\bm{q}_A-\bm{p}_N^\prime-\bm{k}_A)\over
D(E+m_N-H_\Delta) - \Sigma_{\rm pauli} - \Sigma_{\rm spr} }
\psi_j(\bm{p}_N) \nonumber \\
&=& \sum_{j} 
\int {d^3p_\Delta \over (2\pi)^3}
\psi_j^*(\bm{p}_N^\prime) 
{\Gamma_{2A}
N(k_2,q_2) 
\over D(E+m_N-H_\Delta) - \Sigma_{\rm pauli} - \Sigma_{\rm spr} }
\psi_j(\bm{p}_N) \ ,
\end{eqnarray}
where 
$\bm{p}_\Delta=\bm{p}_N+\bm{q}_A=\bm{p}_N^\prime+\bm{k}_A$;
the index $j$ denotes single particle quantum numbers
including the isospin .
The summation ($\sum_j$) is taken over the occupied states
of the nucleus.
The factor $\Gamma_{2A}$ is defined by
\begin{eqnarray}
\Gamma_{2A} = \sqrt{\omega_{\pi,2}E_{N,2}E^\prime_{N,2}\over
\omega_{\pi,A}E_{N,A}E^\prime_{N,A}} \ ,\label{eq:G2A}
\end{eqnarray}
where $\omega_{\pi}$, $E_{N}$ and $E^\prime_{N}$ are the
energies of the pion, the incoming nucleon and 
the outgoing nucleon, respectively,
and the quantities in the numerator (denominator)
refer to 2CM (ACM). 
This factor arises from the fact that
$\bar{F}^{V(A)}_R$ computed in 2CM are to be
embedded in ${\cal M}^A_R$ evaluated in ACM.
To evaluate the numerator in the integrand of \Eq{mr2},
we clearly need a prescription 
for relating variables in 2CM to those in ACM.
Here we use the commonly used 
prescription~\cite{gmitro,chumbalov} 
to fix the nucleon momenta with the lepton momentum transfer
$\bm{q}_A$ and outgoing pion momentum $\bm{k}_A$ as
\begin{eqnarray}
\eqn{p_fix}
\bm{p}_N = - {\bm{q}_A\over A} - {A-1\over 2A}(\bm{q}_A-\bm{k}_A)\ ,
\qquad
\bm{p}_N^\prime = - {\bm{k}_A\over A} + {A-1\over 2A}(\bm{q}_A-\bm{k}_A)
\ ,
\end{eqnarray}
and write the $\pi N$ invariant mass as
\begin{eqnarray}
\eqn{inv_mass}
W = \sqrt{(E_{N\,A}+q_A^0)^2 - (\bm{p}_N+\bm{q}_A)^2
}\ ,
\end{eqnarray}
with $E_{N\,A}=\sqrt{\bm{p}_N^2+m_N^2}$.
Having specified all the relevant variables in ACM, we can derive the
corresponding variables in 2CM via a Lorentz transformation to
obtain $N(k_2,q_2)$ of \Eq{mr2}.
For more details about this Lorentz transformation
(including the discussion of a somewhat different treatment 
of an off-shell pion momentum),
see Appendix \ref{app_2cm}.  
Note that, in treating the wave functions,
$\psi(\bm{p}_N)$ and $\psi(\bm{p}'_N)$, and
the $\Delta$ kinetic term in the denominator
in the integrand of \Eq{mr2},
we do {\it not } use the prescription given in \Eqs{p_fix}{inv_mass};
thus the important recoil effects on 
$\Delta$-propagation are not neglected in our
calculations. 

We incorporate the recoil effect 
on the $\Delta$ self-energy in the first order approximation.
This is done by linearizing the $\Delta$-propagator
with the following expansion\cite{taniguchi}:
\begin{eqnarray}
\eqn{linearize}
D(E+m_N-H_\Delta) &\sim& D(W)-\gamma(W) (H_\Delta - e_\Delta^0) , \\
\eqn{e_delta}
E+m_N &=& W + e_\Delta^0 \ , \\
\gamma(W) &=& \partial D(W) / \partial W , \\
\eqn{delta_H}
 H_\Delta &=& {\bm{p}_\Delta^2\over 2 \mu_\Delta} + V_\Delta + V_\Delta^C +
  e_N , \\
1/\mu_\Delta &=& 1/m_\Delta + 1/(A-1)m_N \ ,
\end{eqnarray}
where $V_\Delta$ ($V_\Delta^C$) is the $\Delta$ (Coulomb) potential in
the nucleus, and $e_N$ is the hole energy.
The $\Delta$ potential is taken to be the same as that for the nucleon;
its explicit expression is given in Appendix \ref{app_misc}.
\Equation{e_delta} defines $e_\Delta^0$.
To carry out the integration over the $\Delta$ momentum $\bm{p}_\Delta$ in
\Eq{mr2}, we express the nucleon wave function $\psi_j(\bm{p})$ in terms of
its  coordinate-space form 
$\phi_j(\bm{r})$.
We note that with the prescription in \Eqs{p_fix}{inv_mass}, the
numerator $N(k_2,q_2)$ of \Eq{mr2} is independent of the
variable $\bm{p}_\Delta$ and can be factorized out of the integration.
With this factorization approximation
and  with the use of the linearized form in \Eq{linearize},
the integration over $\bm{p}_\Delta$ 
leads to the following $r$-space
expression:
\begin{eqnarray}
\eqn{mr3}
{\cal M}^A_{R} &=& - \left(\mu_\Delta \Gamma_{2A} N(k_2,q_2)\over 2\pi\gamma\right)
\sum_{j} 
\int d^3r d^3r^\prime \phi_j^*(\bm{r}^\prime)
e^{-i \bm{k}_A\cdot\bm{r}^\prime}
{e^{i K_\Delta |\bm{r}^\prime-\bm{r}|}
\over |\bm{r}^\prime-\bm{r}|
}
e^{i \bm{q}_A\cdot\bm{r}}\phi_j(\bm{r}) \ ,
\end{eqnarray}
where
\begin{eqnarray}
\eqn{delta_k2}
K_\Delta^2 = {2\mu_\Delta\over \gamma}\left\{
W  -m_\Delta-\Sigma_\Delta(W) + \gamma(E-W+m_N)
- \gamma \left[e_N + V_\Delta + V_\Delta^C \right]
- \Sigma_{\rm pauli} - \Sigma_{\rm spr} 
 \right\} \ .
\end{eqnarray}
Following the procedure described in Ref.~\cite{karaoglu} 
[see Eqs.~(25)--(39) therein],
and subsequently applying the Lorentz transformation 
from ACM to LAB, we obtain the following expression for
the transition matrix element ${\cal M}^L_{R}$ in LAB
\begin{eqnarray}
\eqn{mr4}
&&{\cal M}^L_{R} = {16\sqrt{1+|\lambda|}\pi \over 3} {\mu_\Delta D(W)\over\gamma}
\Gamma_{2L}
\\\nonumber &\times&
\left( -2 \hat{k}_2\cdot\hat{q}_2\times\vec{\epsilon}_\perp M_{R\ 1+}^{(3/2),V}
-2 i \hat{k}_2\cdot \vec{\epsilon}_\perp E_{R\ 1+}^{(3/2),A}
-4 i \hat{k}_2\cdot\hat{q}_2 \epsilon_0 S_{R\ 1+}^{(3/2),A}
+4 i \hat{k}_2\cdot\hat{q}_2 \hat{q}_2\cdot\vec{\epsilon} L_{R\ 1+}^{(3/2),A}
\right)\\\nonumber
&\times& \sum_{N=p,n} (1+{\lambda\tau_N\over 2})
 \int s^2 ds R^2 dR j_0(pR)j_0(Ps)
{e^{i \bar{K}_\Delta s}\over s}
\left\{1 + {i \mu_\Delta s\over \bar{K}_\Delta}
\left[\bar{e_N} + H_N\right]\right\}
\rho_N(R)\hat{j}_1(k_F s) \ ,\label{eq:MLR}
\end{eqnarray}
where $p=|\bm{k}_A-\bm{q}_A|$, $P=|\bm{k}_A+\bm{q}_A|/2$,
 $s=|\bm{r}^\prime-\bm{r}|$,  $R=|\bm{r}^\prime+\bm{r}|/2$,
and $\bar{K}_\Delta$ is obtained 
from $K_\Delta$ by replacing $e_N$ with its average value,
$\bar{e}_N$; we choose $\bar{e}_N = 16$~MeV.
The 2CM variables $k_2$ and $q_2$ are obtained from 
$k_A$ and $q_A$ using the Lorentz transformation as mentioned above.
The variable $\lambda$ denotes the charge state of the outgoing  pion,
 while  $\tau_N=1\, (-1)$ for $N$ = proton (neutron).
The factor $\Gamma_{2L}$ is from the Lorentz transformation from 2CM to LAB
and  is defined by
\begin{eqnarray}
\Gamma_{2L} = \sqrt{\omega_{\pi,2}E_{N,2}E^\prime_{N,2}\over
\omega_{\pi,L}E_{N,L}E^\prime_{N,L}} \ .
\end{eqnarray}
In \Eq{mr4}, $j_\ell(x)$ is the spherical Bessel function 
of order $\ell$, and  $\hat{j}_1(x) \equiv {3\over x} j_1(x)$;
$k_F$ is the Fermi momentum 
\begin{eqnarray}
\eqn{fermi_mom}
k_F^3(R) = {3\pi^2\over 2}\rho_N(R) \ .
\end{eqnarray}
The proton (neutron) matter density is denoted by $\rho_p$ ($\rho_n$),
and is normalized to the total number of protons (neutrons)
inside the target.
For the proton matter form factor
we use the empirical nuclear charge form factor~\cite{charge_formfac}
divided by the proton charge form factor~\cite{nucl_formfac}.
The neutron matter density is assumed 
to be the same as the proton matter density.

The single nucleon Hamiltonian appearing in \Eq{mr4}
is given by 
\begin{eqnarray}
H_N = - {\nabla^2_s\over 2m_N} - {\nabla^2_R\over 8m_N}
+V\left[(R^2+s^2/4)^{1/2}\right] \ ,
\end{eqnarray}
where $V$ is the single particle potential [\Eq{nucl_potential}].

To take account of the final pion-nucleus interactions, 
we convolute the matrix element ${\cal M}^L_{R}$ of \Eq{mr4}
with the pion distorted wave which is expanded in partial waves:
\begin{eqnarray}
\eqn{pi_wave2}
\chi_\lambda^* (\bm{k}_A^\prime) 
= \sum_{l_\pi m_\pi}
\chi_{\lambda\,l_\pi}^* (k_A^\prime) Y_{l_\pi m_\pi}^* (\hat{\bm{k}}_A) 
Y_{l_\pi m_\pi} (\hat{\bm{k}}_A^\prime) \ ,
\end{eqnarray}
where $k_A^\prime$ is the off-shell momentum.
We note that the pion distorted wave also depends 
on the pion charge ($\lambda$).
More details on our calculations of the  pion wave functions
are given  in Sec.~\ref{sec_opt}.

By performing the partial wave decomposition of ${\cal M}^L_{R}$
(now defined by the off-shell pion momentum by setting
$\bm{k}_A \rightarrow \bm{k}_A^\prime$) 
and using \Eq{pi_wave2},
the amplitude ${\cal M}^L_{R}$ with pion-nucleus FSI
takes the following  form: 
\begin{eqnarray}
\eqn{mr5}
{\cal M}^L_{R} &=&  \epsilon_A^\mu
\sum_{l_\pi} \left[ P_{l_\pi}^1 (x_A)\left(
\cos\phi^\pi_A I^{l_\pi\, 1}_{E\, \mu} 
-i\sin\phi^\pi_A I^{l_\pi\, 1}_{M\, \mu} \right)
\right.\\\nonumber
&&\left.
+ P_{l_\pi}^1 (x_A)\left(
\sin\phi^\pi_A I^{l_\pi\, 2}_{E\, \mu}
+i\cos\phi^\pi_A I^{l_\pi\, 2}_{M\, \mu} \right)
-2 P_{l_\pi} (x_A) I^{l_\pi\, 3}_{L\, \mu} 
+2 P_{l_\pi} (x_A) I^{l_\pi\, 0}_{S\, \mu} 
\right] \ ,
\end{eqnarray}
where $x_A=\hat{q}_A\cdot\hat{k}_A$, $\phi^\pi_A$ is the azimuthal
angle of the pion, and
 $\epsilon_A^\mu$ is the lepton current matrix element in ACM.
The associated Legendre function of degree $l_\pi$ and order 0
(1) is denoted by $P_{l_\pi}$ ($P_{l_\pi}^1$).
We have introduced the quantities $I^{l_\pi\, \nu}_{X\, \mu}$ defined by
\begin{eqnarray}
\eqn{i_mtx}
&& I^{l_\pi\, \nu}_{X\, \mu} = - i {32\sqrt{1+|\lambda|}\pi \mu_\Delta\over 3}
\int dk_A^\prime k_A^{\prime 2}\
  \chi_{\lambda\,l_\pi}^*(k_A^\prime)\
\int dx_A^\prime \Lambda^\nu_\mu\Gamma_{AL}^\chi\Gamma_{2AL}
\gamma^{-1} X_{R}\xi^X_{1 l_\pi}(x_A^\prime)
\\\nonumber
&&\times \sum_{N=p,n} (1+{\lambda\tau_N\over 2})\int s^2 ds R^2 dR 
j_0(pR)j_0(Ps)
{e^{i \bar{K}_\Delta s}\over s}
\left\{1 + {i \mu_\Delta s\over \bar{K}_\Delta}
\left[\bar{e_N} + H_N\right]\right\}\rho_N(R) \hat{j}_1(k_F s) \ ,
\end{eqnarray}
where $x_A^\prime= \hat{q}_A\!\cdot\!\hat{k}_A^\prime$, 
$x_2^\prime = \hat{q}_2\!\cdot\!\hat{k}_2^\prime$, \,
and
\begin{eqnarray}
\xi^X_{\ell l_\pi}(x_A^\prime) &=& 
\left\{ \begin{array}{ll}
{\displaystyle 2 l_\pi+1 \over\displaystyle 2 l_\pi (l_\pi+1)} 
P_{\ell}^1 (x_2^\prime)
P_{l_\pi}^1 (x_A^\prime) \ ,& \qquad (X=E, M) \\[5mm]
 {\displaystyle 2 l_\pi+1 \over\displaystyle 2}
P_{\ell} (x_2^\prime)
P_{l_\pi} (x_A^\prime) \ ,&\qquad (X=L, S) \end{array}
\right.
\end{eqnarray}
and
\begin{eqnarray}
{X_{R}\over D(W)} = E_{R\ 1+}^{(3/2),A}\ ,\  M_{R\ 1+}^{(3/2),V}\ ,\  L_{R\
 1+}^{(3/2),A}\ ,\  S_{R\ 1+}^{(3/2),A} \ ,
\end{eqnarray}
for $X=E,M,L,S$.

The Lorentz transformation factors coming
from the electroweak amplitudes ($\Gamma_{2AL}$) and
the wave function ($\Gamma^\chi$) in \Eq{mr5} are respectively
\begin{eqnarray}
\eqn{gam_3}
\Gamma_{2AL} = \sqrt{\omega^\prime_{\pi,2}E^\prime_{N,2}E^{i}_{N,2}
\over \omega^\prime_{\pi,A}E^\prime_{N,L}E^{i}_{N,L}} \ , \qquad
\Gamma^\chi = \sqrt{\omega_{\pi,A}E^{\prime\prime}_{N,A}E^{f}_{N,A}
\over \omega_{\pi,L}E^{\prime\prime}_{N,L}E^{f}_{N,L}} \ ,
\end{eqnarray}
where
$\omega^\prime_{\pi}$ is the pion energy in the intermediate state,
$E^i_{N}$ and $E^f_{N}$ are the nucleon energies in the initial and
final states while
$E^\prime_{N}$ and $E^{\prime\prime}_{N}$ are those in the 
intermediate states; in general,
$E^\prime_{N}$ and $E^{\prime\prime}_{N}$ can be different.
As before, the suffices \{$2,A,L$\} 
attached to the energies specify reference frames.
It is noted that the multipole amplitudes ($X_R^{A}$)
depend on $x_A^\prime$ because the $\pi N$ invariant mass in the
intermediate state depends on it [\Eqs{p_fix}{inv_mass}].
We also have introduced the Lorentz matrix $\Lambda^\nu_\mu$
defined by $\epsilon_2^\nu=\Lambda^\nu_\mu \epsilon_A^\mu$;
$\Lambda^\nu_\mu$ also depends on $x_A^\prime$;
the same Lorentz matrix relates $q_A$ ($k^\prime_A$ )
to $q_2$ ($k^\prime_2$ ).
A procedure for deriving the Lorentz matrix
and the transformation factors in \Eq{gam_3}  are explained in Appendix
\ref{app_2cm}.

\subsubsection{transition matrix element: non-resonant part}

We assume that there is no medium effect on the non-resonant
part, $\bar{F}^{V}_{nr}-\bar{F}^{A}_{nr}$,
of the weak pion production amplitude on a nucleon in nuclei.
Including the final pion-nucleus interactions and using
the same factorization approximation based on the choice \Eq{p_fix}
of the nucleon momenta to evaluate $\bar{F}^{V}_{nr}-\bar{F}^{A}_{nr}$,
the non-resonant coherent pion production matrix element ${\cal M}^L_{nr}$
can be written as 
\begin{eqnarray}
\eqn{mnr1}
{\cal M}^L_{nr} = \sum_{N=p,n}
\int d^3k_A^\prime \chi_\lambda^*(\bm{k}_A^\prime) 
\Gamma_{AL}^\chi\Gamma_{2AL}
F_N(\bm{k}_A^\prime-\bm{q}_A) (\bar{F}^{V,\zeta}_{nr}-\bar{F}^{A,\zeta}_{nr}) \ ,
\end{eqnarray}
where $\bar{F}^{V,\zeta}_{nr}$ ($\bar{F}^{A,\zeta}_{nr}$) is the non-resonant part of 
$\bar{F}^V$ ($\bar{F}^A$) given in \Eqs{bar_fv}{bar_fa}.
$\bar{F}^{V (A)}_{nr}$ depends on $N$ and $\lambda$ [\Eq{amp_iso}],
and the set $(N, \lambda)$ is collectively denoted by $\zeta$.
The nuclear form factor $F_N(\bm{p})$ is given by
\begin{eqnarray}
F_N(\bm{p}) = \int d^3r \rho_N (\bm{r}) e^{i \bm{p}\cdot \bm{r}} \ .
\end{eqnarray}
After the partial wave expansion of the pion distorted wave, 
we arrive at 
\begin{eqnarray}
\eqn{mnr2}
{\cal M}^L_{nr} &=&  \epsilon_A^\mu
\sum_{l_\pi} \left[ P_{l_\pi}^1 (x_A)\left(
\cos\phi^\pi_A J^{l_\pi\, 1}_{E\, \mu} 
-i\sin\phi^\pi_A J^{l_\pi\, 1}_{M\, \mu} \right)
\right.\\\nonumber
&&\left.
+ P_{l_\pi}^1 (x_A)\left(
\sin\phi^\pi_A J^{l_\pi\, 2}_{E\, \mu}
+i\cos\phi^\pi_A J^{l_\pi\, 2}_{M\, \mu} \right)
- P_{l_\pi} (x_A) J^{l_\pi\, 3}_{L\, \mu} 
+ P_{l_\pi} (x_A) J^{l_\pi\, 0}_{S\, \mu} 
\right] \ ,
\end{eqnarray}
where we have introduced $J^{l_\pi\, \nu}_{X\, \mu}$
defined by
\begin{eqnarray}
\eqn{j_mtx}
 J^{l_\pi\,\nu}_{X\,\mu} &=&-4\pi i \int dk_A^\prime k_A^{\prime 2}\
  \chi_{\lambda\,l_\pi}^*(k_A^\prime)\
\int dx_A^\prime \Lambda^\nu_\mu\Gamma_{AL}^\chi\Gamma_{2AL} \nonumber \\
&\times &
\sum_\ell \xi^X_{\ell l_\pi}(x_A^\prime)
\sum_{N=p,n}
X^{\ell,\zeta}_{nr}
\int r^2 dr \rho_N(r)j_0(pr) \ ,
\end{eqnarray}
for $X=E,M,L,S$.
The multipole amplitudes are included in $X^{\ell,\zeta}_{nr}$ as
\begin{eqnarray}
\eqn{x_ell}
 X^{\ell,\zeta}_{nr} &=& (\ell+1)^2 X_{nr\;\ell +}^{A,\zeta} + \ell^2 X_{nr\;\ell -}^{A,\zeta} \
  , 
\end{eqnarray}
for $X=L,S$,  and
\begin{eqnarray}
\eqn{e_ell} 
E^{\ell,\zeta}_{nr} &=& (\ell+1)E_{nr\;\ell +}^{A,\zeta} - \ell E_{nr\;\ell -}^{A,\zeta} \
  ,  \\
   \eqn{m_ell}
 M^{\ell,\zeta}_{nr} &=& (\ell+1)M_{nr\;\ell +}^{V,\zeta} + \ell M_{nr\;\ell -}^{V,\zeta} \ ,
\end{eqnarray}
for $X=E,M$.
The $\zeta$ dependence of the multipole amplitudes
is indicated explicitly.
For example, $E_{nr\;\ell +}^{A,\zeta}$ is the non-resonant
part of $E_{\ell +}^{A}$ which has been introduced previously. 
The same rule applies to the other multipole amplitudes.

\subsubsection{Cross Section}
Having written the transition amplitude for the coherent process 
in terms of the SL multipole amplitudes,
we can proceed to calculate the cross section
for the CC process. First, 
we write the transition
amplitudes in \Eqs{mr5}{mnr2} as
\begin{eqnarray}
{\cal M}^L_{R}&=&\bar{{\cal M}}^L_{R,\mu}\epsilon^\mu_A \nonumber \\
{\cal M}^L_{nr}&=&\bar{{\cal M}}^L_{nr,\mu}\epsilon^\mu_A \ . \nonumber
\end{eqnarray}
In the Laboratory frame, the differential
cross sections for 
$\nu_\ell (p_\nu)+ t (p_t) \to \ell^- (p_\ell^\prime) +
\pi^+ (k) + t (p_t^\prime)$ 
is then given by
\begin{eqnarray}
\eqn{d5s}
\frac{d^5\sigma}{d E_\ell^\prime d\Omega_\ell^\prime d\Omega_\pi}
&=&\frac{G_F^2\cos^2\theta_c}{2}
\left({ {|\bm{k}| \over \omega_\pi} +
 {|\bm{k}| - {\hat{\bm{k}}\cdot (\bm{p}_\nu -
 \bm{p}_\ell^\prime) }\over E_t^\prime}}\right)^{-1}\!\!
\frac{|\bm{p}_\ell^\prime||\bm{k}|^2}{(2\pi)^5|\bm{p}_\nu|}
{E_{\ell,A}^\prime p_{\nu,A}}\label{CrossSec}
\\\nonumber&\times& 
\sum_{s_\ell^\prime} |(\bar{{\cal M}}^L_{R,\mu} +\bar{{\cal M}}^L_{nr,\mu})
\epsilon^\mu_A |^2 \ ,
\end{eqnarray}
where $E_t^\prime \left(= \sqrt{\bm{p}^2_t + (A m_N)^2}\right)$ is the
total energy of the nucleus in the final state in LAB,
and $E_{\ell,A}^\prime$ and $p_{\nu,A} $ are the energies of the final
lepton and the initial neutrino in ACM. 
Note that the calculation of
$\sum_{s_\ell^\prime} |(\bar{{\cal M}}^L_{R,\mu} +\bar{{\cal M}}^L_{nr,\mu})
\epsilon^\mu_A |^2$ 
of  \Eq{d5s} can make use of the following property:
\begin{eqnarray}
\eqn{lepton_tensor}
L^{\mu\nu}_A \equiv  {E_{\ell,A}^\prime p_{\nu,A}\over 2} 
\sum_{s_\ell^\prime} \epsilon^\mu_A\epsilon^{\nu *}_A
= p^\mu_{\nu,A}p^{\prime\,\nu}_{\ell,A} + p^\nu_{\nu,A}p^{\prime\,\mu}_{\ell,A}
-g^{\mu\nu}p_{\nu,A}\cdot p^\prime_{\ell,A} 
\pm i\epsilon^{\mu\nu\rho\sigma}p_{\nu,A\, \rho}p^\prime_{\ell,A\, \sigma} \ ,
\end{eqnarray}
where $g^{\mu\nu}$ is the geometric tensor and
$\epsilon^{\mu\nu\rho\sigma}$ is the antisymmetric tensor with
$\epsilon^{0123}=1$. 
The plus (minus) sign in the last term 
is for the (anti-)neutrino process. 

To obtain the cross section formula
for the neutrino NC process, 
$\nu + t \to \nu + \pi^0 + t$,
we make the following changes in Eq.(\ref{CrossSec}):
Remove the Cabbibo angle.  Set the lepton mass equal to zero.
Set the pion charge index $\lambda$ (and $\zeta$) to zero in 
$I^{l_\pi\, \nu}_{X\, \mu}$ and
$J^{l_\pi\, \nu}_{X\, \mu}$ ($X=E,M,L,S$) in \Eqs{i_mtx}{j_mtx}.
(Note that the pion wave function ($\chi_{\lambda\,l_\pi}$) 
also contains $\lambda$-dependence.)
Finally, multiply the multipole amplitudes $M_{\ell +}^{(3/2,1/2),V}$
with $(1-2\sin^2\theta_W)$, 
where $\theta_W$ is the Weinberg angle ($\sin^2\theta_W = 0.23$), and
multiply $M_{\ell +}^{(0),V}$ with $(-2\sin^2\theta_W)$.

For the anti-neutrino CC process, 
the result for the neutrino CC process is modified as follows.
Set the pion charge index $\lambda$ (and $\zeta$) to $-1$ in 
$I^{l_\pi\, \nu}_{X\, \mu}$ and $J^{l_\pi\, \nu}_{X\, \mu}$.
Replace the lepton current by the one for the anti-neutrino process,
which amounts to adopting the negative sign
in the leptonic tensor, \Eq{lepton_tensor}.
What modifications are needed for getting the cross section
for the anti-neutrino NC process is now obvious.

\subsection{Coherent Pion Photo-Production}

With the same derivation given above, we can also get an
expression for the differential cross section of
 the  coherent $\pi^0$  photo-production process 
\,$\gamma(q)+ t (p_t) \to \pi^0 (k) + t (p_t^\prime)$
in the LAB frame:
\begin{eqnarray}
\eqn{xs4}
\frac{d^2\sigma}{d\Omega_\pi}
=\frac{\alpha}{2\pi}
\left( {|\bm{k}|\over \omega_\pi} + {|\bm{k}| - \hat{k}\cdot\bm{q}\over E_t^\prime}
\right)^{-1}
{|\bm{k}|^2 \over |\bm{q}|\omega_\pi}
{1\over 2} \sum_\epsilon |{\cal M}^\gamma_R + {\cal M}^\gamma_{nr}|^2 \ ,
\end{eqnarray}
where $\alpha$ is the fine structure constant, 
and ${1\over 2}\sum_\epsilon$ stands for averaging
over the photon polarization.
The transition amplitudes ${\cal M}^\gamma_R$ and ${\cal M}^\gamma_{nr}$
for the photo-process are obtained from \Eqs{mr5}{mnr2} 
by retaining only the vector current, setting $\phi^\pi_A = 0$, 
and regarding $\epsilon^\mu_A$ as the polarization
vector of the incident photon.
Finally, the pion charge index ($\lambda$) is set to zero in 
$I^{l_\pi\, \nu}_{X\, \mu}$ and $J^{l_\pi\, \nu}_{X\, \mu}$
[\Eqs{i_mtx}{j_mtx}].

\subsection{Optical Potential for Pion-Nucleus Scattering}
\label{sec_opt}

We calculate the pion-nucleus scattering 
using the computer code, PIPIT~\cite{pipit} by appropriately
modifying the optical potential there to
accommodate the dynamical features of the $\Delta$-hole model and the
SL model.
In the original PIPIT, the optical potential ($U$), which is 
derived within the multiple scattering formalism
by Kerman, McManus and Thaler (KMT)~\cite{kmt}, is given 
by\footnote{PIPIT also includes a finite-range Coulomb interaction, and 
corrections from the truncated part of the Coulomb interaction 
are taken into account using the Vincent-Phatak method~\cite{vincent}.
}
\begin{eqnarray}
U(\bm{k}_A^\prime,\bm{k}_A)={A-1\over A}
\left\{\rho_p(\bm{q}) t_{\pi p}(\bm{k}_A^\prime,\bm{k}_A;k^o_A)
+\rho_n(\bm{q})t_{\pi n}(\bm{k}_A^\prime,\bm{k}_A;k^o_A)
\right\} \ ,
\end{eqnarray}
where $\bm{k}_A$ ($\bm{k}_A^\prime$) is
the incoming (outgoing) pion momentum in ACM, and
$k^o_A$ the magnitude of the on-shell momentum.
The quantities
$\rho_p(\bm{q})$ ($\rho_n(\bm{q})$) is the form factor of
the proton (neutron) matter distribution for 
$\bm{q}=\bm{k}_A-\bm{k}_A^\prime$, and 
$t_{\pi p}$ ($t_{\pi n}$) is the pion-proton (pion-neutron) 
scattering T-matrix whose normalization has been defined in \Eq{phase}.
It is to be noted that this original optical potential 
does not take account of $\Delta$-propagation in nuclei.
In Ref.~\cite{karaoglu}, KM separated 
$t_{\pi p}$ ($t_{\pi n}$)
into the resonant and non-resonant parts, 
took the non-resonant and the Coulomb parts
of the optical potential from the PIPIT code,
and combined it with the resonant part derived 
from a simplified $\Delta$-hole model.
A phenomenological s-wave potential
which is proportional to the square of the nuclear density
($\rho_t=\rho_p+\rho_n$) was also included to account for the
pion absorption by two nucleons through non-$\Delta$ mechanisms.
Thus the KM optical potential is given by 
\begin{eqnarray}
\eqn{separation}
U(\bm{k}_A^\prime,\bm{k}_A)=
U_{nr}+U_{R}+U_{ph}(\rho_t^2) \ ,
\end{eqnarray}
where $U_{nr}$, $U_{R}$ and $U_{ph}$ are the non-resonant, resonant and
phenomenological parts, respectively.

In constructing our optical potential,
we follow the same separation as in \Eq{separation}.
The non-resonant part of the optical potential is obtained 
from the PIPIT code by replacing
the non-resonant T-matrices in the original code with 
those derived from the SL model.
It is worth emphasizing that the SL model provides 
both on-shell and off-shell T-matrix elements.
Another difference from the original PIPIT code is 
that we use a different prescription 
for the Lorentz transformation from ACM to 2CM,
as explained in Appendix \ref{app_2cm}.

Regarding the resonant part, we use the resonant part of $\pi N$
T-matrix from the SL model, basically following the procedure used 
in Ref.~\cite{karaoglu} 
(apart from a more elaborate treatment
of kinematics (Lorentz transformation, etc.)).
First, we expand the optical potential into partial waves as
\begin{eqnarray}
\eqn{partial_potential}
U(\bm{k}_A^\prime,\bm{k}_A)= {2\over \pi} \sum_{l_\pi m_\pi}
V^{l_\pi}(k_A^\prime,k_A) Y^*_{l_\pi m_\pi}(\hat{k}_A^\prime)
Y_{l_\pi m_\pi}(\hat{k}_A) \ .
\end{eqnarray}
The resonant part of the potential is (cf. Eq.~(39) of Ref.~\cite{karaoglu})
\begin{eqnarray}
\eqn{v_res}
&&V^{l_\pi}_R(k_A^\prime,k_A) = {A-1\over A}
{8\pi^2 \mu_\Delta\over 3}
\int dx_A \Gamma_{A2}\gamma^{-1}
x_2 P_{l_\pi}(x_A) 
F_{\pi N\Delta}(k_2^\prime)F_{\pi N\Delta}(k_2) \\\nonumber
&\times&
\sum_{N=p,n} (1+{\lambda\tau_N\over 2})\int s^2 ds R^2 dR 
j_0(pR)j_0(Ps)
{e^{i \bar{K}_\Delta s}\over s}
\left\{1 + {i \mu_\Delta s\over \bar{K}_\Delta}
\left[\bar{e_N} + H_N\right]\right\}\rho_N(R) \hat{j}_1(k_F s) \ ,
\end{eqnarray}
where $\bm{k}_2$ ($\bm{k}_2^\prime$) is
the incoming (outgoing) pion momentum in 2CM, and
$x_A=\hat{k}_A\cdot\hat{k}_A^\prime$,
$x_2=\hat{k}_2\cdot\hat{k}_2^\prime$,
$P=|\bm{k}_A+\bm{k}_A^\prime|/2$,
$p=|\bm{k}_A-\bm{k}_A^\prime|$.
The dressed $\pi N\Delta$ coupling ($F_{\pi N\Delta}$) has been
introduced in \Eq{t-res}.
The Lorentz transformation of the T-matrix from 2CM
to ACM gives rise to the factor $\Gamma_{A2}$ defined by
\begin{eqnarray}
\Gamma_{A2} = \sqrt{\omega_{\pi,2}\,\omega_{\pi,2}^\prime\,
 E_{N,2}E_{N,2}^\prime \over
\omega_{\pi,A}\,\omega_{\pi,A}^\prime\, E_{N,A}E_{N,A}^\prime} \ ,
\end{eqnarray}
with $\omega_{\pi,2}^{(\prime)}=\sqrt{\bm{k}_2^{(\prime)\,2}+m_\pi^2}$,
$\omega_{\pi,A}^{(\prime)}=\sqrt{\bm{k}_A^{(\prime)\,2}+m_\pi^2}$,
$E_{N,2}^{(\prime)}=\sqrt{\bm{k}_2^{(\prime)\,2}+m_N^2}$ and
$E_{N,A}^{(\prime)}=\sqrt{\bm{p}_{N,A}^{(\prime)\,2}+m_N^2}$.
The values of $\bm{k}_2^{(\prime)}$ and $\bm{p}_{N,A}^{(\prime)}$ are fixed 
according to the prescription explained in Appendix \ref{app_2cm}.
The other quantities have already been introduced in Sec.~\ref{sec_res}.

Finally, we discuss the phenomenological term, $U_{ph}$.
We assume that in coordinate space $U_{ph}$
can be parametrized as
\begin{eqnarray}
\eqn{rho2-term}
U_{ph}(\bm{r}) = B \left({\rho_t(r)\over\rho_t(0)}\right)^2 \ ,
\end{eqnarray}
where $B$ is the partial wave dependent strength of the potential.
The corresponding partial wave potential in momentum space 
is given by
\begin{eqnarray}
\eqn{wave}
V^{l_\pi}_{ph}(k_A^\prime, k_A) = {A-1\over A} 4\pi^3 B_{l_\pi} 
\int^1_{-1}dx_A P_{l_\pi}(x_A)\int dr r^2 j_0(pr) 
\left({\rho_t(r)\over\rho_t(0)}\right)^2 \ .
\end{eqnarray}
In the present calculation
we include $V^{0}_{ph}$ and $V^{1}_{ph}$ and
treat their strengths $B_0$ and $B_1$ as adjustable parameters. 
Thus our model contains as free parameters 
$B_0$ and $B_1$ (complex numbers)  in addition to the
couplings in the spreading potential.

Given the optical potential, we solve the Lippmann-Schwinger equation
\begin{eqnarray}
\eqn{lippman}
T^\prime_{l_\pi}(k_A^\prime,k_A;k_A^o)
= V_{l_\pi}(k_A^\prime,k_A;k_A^o)
+{2\over\pi}\int 
{V_{l_\pi}(k_A^\prime,\bar{k}_A;k_A^o)T^\prime_{l_\pi}(\bar{k}_A,k_A;k_A^o)
\bar{k}_A^2 d\bar{k}_A 
\over \omega_\pi(k_A^o)+E_t(k_A^o)-\omega_\pi(\bar{k}_A)-E_t(\bar{k}_A)
+ i\epsilon} \ .
\end{eqnarray}
The solution to this equation
will be used in two contexts.
First, we use it to calculate pion-nucleus elastic and total scattering
cross sections, and compare them with data 
to find the optimal values of the free parameters in our model.
The solution to \Eq{lippman} is also used
to compute the pion distorted wave function that features
in the matrix elements in \Eqs{i_mtx}{j_mtx}.
For the former purpose, we obtain the full T-matrix
of pion-nucleus scattering from $T^\prime$ in \Eq{lippman} 
using the relation
\begin{eqnarray}
T = {A\over A-1}T^\prime \ .
\end{eqnarray}
For charged-pion scattering, corrections for the finite range
Coulomb potential are incorporated with the use of 
the Vincent-Phatak method~\cite{vincent}.
The procedure for calculating scattering observables 
from $T$ is detailed in Ref.~\cite{pipit}.
For the latter purpose, we calculate 
the pion distorted wave $\chi_{l_\pi}^*(k_A)$
associated with $T^\prime$ using the relation
\begin{eqnarray}
\eqn{pi_wave}
 \chi_{l_\pi}^*(k_A) = {\delta(k_A-k_A^o)\over k_A^2 }
+ {T^\prime_{l_\pi}(k_A^o,k_A;k_A^o)\over 
\omega_\pi(k_A^o)+E_t(k_A^o)-\omega_\pi(k_A)-E_t(k_A)
+ i\epsilon} \ ,
\end{eqnarray}
where, for notational simplicity,
dependence on the pion charge ($\lambda$) is suppressed.
Following the KMT formalism~\cite{kmt}, 
we use $ \chi_{l_\pi}^*(k_A) $ in evaluating the
matrix elements in \Eqs{i_mtx}{j_mtx}.
This wave function is related to the full wave function by
\begin{eqnarray}
 \chi_{l_\pi}^{\rm (full)*} = 
-{1\over A-1} + {A\over A-1} \chi_{l_\pi}^*  \ .
\end{eqnarray}
For charged-pion scattering,
$\chi_{l_\pi}^{\rm (full)*}$ does not have
the correct normalization,
because the Coulomb potential
has been cut off at a finite distance;
this entails the necessity of
multiplying $\chi_{l_\pi}^{\rm (full)*}$
with a normalization factor (call it $\kappa$).
We note that it is $\chi_{l_\pi}^*$ rather than
$\chi_{l_\pi}^{\rm (full)*}$
that enters into our calculation,
and we choose to use the same normalization
factor $\kappa$ for $\chi_{l_\pi}^*$
as for $\chi_{l_\pi}^{\rm (full)*}$.
Thus, in evaluating the matrix elements in \Eqs{i_mtx}{j_mtx},
we use $\kappa\chi_{l_\pi}^*$ instead of  $\chi_{l_\pi}^*$.
In the $\Delta$ resonance region of our interest, 
it turns out that $|\kappa -1| \ltap 0.01$.
(For neutral pion scattering, $\kappa = 1$.)

\section{Numerical Results}
\label{sec_result}

\subsection{Pion-Nucleus Scattering}

\begin{figure}[t]
\begin{center}
 \includegraphics[width=90mm]{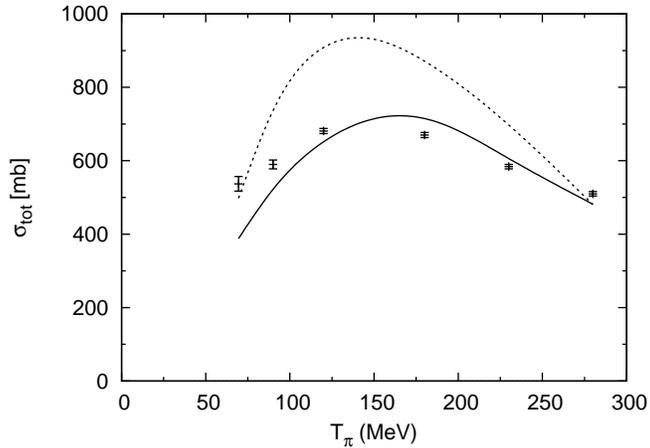}
\caption{\label{fig_total}
Total cross sections for $\pi^- - {}^{12}$C scattering. 
The solid curve is obtained with our full calculation, while the dashed
curve is obtained without the spreading potential.
The data are from Ref.~\cite{pi-nucleus1}.
}
\end{center}
\end{figure}

\begin{figure}
\begin{minipage}[t]{80mm}
 \includegraphics[width=80mm]{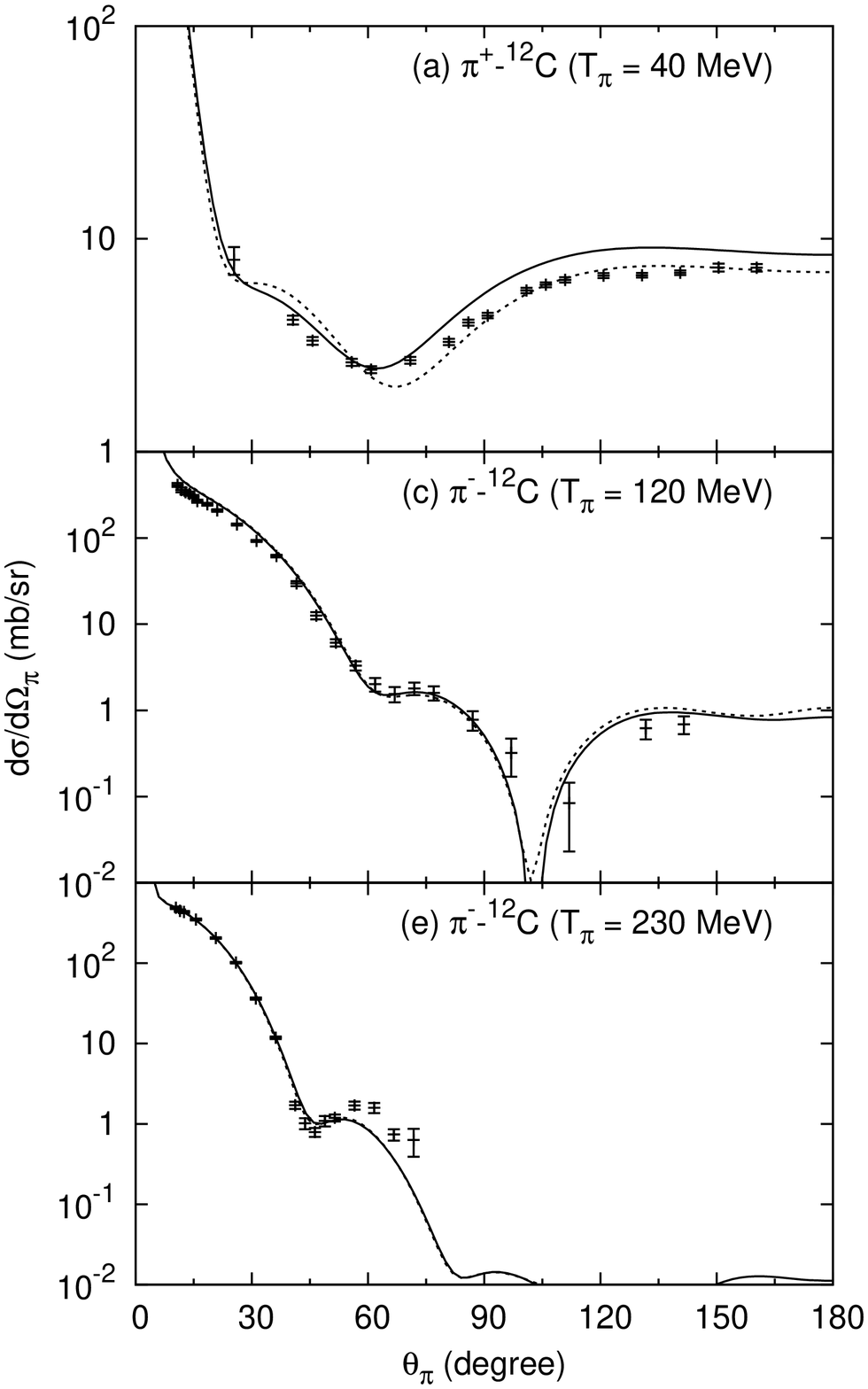}
\end{minipage}
\hspace{2mm}
\begin{minipage}[t]{80mm}
 \includegraphics[width=80mm]{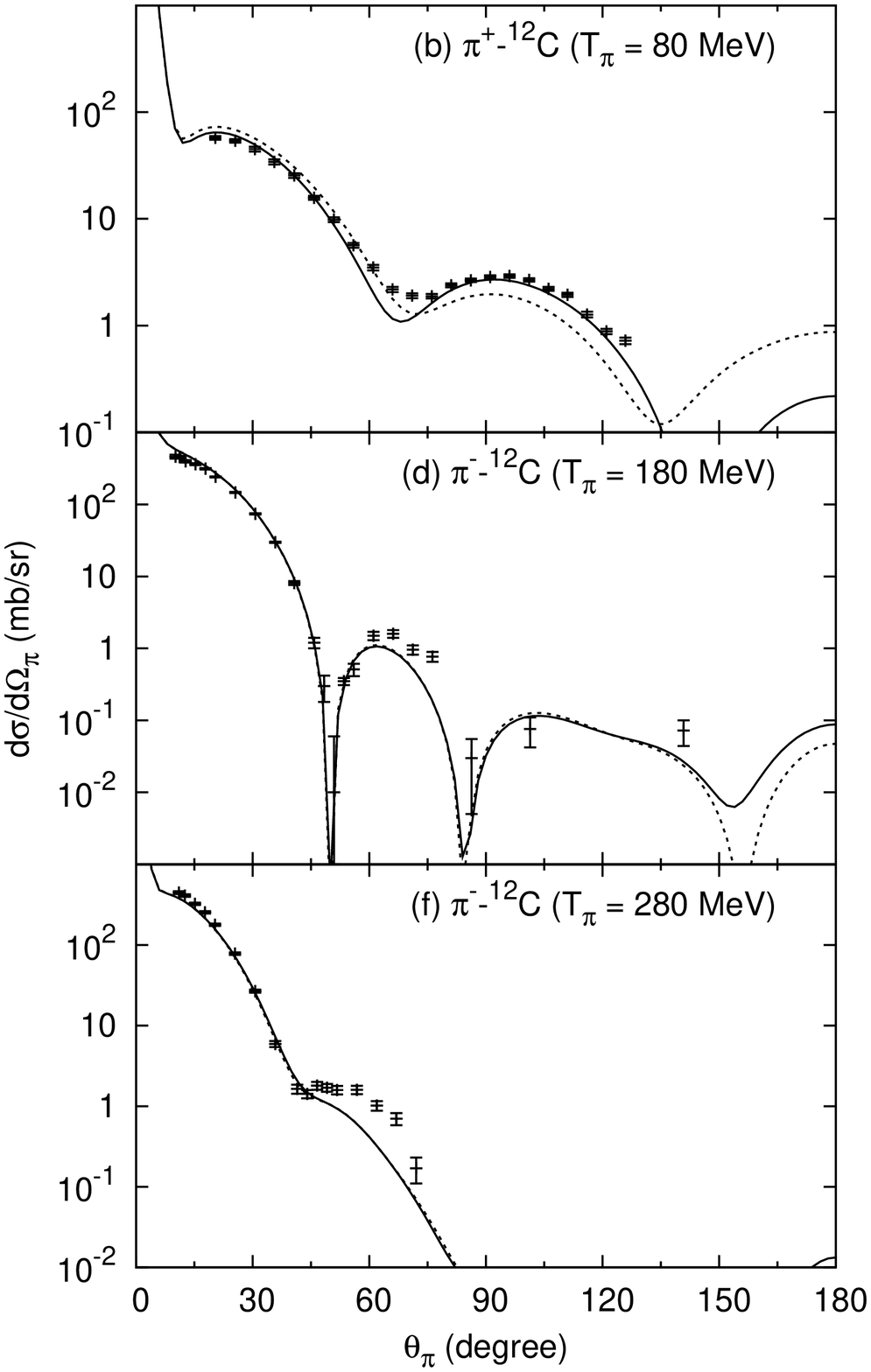}
\end{minipage}
\caption{\label{fig_elastic} $\pi - {}^{12}$C elastic differential cross
 sections.
The solid curve is obtained with our full calculation while the dashed
curve is obtained without the phenomenological terms in \Eq{rho2-term}.
The data are from Ref.~\cite{pi-nucleus2} for (a),
Ref.~\cite{pi-nucleus3} for (b) and
Ref.~\cite{pi-nucleus1} for (c)-(f).
}
\end{figure}

As explained in the previous sections,
our model contains four complex free parameters.
Two of them are the central ($V_C$) and LS ($V_{LS}$) parts of the
spreading potential [see \Eq{spr}], and the other two are 
the strengths, $B_0$ and $B_1$,
of the s-wave and p-wave phenomenological terms 
in the optical potential [see \Eq{wave}].
These free parameters are optimized to fit the pion-nucleus scattering data.
Since our aim here is to calculate coherent pion production off $^{12}$C, 
we should use the $\pi\! -\!^{12}$C scattering data 
to fix these parameters. 
Adjusting them to reproduce
the total cross sections and the elastic differential cross sections
for $\pi\! -\!^{12}$C scattering,
we obtain:
\begin{eqnarray}
V_C =  48.0 - 34.5 i\ {\rm MeV} \ ,\qquad
V_{LS} = -3.0 - 2.0 i\ {\rm MeV} \nonumber\\
B_0 = 5.1 + 5.2 i\ {\rm MeV} \ ,\qquad
B_1 = 2.8 - 5.7 i\ {\rm MeV} \ .
\end{eqnarray}
We note that our calculations include the pion-nucleus partial waves up to 
$l_\pi \le 9$ [\Eq{lippman}], and $s$- and $p$-waves (and all possible
spin-isospin states) for
the elementary $\pi N$ scattering.\footnote{
Hereafter, we include the same set of partial waves
($l_\pi$) in the amplitudes for
both pion-nucleus scattering and pion production off a nucleus.
For the non-resonant elementary pion production amplitudes,
we include the partial waves up to $\ell\le 4$ in \Eq{j_mtx}.
}
Figures.~\ref{fig_total} and \ref{fig_elastic} illustrate
the quality of fit to the $\pi-{}^{12}$C scattering data
achieved in our model (with our optical potential).
In Fig.~\ref{fig_total}, the total cross sections for $\pi^-$-$^{12}$C
scattering are shown as a function of the pion kinetic energy 
$T_\pi$ in the laboratory frame.
The results of our full calculation
are given by the solid curve and, for comparison,
the results obtained without the spreading potential
are also shown in the dashed curve.
We observe a large reduction in the total cross section
as we go from the dashed to solid lines, which is
mainly caused by the strong pion absorption simulated 
by the spreading potential.
In connection with fitting to the pion-nucleus scattering data,
it is worthwhile to make the following comment.
In the calculation of coherent pion production,
the final-state interaction is nothing but
elastic scattering between the pion and nucleus.
One might therefore think that a phenomenological adjustment
of the pion-nucleus optical potential
to fit the elastic pion-nucleus scattering data will be good enough.
However, in our consistent model building,  
the spreading potential enters not only into
the optical potential but also into the pion production operators,
and hence it is important to control its
 strength  using the total cross section data.
The fact that the spreading potential has a very large effect 
on the total cross sections makes this point particularly important.

Our results for the differential cross sections are shown in 
Fig.~\ref{fig_elastic}.
In addition to our full calculation shown in the solid curve, we also
show in the dashed curve the results
obtained without the phenomenological term $U_{ph}$
[see \Eq{rho2-term}].
We see that this phenomenological $\rho^2$ term, which simulates absorption 
of $s$-wave and $p$-wave pions by two-nucleons within our model,
is not large in the considered $T_\pi > $ 40~MeV region for 
$\pi-^{12}$C elastic scattering.  
However it is known that $U_{ph}$
can play an important role for many observables in low-energy 
pion-nucleus scattering.
As an example to shed light on this point, 
we have calculated $\pi-{}^{16}O$ elastic scattering 
at $T_{\pi}$ = 50 MeV using the same model (only the nuclear density is
different).
We have found that, in reproducing the data satisfactorily in our approach,
the $\rho^2$ term plays an important role,
its size being almost as large as that
found in Fig.~4(a) of Ref.~\cite{karaoglu}.
Overall, the results of our full calculation
satisfactorily reproduce the data
for both the total and elastic cross sections.

\subsection{Coherent Pion Photo-Production}

\begin{figure}[t]
\begin{center}
\begin{minipage}[t]{77mm}
 \includegraphics[width=77mm]{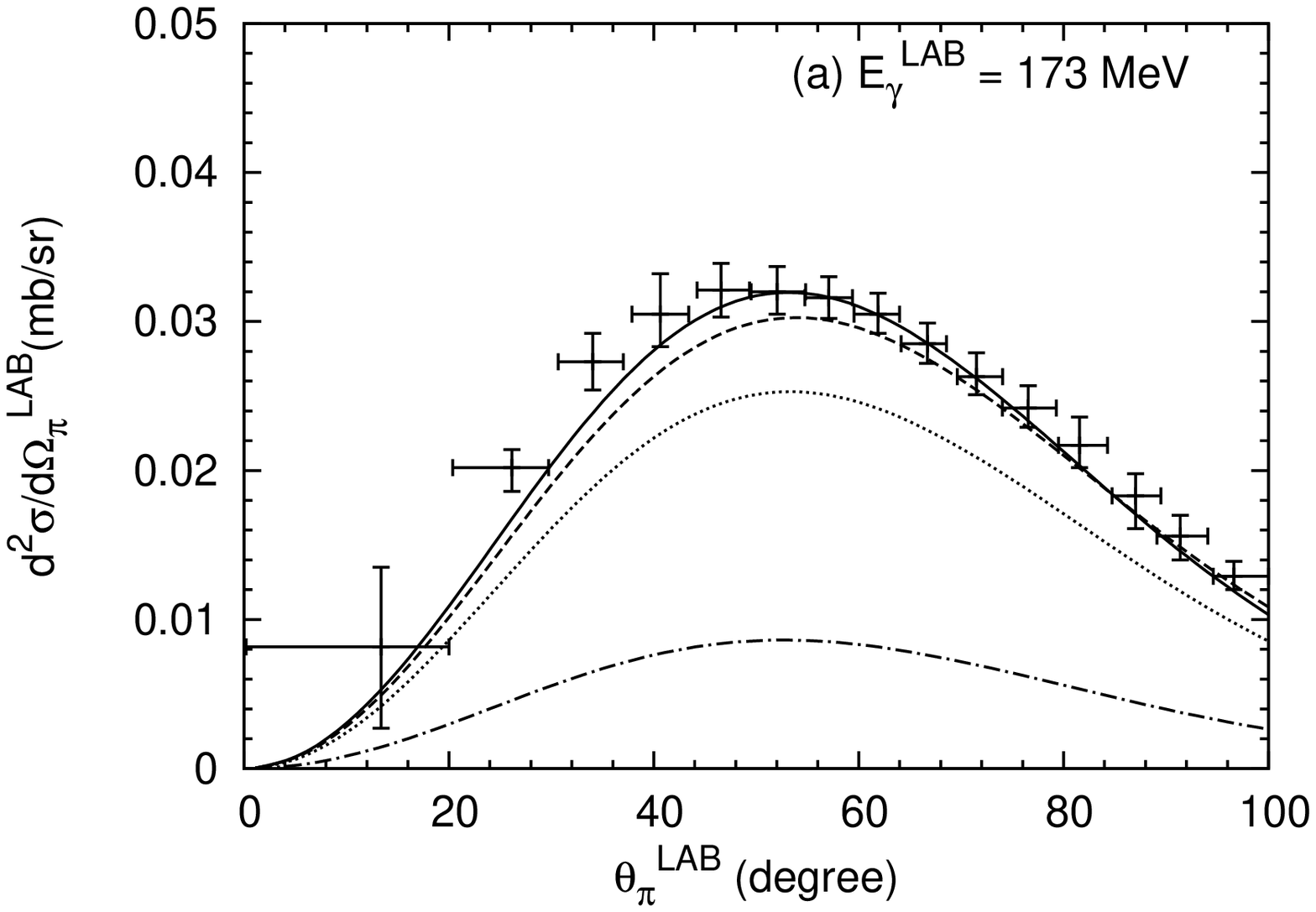}
\end{minipage}
\hspace{5mm}
\begin{minipage}[t]{77mm}
 \includegraphics[width=77mm]{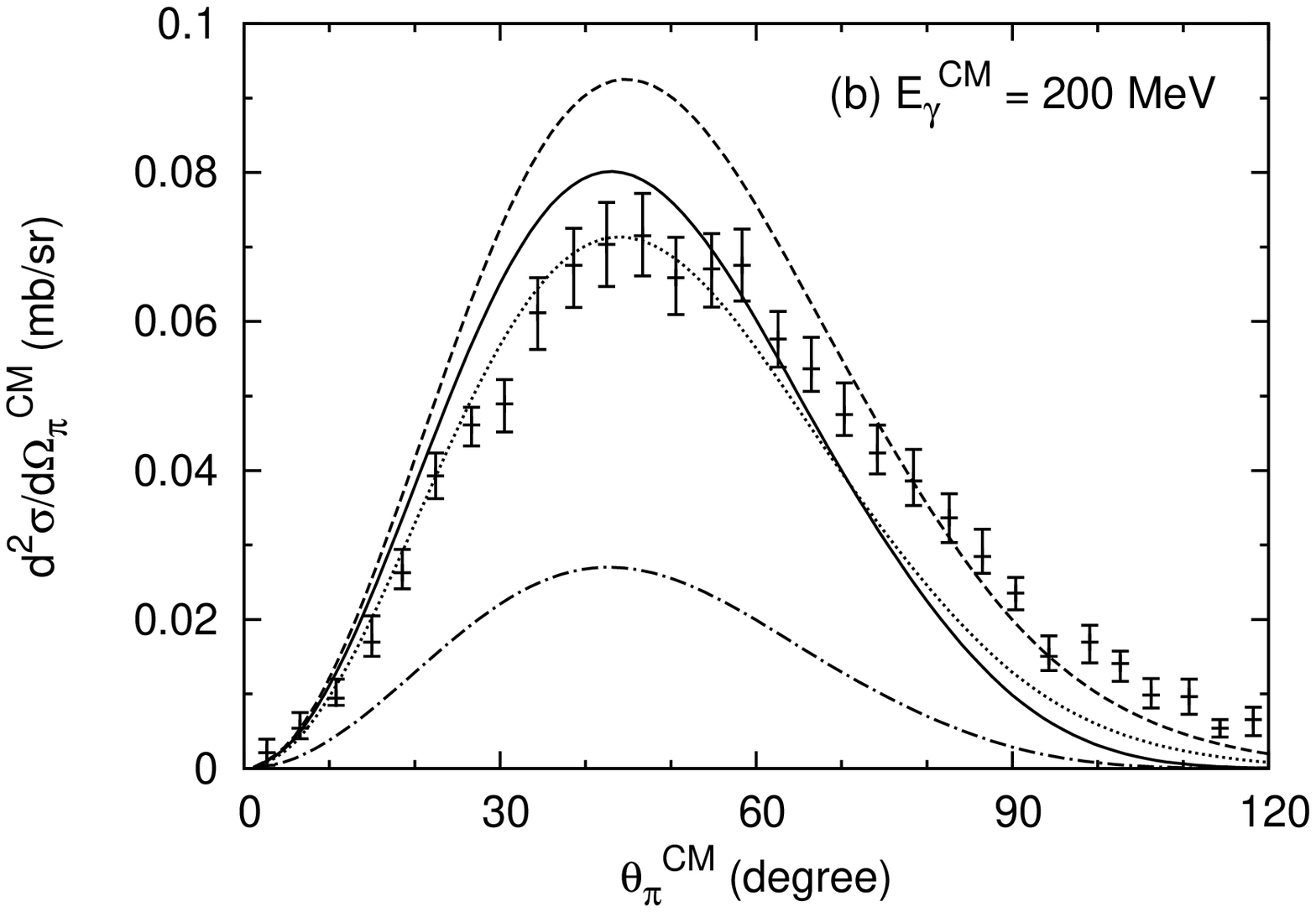}
\end{minipage}

\begin{minipage}[t]{77mm}
 \includegraphics[width=77mm]{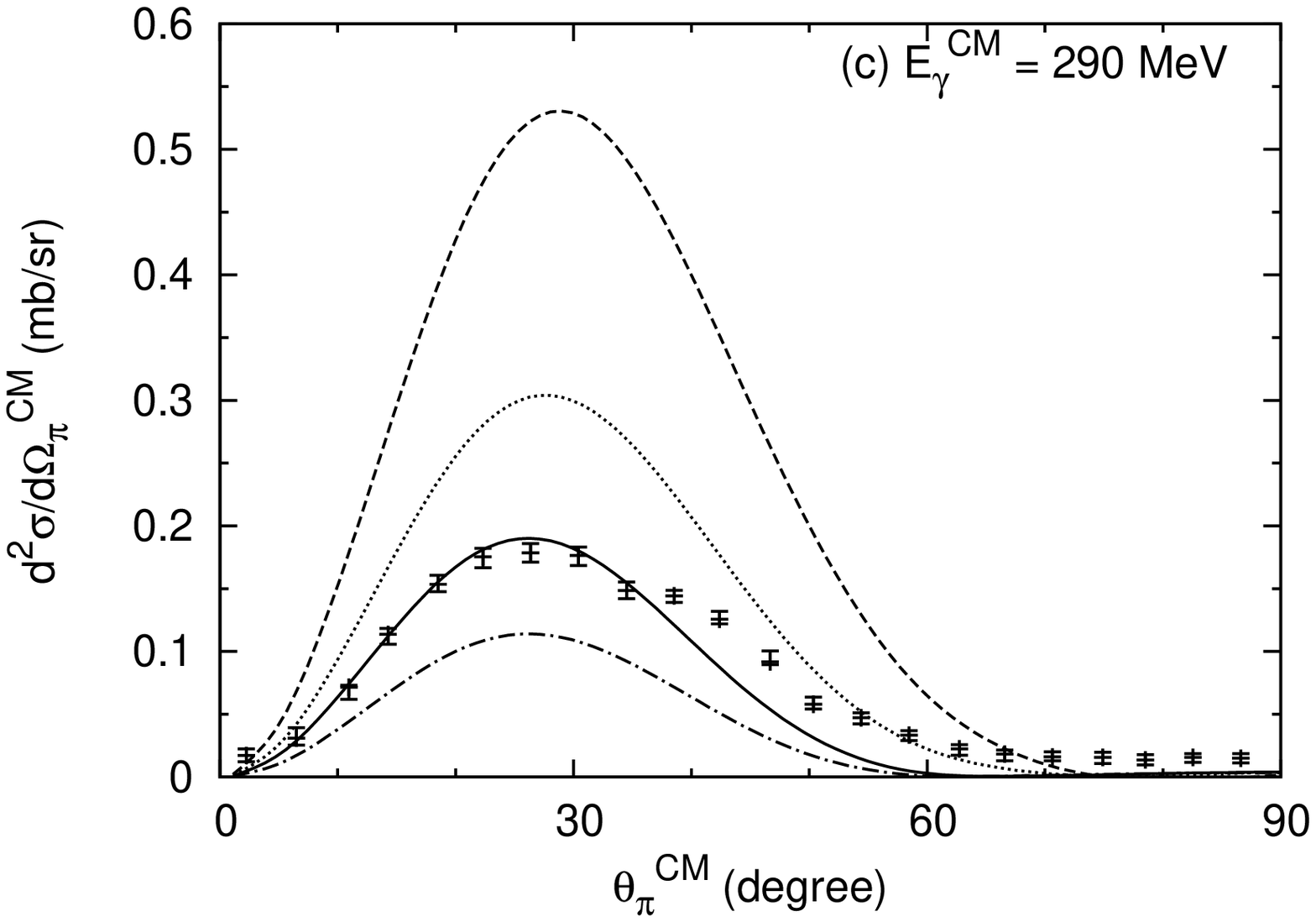}
\end{minipage}
\hspace{5mm}
\begin{minipage}[t]{77mm}
 \includegraphics[width=77mm]{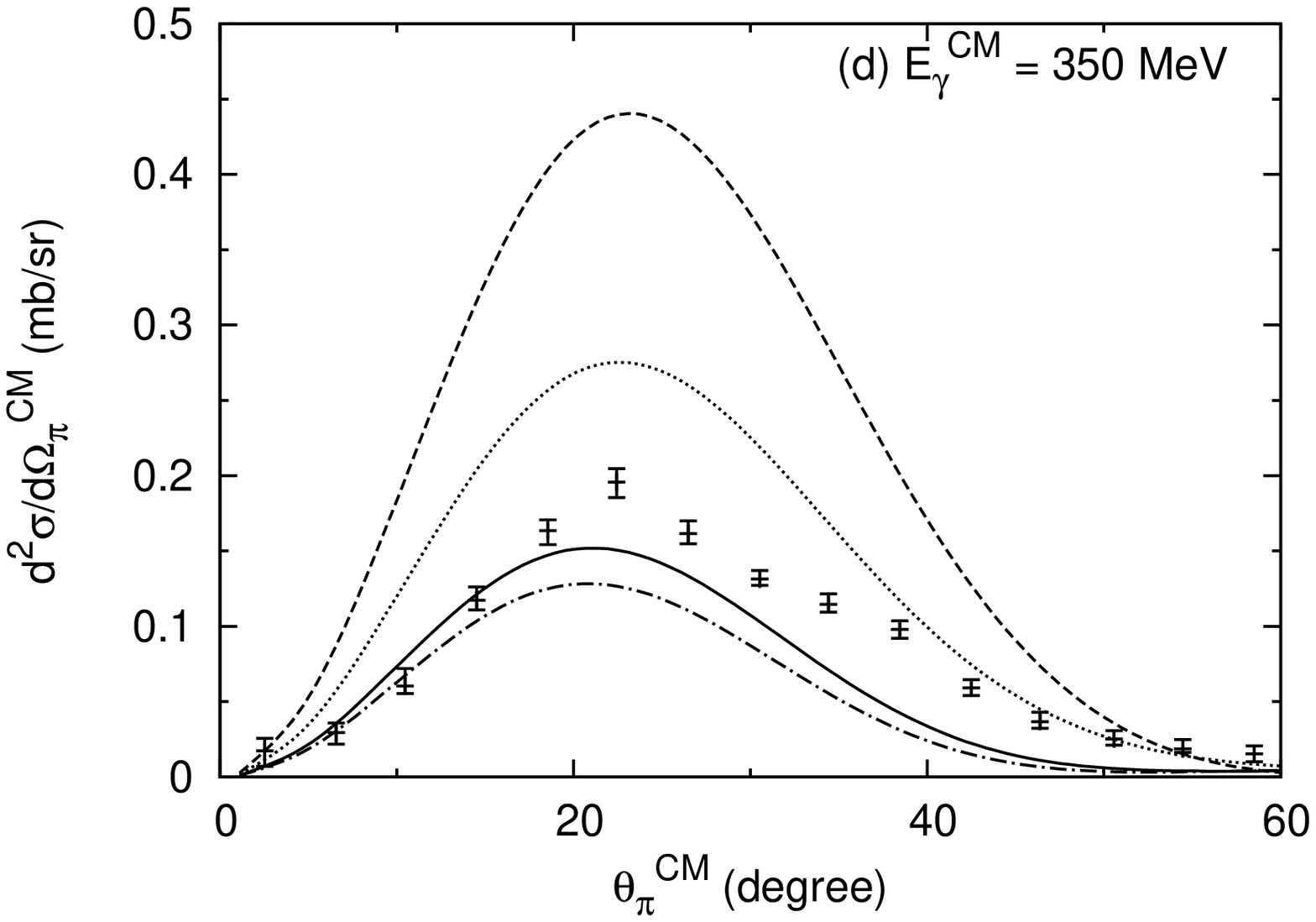}
\end{minipage}
\caption{\label{fig_krusche}
Differential cross sections for 
$\gamma + {}^{12}{\rm C}_{g.s.} \to \pi^0 + {}^{12}{\rm C}_{g.s.}$
for different incident photon energies (indicated in each panel).
The solid lines represent the results of the full calculation.
The dashed lines are obtained without the FSI and without
the medium effects on the $\Delta$-propagation,
while the dotted lines are obtained with the medium effects 
on the $\Delta$ included.
The dash-dotted curves correspond to
a case in which the pion production operator 
includes only the $\Delta$ mechanism.
For more detailed explanations for the different cases, 
see the text.
The data are from Ref.~\cite{gothe} for (a) and 
from Ref.~\cite{krusche} for (b)-(d).
}
\end{center}
\end{figure}

We are now in a position to perform a parameter-free calculation of
the cross sections for coherent pion production.  
The photo-process, for which extensive data are available,
provides a good testing ground
for checking the reliability of our approach.
We compare in Fig.~\ref{fig_krusche}
our numerical results for the differential cross sections for
$\gamma + {}^{12}{\rm C}_{g.s.} \to \pi^0 + {}^{12}{\rm C}_{g.s.}$
with the existing data~\cite{gothe,krusche}. 
The long-dash lines are obtained without FSI and without 
the medium effects on $\Delta$-propagation.\footnote{
\label{foot:medium}
The ``medium effects on the $\Delta$'' here 
refer to the combined effects of the Pauli blocking of 
$\Delta$-decay ($\Sigma_{\rm pauli}$), the spreading potential
($\Sigma_{\rm spr}$), and the terms in the square bracket in \Eq{delta_k2}.}
With the medium effects on the
$\Delta$ included, the short-dash lines are obtained, 
and the results of our full calculation
are given by the solid lines.
Figure~\ref{fig_krusche} indicates that the medium effects are quite sizable,
and they play an important role in bringing the calculated 
differential cross sections in agreement with the data.
Particularly noteworthy is the drastic reduction of the cross section 
in the $\Delta$ region [Fig.~\ref{fig_krusche} (c)],
a feature that reflects the fact
that a significant part of the medium effects 
simulate pion absorption.
The good general agreement seen in Fig.~\ref{fig_krusche}
indicates the basic soundness of the method
we have used in determining the spreading potential.

It is true that, for higher incident energies, 
in the large angle region beyond the peak position,
there are noticeable discrepancies
between the results of our full calculation and the data. 
However, as noted in Ref.~\cite{krusche},
the data in this region are likely to be substantially contaminated 
by incoherent processes in which the final
nucleus is in its low-lying excited states. 
The effects of this type of contamination are expected to grow
for higher incident photon energies
and for larger momentum transfers (the large angle region)
because of increased nuclear excitations. 
We therefore take the viewpoint that
the discrepancy found in
Figs.~\ref{fig_krusche} (b)-(d) does not
necessarily signal a failure of our model, 
and that our model describes coherent pion
photo-production reasonably well.

Figure~\ref{fig_krusche} also shows (in the dash-dotted lines)
the results corresponding to a case in which 
the pion production operator includes only the $\Delta$ mechanism
(the non-resonant mechanism turned off); \footnote{\label{footnote:non-res}
In the SL model, the resonant amplitude itself contains the non-resonant
mechanism. We refer to the purely non-resonant amplitudes as 
``non-resonant amplitudes'', and 
it is only these non-resonant amplitudes 
that we turn off here and later in 
Figs.~\ref{fig_tpi.1gev}-\ref{fig_q2.nc} and \ref{fig_non_local}.
}
the distorted pion wave function incorporating FSI 
is the same as that used for the full calculation.
These results serve to demonstrate the importance
of the non-resonant mechanism. 
Fig.~\ref{fig_krusche} (a) indicates that, near threshold, 
the contributions from the resonant and non-resonant mechanisms 
are comparable, a feature that is not surprising away from 
the resonance peak.
A remarkable feature is that even near the resonance energy
[see Fig.~\ref{fig_krusche} (c)] the contribution from 
the non-resonant mechanism is quite significant.
This is partly because the resonant contribution is considerably
suppressed by pion absorption (the spreading potential) and the
non-local effect of $\Delta$ propagation (the $\Delta$ kinetic
term).\footnote{
We come back to the non-local effect due to the $\Delta$ kinetic term 
later when we discuss the neutrino-induced processes.
}

To summarize this section, the results for the coherent 
photo-pion production process establish 
to a satisfactory degree the reliability
of our present approach
({\it i.e.,} combined use of the SL model 
and the $\Delta$-hole model)
and motivate us to apply the same approach
to neutrino-induced coherent pion production.

\subsection{Neutrino-Induced Coherent Pion Production}

\begin{figure}[t]
\begin{center}
 \includegraphics[width=85mm]{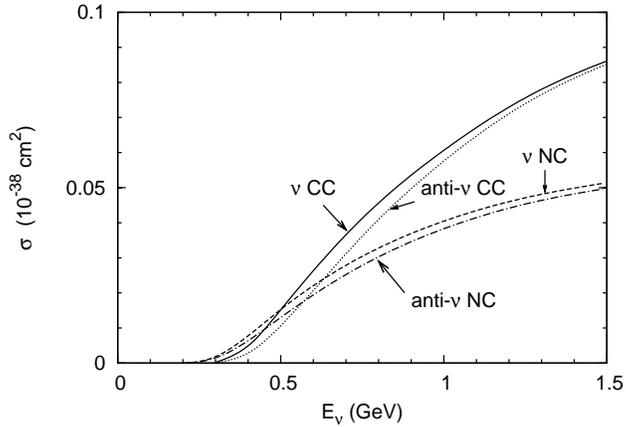}
\caption{\label{fig_tot}
The $E_\nu$-dependence of the total cross section for
$\nu_\mu + {}^{12}{\rm C}_{g.s.} \to \mu^- + \pi^+ + {}^{12}{\rm
 C}_{g.s.}$ (solid line),
$\nu + {}^{12}{\rm C}_{g.s.} \to \nu + \pi^0 + {}^{12}{\rm C}_{g.s.}$
 (dashed line),
$\bar{\nu}_\mu + {}^{12}{\rm C}_{g.s.} \to \mu^+ + \pi^- + {}^{12}{\rm
C}_{g.s.}$ (dotted line) and
$\bar{\nu} + {}^{12}{\rm C}_{g.s.} \to \bar{{\nu}} + \pi^0 + {}^{12}{\rm
 C}_{g.s.}$ (dash-dotted line).
}
\end{center}
\end{figure}
We now present the numerical results of our calculations
for neutrino-induced coherent pion production
on the $^{12}$C target.
We consider the CC and NC processes induced by
a neutrino or an anti-neutrino:
\begin{eqnarray}
\nu_\mu + {}^{12}{\rm C}_{g.s.} &\to&
 \mu^- + \pi^+ + {}^{12}{\rm C}_{g.s.}\nonumber\\
\nu + {}^{12}{\rm C}_{g.s.} &\to&
 \nu + \pi^0 + {}^{12}{\rm C}_{g.s.}\\
 \bar{\nu}_\mu + {}^{12}{\rm C}_{g.s.} &\to&
  \mu^+ + \pi^- + {}^{12}{\rm C}_{g.s.} \nonumber\\
 \bar{\nu} + {}^{12}{\rm C}_{g.s.} &\to&
  \bar{{\nu}} + \pi^0 + {}^{12}{\rm C}_{g.s.} \nonumber
\end{eqnarray}
Figure~\ref{fig_tot} gives the total cross sections for these processes
as functions of the incident neutrino (anti-neutrino) energy 
in the laboratory system, $E_\nu$. 
It is seen that, for higher incident energies, 
the ratio $\sigma_{CC}/ \sigma_{NC}$ approaches 2,
a value expected from the isospin factor.
For lower incident energies ($E_\nu$\ltap 500~MeV), however, 
$\sigma_{NC}$ is larger than $\sigma_{CC}$,
reflecting the fact that the phase space for the CC process
is reduced significantly by the muon mass.
It is well known that interference 
between the vector and axial-vector currents
can lead to different cross sections 
for the neutrino and anti-neutrino processes.
However, since the coherent process is dominated
by the contribution of the axial current
(see Fig.~\ref{fig_v-a}),
the role of the interference term is diminished drastically.
This explains why in Fig.~\ref{fig_tot}
the cross sections for the neutrino and anti-neutrino processes
are almost the same. 

\begin{figure}[t]
\begin{minipage}[t]{77mm}
 \begin{center}
 \includegraphics[width=77mm]{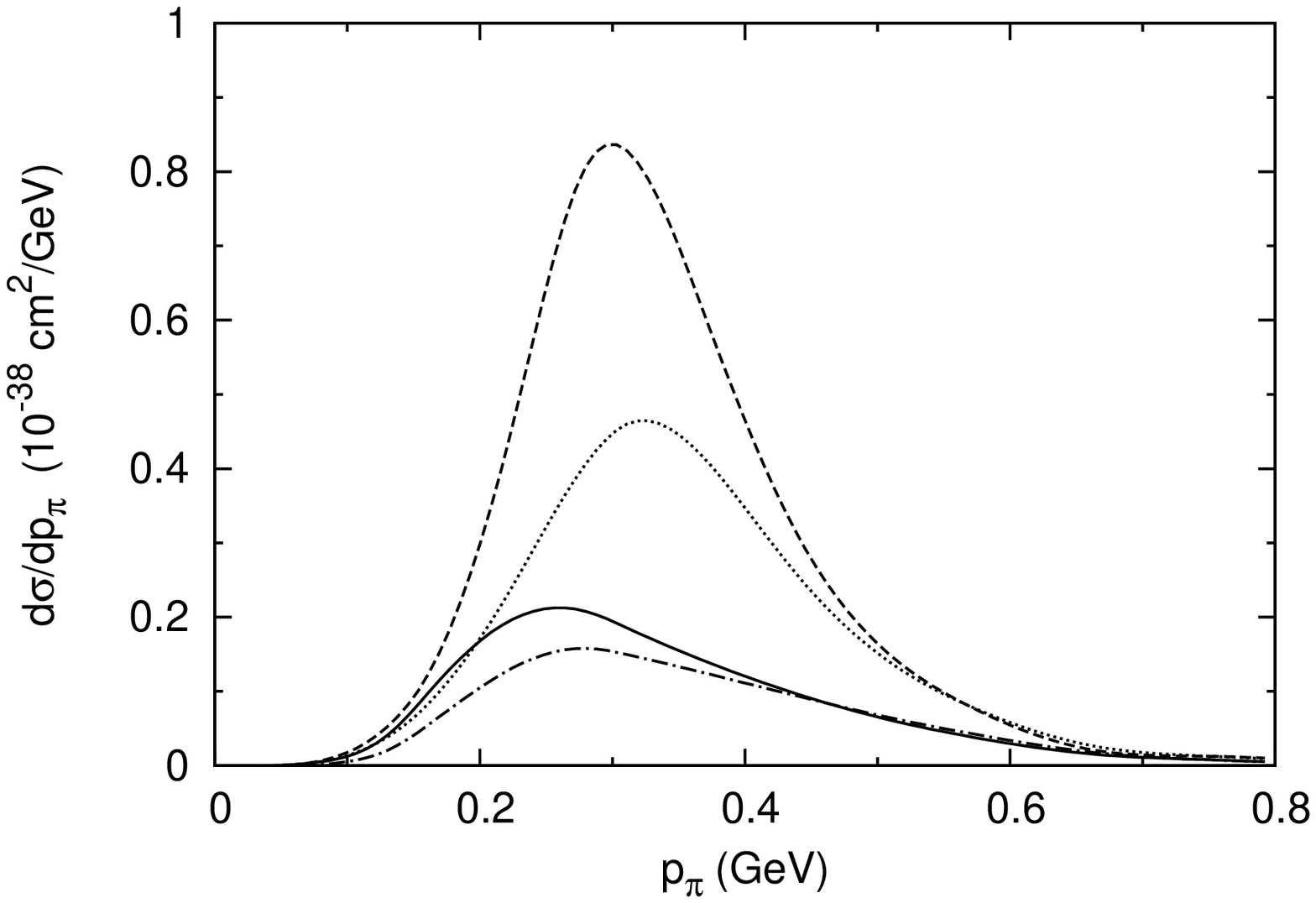}
 \caption{\label{fig_tpi.1gev}
The pion momentum distribution for
 $\nu_\mu +\! {}^{12}{\rm C}_{g.s.} \to \mu^- \!+ \!\pi^+\!+ \!{}^{12}{\rm
  C}_{g.s.}$ at $E_\nu$ = 1 GeV;
$p_\pi$ is the pion momentum in the laboratory frame.
The use of the solid, dashed, dotted and dash-dotted lines
follows the same convention as in Fig.~\ref{fig_krusche}.}
 \end{center}
\end{minipage}
\hspace{5mm}
\begin{minipage}[t]{77mm}
 \begin{center}
 \includegraphics[width=77mm]{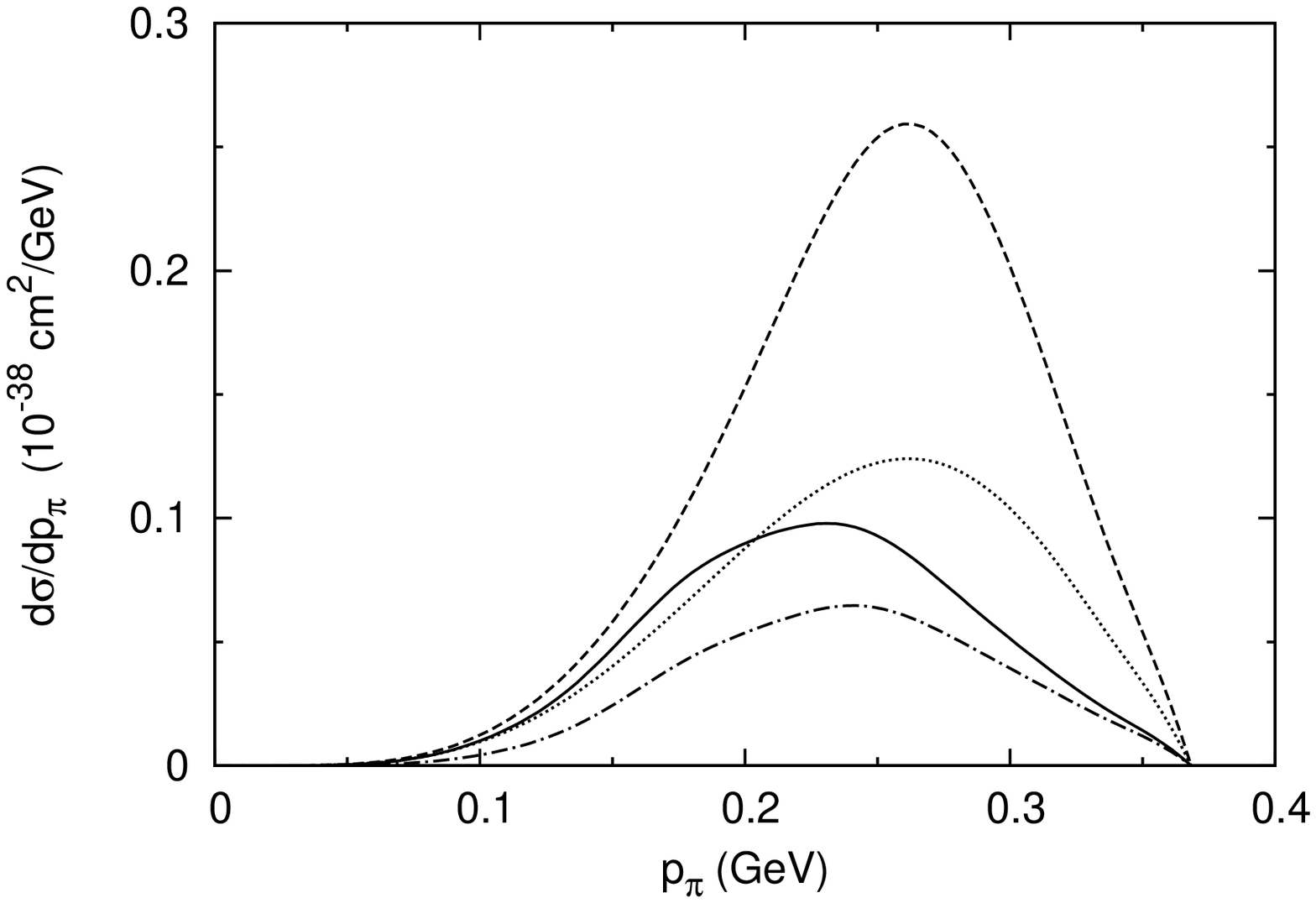}
 \caption{\label{fig_tpi.0p5gev}
Same as in Fig.~\ref{fig_tpi.1gev} but for $E_\nu$ = 0.5 GeV.
 }
 \end{center}
\end{minipage}
\end{figure}

To compare our results with data,
we need to evaluate the total cross sections averaged
over the neutrino fluxes that pertain to the relevant experiments.
We choose to use the fluxes up to $E_\nu \le$ 2 GeV 
and neglect the fluxes beyond that limit 
based on the following consideration.
Since our model includes no resonances other than the $\Delta$,
it is expected to be reliable only for $W$\ltap 1.4 GeV.
The fact that even at $E_\nu$ = 1 GeV
coherent pion production can involve contributions
coming from the $W \!>\!$ 1.4 GeV region is disquieting,
but we can still expect that the $\Delta$-excitation contribution
is predominant for the total cross section for the coherent process.
[This feature can be seen in, {\it e.g.}, Fig.~\ref{fig_tpi.1gev} 
to be discussed later.]
For $E_\nu \sim$ 2 GeV, we do expect
that $\Delta$ dominance gets significantly less pronounced
but that $\Delta$ still gives the most important contribution.
Meanwhile, the region $E_\nu$\gtap 1.5 GeV belongs 
to the tail of the neutrino flux used in MiniBooNE.
We therefore consider it reasonable to compare with data
our theoretical cross section averaged
over the neutrino flux up to $E_\nu$ = 2 GeV.
For the CC process, we use the flux reported in Ref.~\cite{K2K} 
and deduce 
\begin{eqnarray}
 \sigma_{\rm ave}^{CC} = 6.3 \times 10^{-40} {\rm cm}^2 \ .
 \label{SCCave}
\end{eqnarray}
A K2K experiment~\cite{hasegawa} reports the upper limit
\begin{eqnarray}
\sigma_{\rm K2K}  <  7.7 \times 10^{-40} {\rm cm}^2 \ .
\end{eqnarray}
In fact, this upper limit corresponds to events satisfying
the muon momentum cut, $p_\mu > 450$ MeV and the cut on the momentum
transfer squared, $Q_{\rm rec}^2 < 0.1$ GeV$^2$; 
$Q_{\rm rec}^2$ is calculated as
\begin{eqnarray}
\eqn{q_rec}
Q_{\rm rec}^2 = 2 E_\nu^{\rm rec} (E_\mu - p_\mu\cos\theta_\mu) -
m_\mu^2 \ ,
\end{eqnarray}
where the reconstructed neutrino energy ($E^{\rm rec}_\nu$) is
calculated from the muon kinematics
[the energy ($E_\mu$) and the scattering angle ($\theta_\mu$)]
assuming the quasi-elastic kinematics: 
\begin{eqnarray}
\eqn{e_rec}
E^{\rm rec}_\nu = {1\over 2}{(m_p^2 - m_\mu^2) - (m_n - V)^2 + 2E_\mu
(m_n - V)\over (m_n - V) - E_\mu + p_\mu\cos\theta_\mu} \ ,
\end{eqnarray}
where $m_p$, $m_n$ and $m_\mu$ are the masses of the proton, neutron and
muon, respectively and the nuclear potential ($V$) is set to 27~MeV.
Our result in Eq.(\ref{SCCave}) is also obtained with these cuts,
and is consistent with the K2K data.
We note that a recent report from SciBooNE~\cite{hiraide} gives
a similar empirical upper limit.

For the NC process, we use the flux reported 
by MiniBooNE in Ref.~\cite{miniboone_flux} and arrive at
\begin{eqnarray}
\sigma_{\rm ave}^{NC} = 2.8 \times 10^{-40} {\rm cm}^2 \ .
\end{eqnarray}
This is to be compared with
\begin{eqnarray}
\sigma_{\rm MiniBooNE}  =  7.7 \pm 1.6\pm 3.6 \times 10^{-40} {\rm cm}^2  \ , 
\end{eqnarray}
given in Ref.~\cite{raaf}.
Our result is consistent with the empirical value 
within the large experimental errors, even though 
the theoretical value is rather visibly smaller 
than the empirical central value.
It is to be noted however that Ref.~\cite{raaf}
is a preliminary report,  
and that, as discussed in great detail in Ref.~\cite{amaro}, 
$\sigma_{\rm MiniBooNE}$ may be overestimated 
due to the use of the RS model\cite{RS} in the analysis. 

\begin{figure}
\begin{minipage}[t]{77mm}
\begin{center}
 \includegraphics[width=77mm]{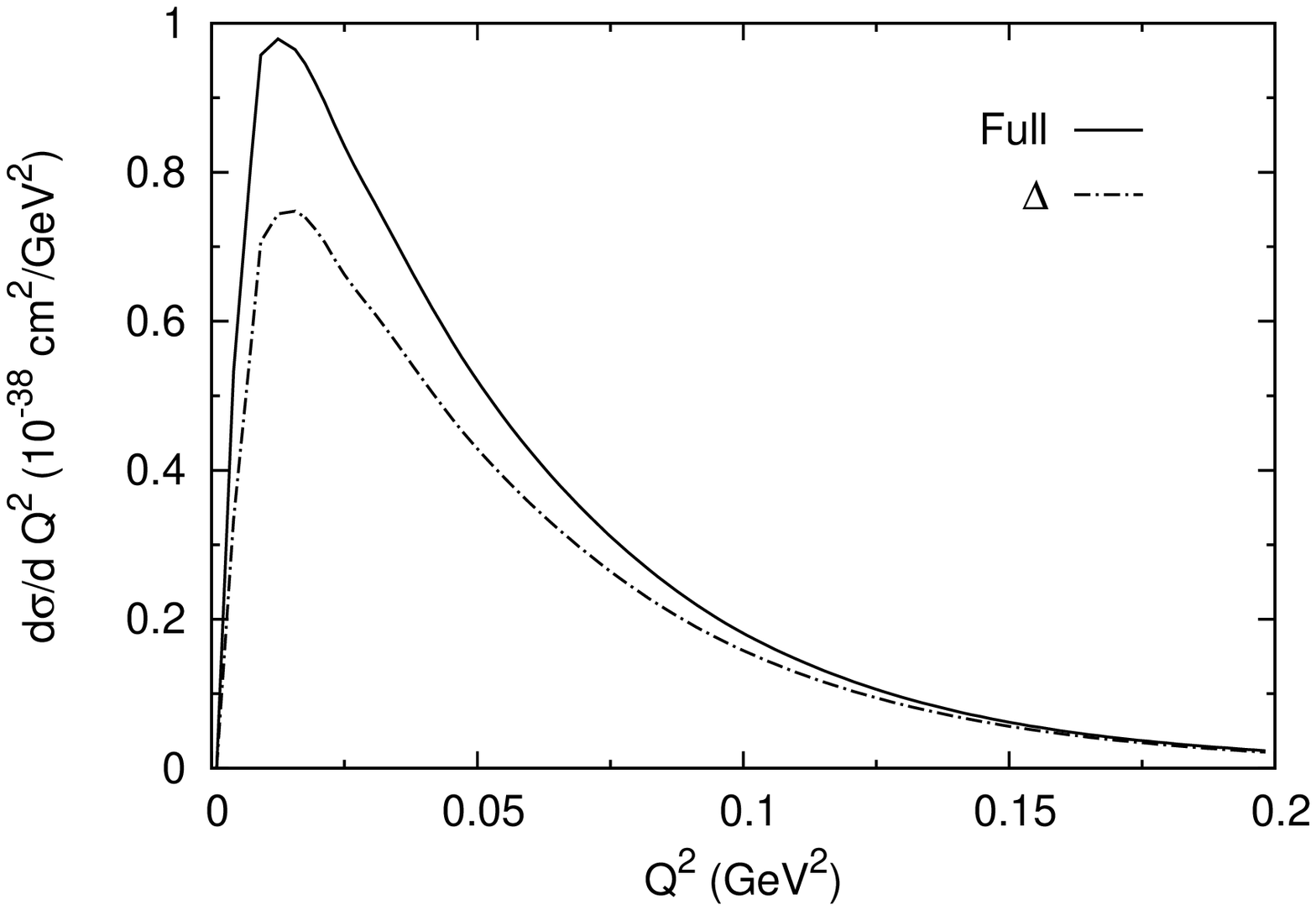}
\caption{\label{fig_q2.cc}
The $Q^2$-spectrum for
$\nu_\mu + {}^{12}{\rm C}_{g.s.} 
\to \mu^- + \pi^+ + {}^{12}{\rm C}_{g.s.}$ at $E_\nu = 1$ GeV.
}
\end{center}
\end{minipage}
\hspace{5mm}
\begin{minipage}[t]{77mm}
\begin{center}
 \includegraphics[width=77mm]{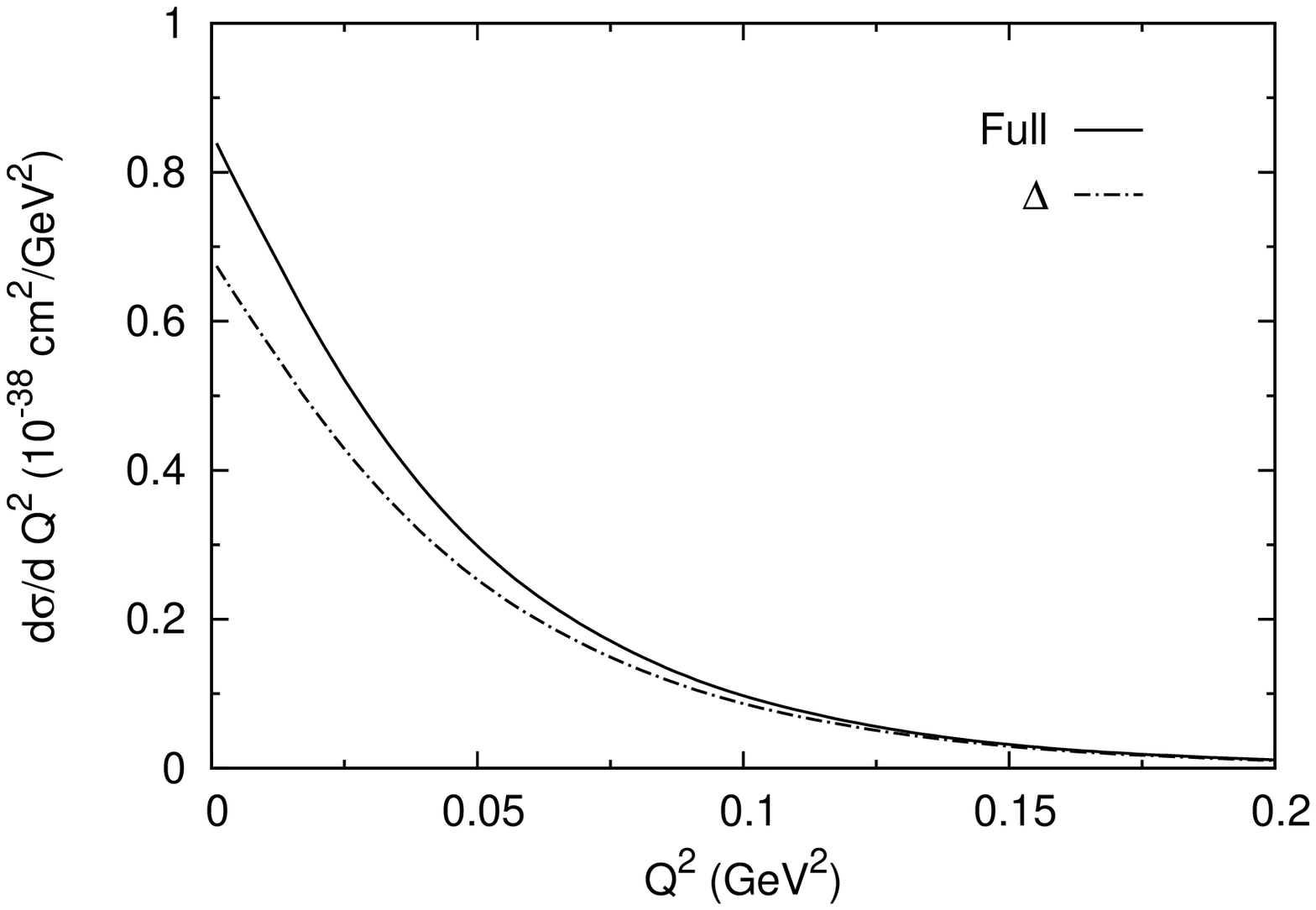}
\caption{\label{fig_q2.nc}
Same as in Fig.~\ref{fig_q2.cc} but for the NC process
at $E_\nu = 1$ GeV.
}
\end{center}
\end{minipage}
\end{figure}

We now proceed to present our results for differential observables.
In view of the fact that the event rates (cross section times flux)
in the K2K, MiniBooNE and SciBooNE experiments~\cite{K2K,miniboone}
have been reported to have a peak around $E_\nu \sim$ 1 GeV, 
we shall often use this energy as a representative in the following
presentation.
Meanwhile, since the neutrino flux 
in the planned T2K experiment is expected to have a
peak around $E_\nu =$ 0.6 $\sim$ 0.7 GeV\cite{kato},
we shall also present results for lower neutrino energies
when that seems useful.

The pion momentum spectrum for CC
neutrino-induced coherent pion production
is shown in Fig.~\ref{fig_tpi.1gev} (Fig.~\ref{fig_tpi.0p5gev})
for $E_\nu$ = 1 GeV (0.5 GeV).
The importance of the medium effects 
manifests itself here in the same manner as 
in the photo-process (Fig.~\ref{fig_krusche}).
In the $\Delta$ region, strong pion absorption 
is seen to reduce the cross sections significantly,
and FSI shifts the peak position.
The dash-dotted line corresponds to a case
in which the pion production operator
contains only the $\Delta$ mechanism 
(without non-resonant contributions), 
while the pion optical potential is kept unchanged. 
We note that, at $E_\nu$ = 1 GeV (0.5 GeV),
the dash-dotted line corresponds to 
82\% (64\%) of the solid line (the results of the full calculation). 
We have seen in the photo-process that the non-resonant mechanism is
more important for a smaller energy transfer. 
To what extent the neutrino case should share this feature
is not obvious because the axial-vector current contributions
dominate here 
(see Fig.~\ref{fig_v-a}).
However, we can see in Figs.~\ref{fig_tpi.1gev} and \ref{fig_tpi.0p5gev}
that, in the neutrino case as well,
the differential cross sections with smaller pion momenta 
are more enhanced by the non-resonant mechanism, and that 
this feature is more prominent for a smaller value of $E_\nu$.
A similar tendency is seen for the NC process also.
These results indicate that the non-resonant amplitudes
in our model, which are dressed by the rescattering, 
play a significant role in coherent pion production; 
their role is particularly important for $E_\nu$\ltap 0.5 GeV.
This characteristic feature of our model 
should be contrasted with the fact that (tree-level) non-resonant mechanisms
play essentially no role in any of the previous microscopic calculations
for neutrino-induced coherent pion production.
A more detailed comparison of the elementary amplitudes used in 
our present calculation 
and the previous microscopic-model calculations 
will be given later in Sec.~\ref{sec_comp}.

We show in Fig.~\ref{fig_q2.cc} (Fig.~\ref{fig_q2.nc})
the $Q^2$-distribution for the CC (NC) process.
Note that $Q^2$ defined by 
$Q^2 \equiv -q^2 \equiv- (p_\nu-p_\ell^\prime)^2$
is different from $Q^2_{\rm rec}$ defined in \Eq{q_rec}.
Because of the nuclear form factor effect, 
the distribution rises sharply as $Q^2$ approaches 0; 
for the CC process, however, $Q^2$-distribution 
becomes zero at $Q^2 = 0$
due to the finite muon mass.
Here again we show the results corresponding to a case
in which the pion production operator
contains only the $\Delta$ effect 
(with non-resonant contributions turned off).   
The non-resonant mechanism is seen to change
the spectrum shape significantly and lead to a sharper peak.

It is informative to examine the individual contributions 
of the vector and axial-vector currents.
We show in Fig.~\ref{fig_v-a} these individual contributions
to the neutrino CC process.
We find strong dominance of the axial-vector current.
The nuclear form factor causes the drastic suppression 
of non-forward pion production.
This aspect combined with the fact that
the transverse photon coupling of the vector current [\Eq{bar_fv}]
forbids forward pion production 
leads to strong suppression of the vector current contribution.
By contrast, since the vertex structure of the axial-vector current
favors forward pion production,
the strong suppression mechanism at work for the vector current
does not apply here.
This is the reason why the axial-vector current dominates.
This result may be used to argue that incoherent pion production processes
in which a nucleus does not break up but transits to excited states,
are much less important than coherent pion production in the
neutrino-nucleus scattering.
As seen in Fig.~\ref{fig_krusche}, the incoherent processes give
considerable contributions to the total pion production in the
photo-process,\footnote{
The contributions from the incoherent processes are larger than
they appear in Fig.~\ref{fig_krusche} because $\sin\theta_\pi$ needs to be
multiplied in integrating over $\theta_\pi$.
}
a feature that may lead to the expectation that the incoherent processes 
are considerable in the neutrino process as well.
However, the mechanism responsible for the axial-vector dominance
in the neutrino process works for the photo process
in such a manner that coherent photo-pion production
is strongly  suppressed.
Also, the inelastic transition form factor has a peak 
at a non-zero momentum transfer.
As a result, for the photo reaction,
the contributions from the incoherent processes 
become comparable to those from the coherent process.
Thus the importance of the incoherent processes relative 
to the coherent process can be very different 
between the photo and neutrino processes.
Takaki et al.~\cite{takaki} used a similar argument to explain
a significant (very small) contribution from the incoherent processes in
the photo-pion production (pion-nucleus scattering),
compared to the coherent process.
This argument may serve as a justification for the assumption
currently used in data analyses
that the incoherent processes need not be
taken into account explicitly .

\begin{figure}[t]
\begin{minipage}[t]{77mm}
 \begin{center}
 \includegraphics[width=77mm]{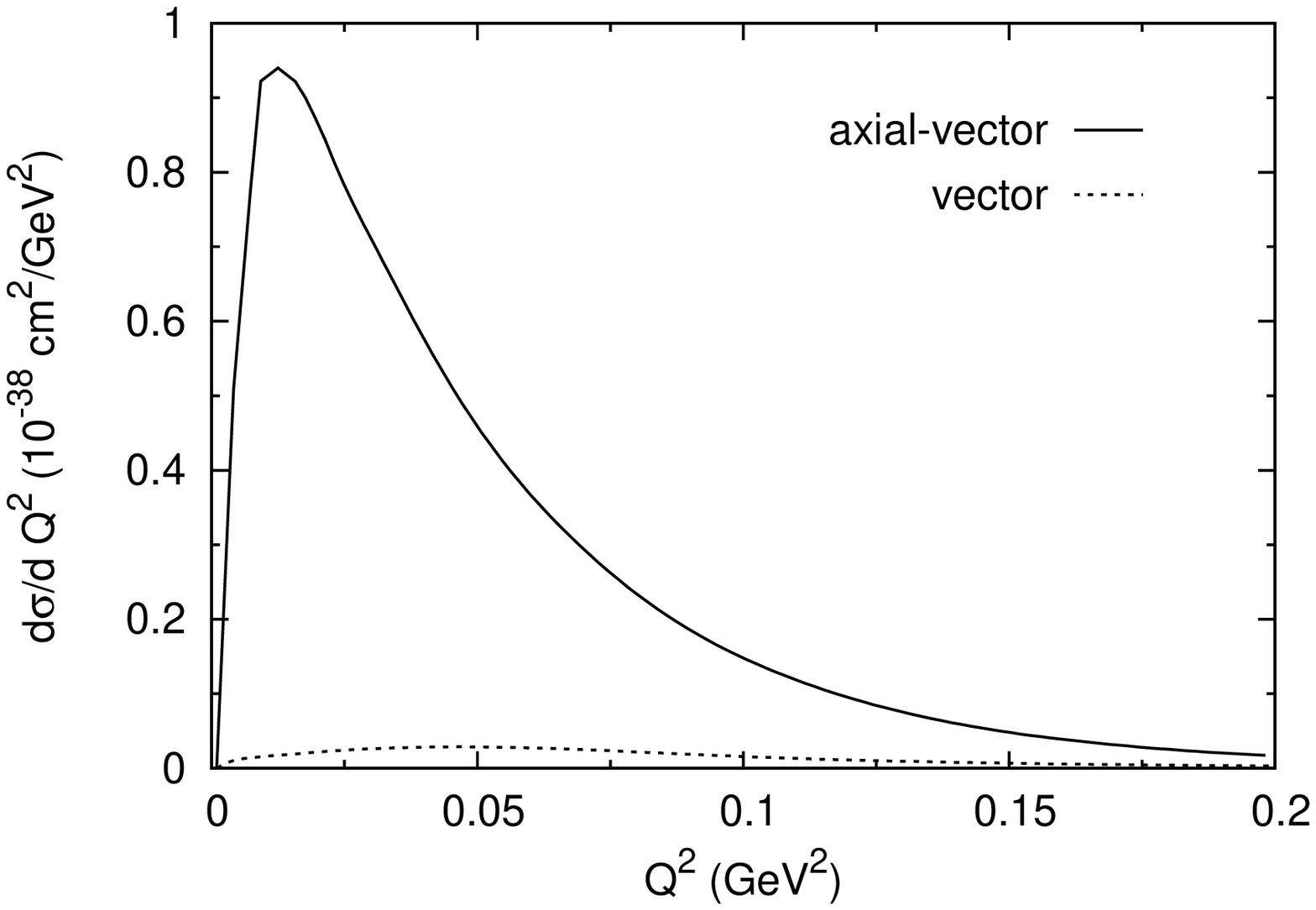}
 \caption{\label{fig_v-a}
Contributions from the axial-vector (solid) and vector (dashed) currents for
$\nu_\mu + {}^{12}{\rm C}_{g.s.} \to \mu^- + \pi^+ + {}^{12}{\rm C}_{g.s.}$
at $E_\nu = 1$ GeV.
 }
 \end{center}
\end{minipage}
\hspace{5mm}
\begin{minipage}[t]{77mm}
\begin{center}
 \includegraphics[width=77mm]{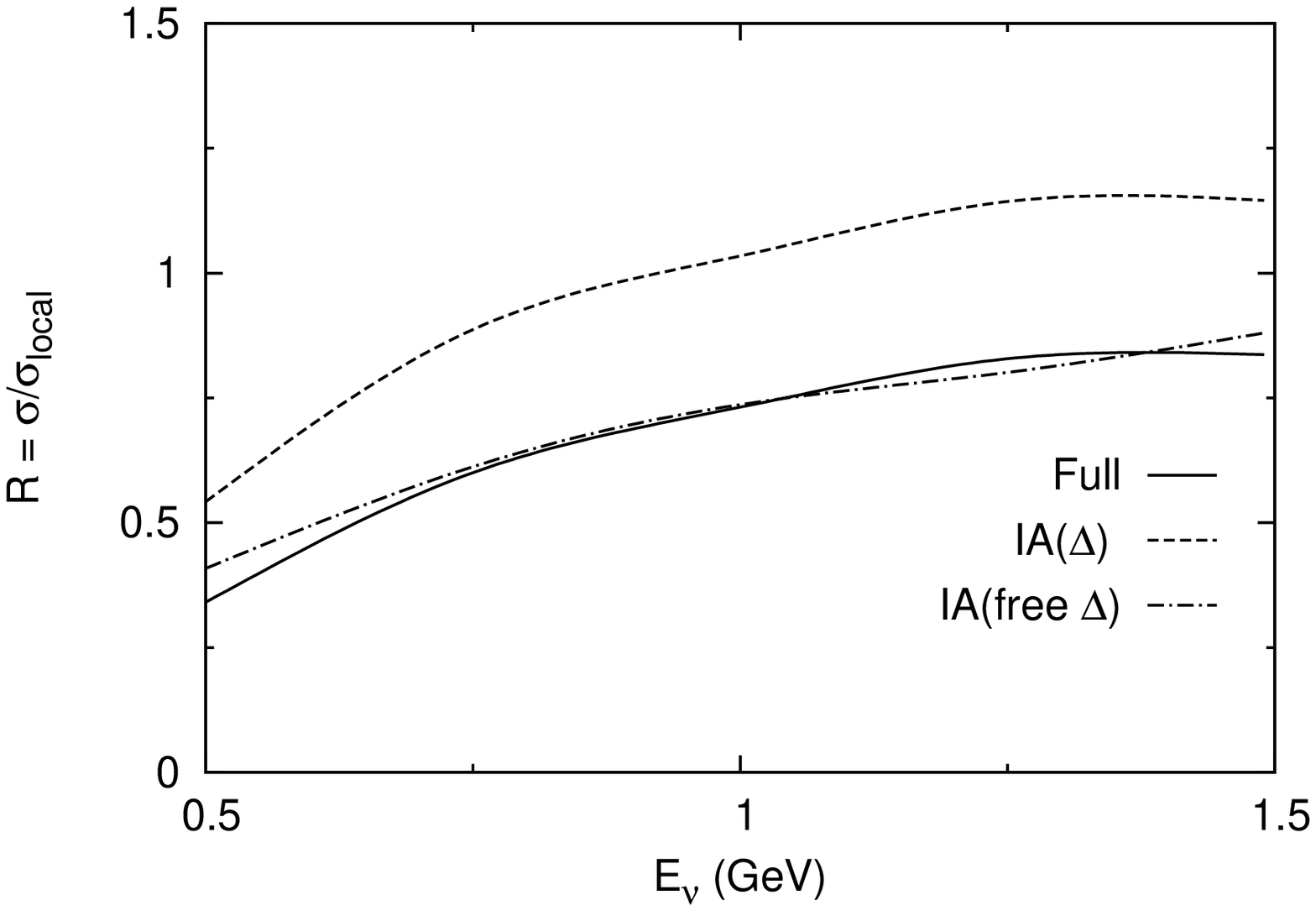}
 \caption{\label{fig_non_local}
The effect of the non-locality of the $\Delta$-propagation for
$\nu_\mu \!+\! {}^{12}{\rm C}_{g.s.}
\to \mu^- \!+\! \pi^+\! +\! {}^{12}{\rm C}_{g.s.}$.
The ratio ${\cal R}$ of the total cross sections 
obtained with and without taking
 account of the non-local effect.
 }
\end{center}
\end{minipage}
\end{figure}

Finally, we examine the effect of the non-locality of 
$\Delta$-propagation in nuclei;
because we employed the local density approximation for evaluating 
the $\Delta$ Green function [\Eq{nuclear-delta-prop}],
this effect arises only from the $\Delta$ kinetic term in the 
$\Delta$ Hamiltonian [\Eq{delta_H}].
Although, as mentioned in the introduction, 
this subject has been studied in Ref.~\cite{non-local},
that study only included the $\Delta$-mechanism 
without considering FSI or the medium effects on the $\Delta$.
It is thus interesting to revisit this problem 
in the framework of our significantly extended treatment.
In the local approximation, we neglect the kinetic term in the
$\Delta$-Hamiltonian [\Eq{delta_H}], which means that the $\Delta$ 
is considered to be so heavy that it does not propagate 
in nuclear medium.
To facilitate our discussion,
we introduce the ratio ${\cal R}(E_\nu)$ defined by
\be
{\cal R}(E_\nu)\equiv \sigma(E_\nu)
/\sigma_{\rm local}(E_\nu)\,,
\ee
where $\sigma(E_\nu)$ represents the total cross section for 
$\nu_\mu \!+\! {}^{12}{\rm C}_{g.s.}
\to \mu^- \!+\! \pi^+\! +\! {}^{12}{\rm C}_{g.s.}$
calculated with the $\Delta$-propagator including the $\Delta$ kinetic
term, whereas
$\sigma_{\rm local}(E_\nu)$ is that obtained in the local approximation.
Figure~\ref{fig_non_local} shows ${\cal R}(E_\nu)$
calculated for the various cases.
The long-dash curve corresponds to the $\Delta$-only case
(without FSI or the medium effects on the $\Delta$; see 
footnote~\ref{foot:medium}
)
and the solid line to the case that includes the non-resonant components,
medium effects on the $\Delta$ and FSI.
To make comparison with Ref.~\cite{non-local},
we first consider the long-dash line;
${\cal R}(E_\nu)$ in this case 
is found to be $0.55$, $1.03$ and $1.14$
at $E_\nu$ = 0.5, 1.0 and 1.5 GeV.
Meanwhile,  Ref.~\cite{non-local} reports
${\cal R}(E_\nu)$ = \ltap $0.5$, $0.6$ and \ltap 1
at $E_\nu$ = 0.5, 1.0 and 1.5 GeV.
Although both calculations indicate 
that the non-local effects are important,
our results are qualitatively different 
from those of Ref.~\cite{non-local}.
This difference originates from different ways 
of treating the energy in the $\Delta$-propagator.
In Ref.~\cite{non-local}, the in-medium $\Delta$-propagator 
is assumed to be the same as
the free $\Delta$-propagator, whereas our $\Delta$-propagator
[$G_{\Delta h}$, \Eq{nuclear-delta-prop}] is a
nuclear many-body operator~\cite{taniguchi} 
(with some of the medium effects switched off).
To illustrate this point, 
we include in Fig.~\ref{fig_non_local} (dash-dotted line)
the results obtained with the use of the free $\Delta$-propagator.
In this case, we find
${\cal R}(E_\nu)$ = $0.4$, $0.76$ and $0.88$
at $E_\nu$ = 0.5, 1.0 and 1.5 GeV,
which is fairly close to the results in Ref.~\cite{non-local}.
The result shown by the solid line indicates
that, after the sophistication of the calculation, 
 the non-locality due to the kinetic term is still important
over the entire range of $E_\nu$ under consideration.
In the previous microscopic calculations for neutrino-induced
coherent pion production, the non-locality has not been explicitly taken
into account.
However, this does not necessarily mean that the earlier results
are off  by an amount suggested
by comparison of the curves in Fig.~\ref{fig_non_local},
for it is possible that the non-locality effects are
partly included with the use of the spreading potential fitted to observables. 
In view of the importance of the non-local effect, however,
we consider it preferable to take it into account explicitly, 
rather than include it operationally in the $\Delta$ mass shift.

An additional point of interest is that it was reported
in Ref.~\cite{non-local} that the non-locality changes 
the shapes of the differential cross sections.
We remark that our results (not shown here) agree with that finding.

\subsection{Comparison with SciBooNE and MiniBooNE data}

\begin{figure}[t]
\begin{minipage}[t]{77mm}
 \begin{center}
 \includegraphics[width=77mm]{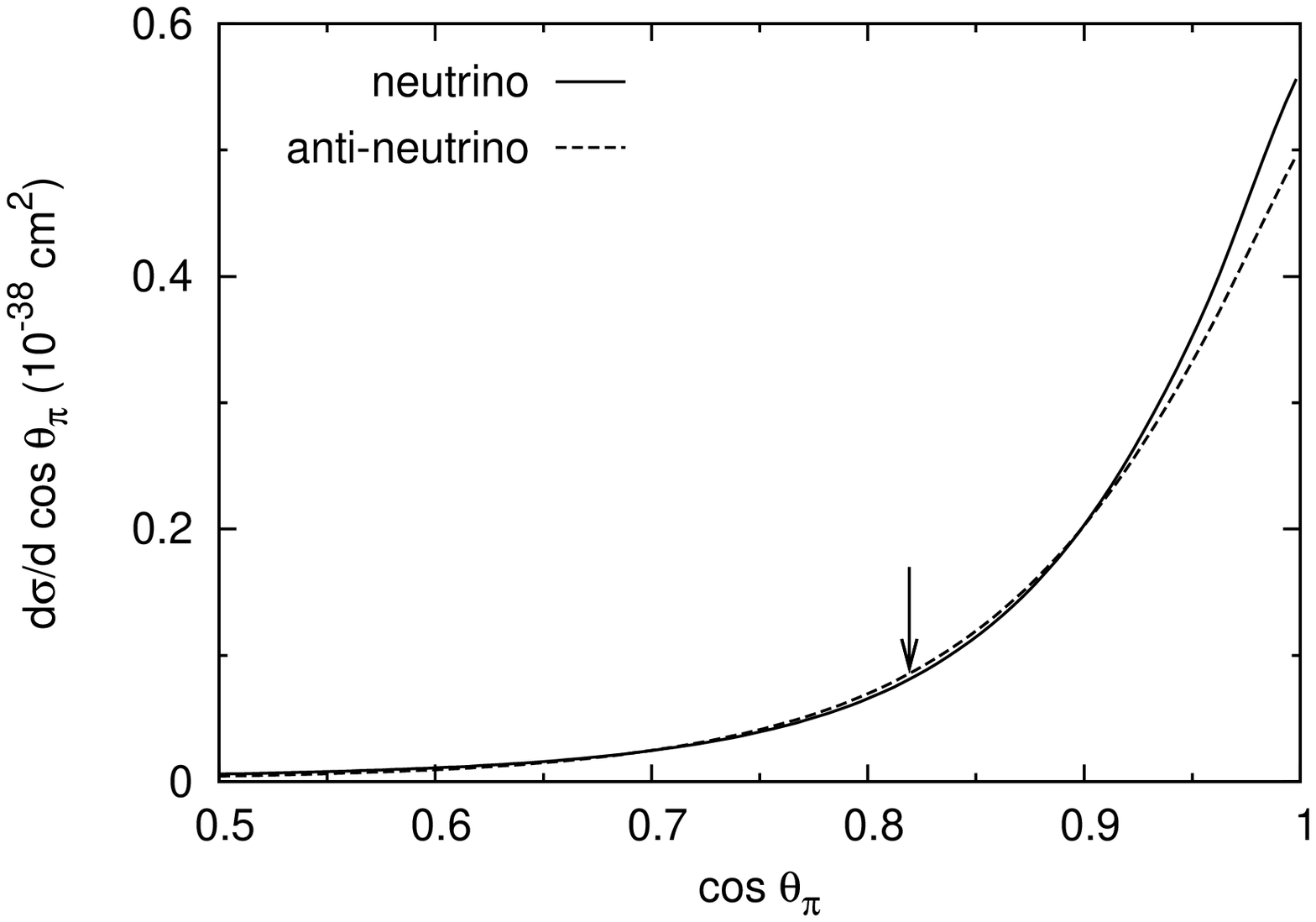}
 \caption{\label{fig_theta_p.sci}
The $\cos\theta_\pi$-distribution 
for $\nu_\mu + {}^{12}{\rm C}_{g.s.} \to \mu^- + \pi^+ + {}^{12}{\rm C}_{g.s.}$
and $\bar{\nu}_\mu + {}^{12}{\rm C}_{g.s.} \to \mu^+ + \pi^- + {}^{12}{\rm C}_{g.s.}$
at $E_\nu = 1$ GeV.
The position of $\theta_\pi = 35^\circ$ is indicated by the arrow.
 }
 \end{center}
\end{minipage}
\hspace{5mm}
\begin{minipage}[t]{77mm}
 \begin{center}
 \includegraphics[width=77mm]{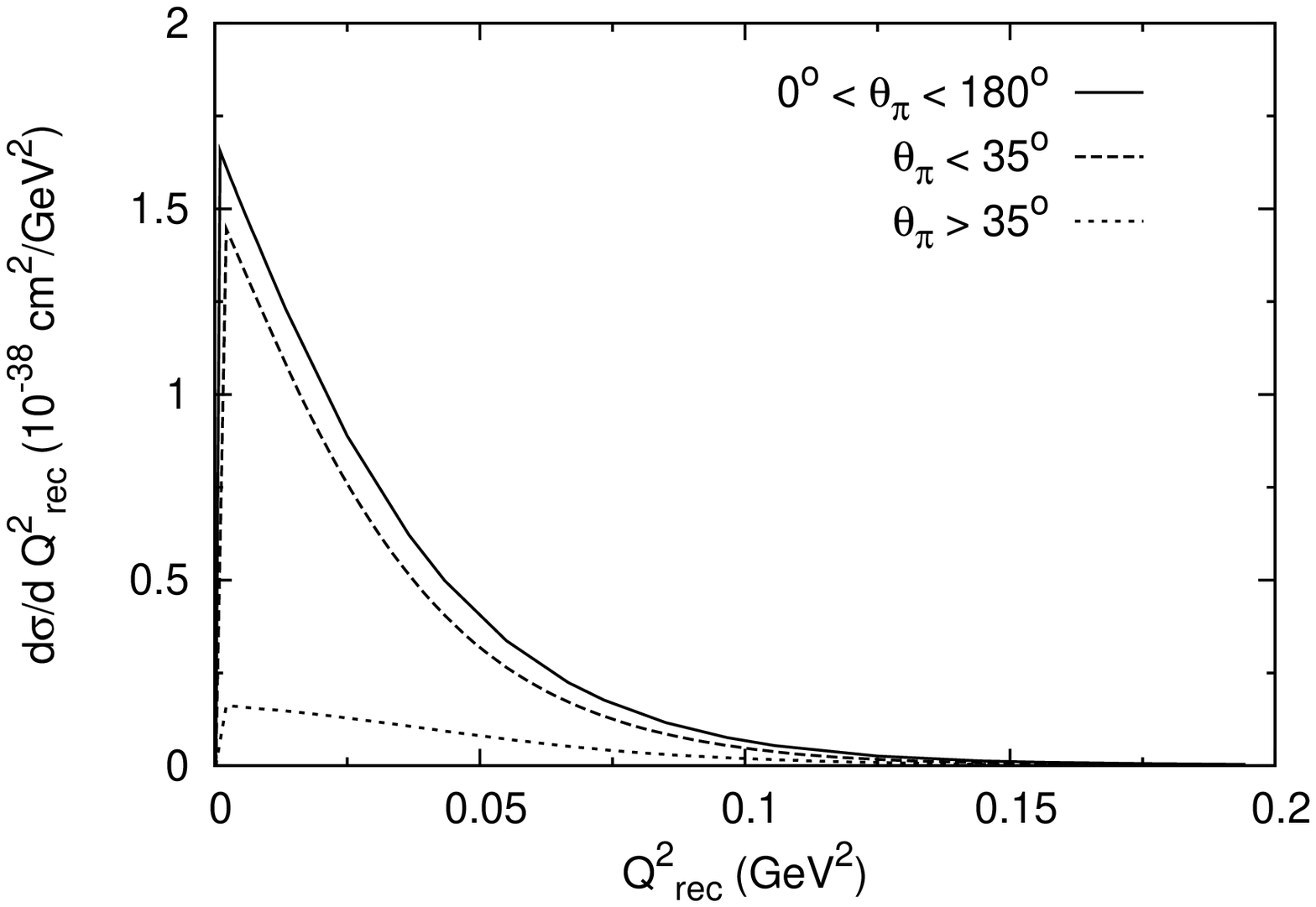}
 \caption{\label{fig_q2_rec.sci}
The $Q^2_{\rm rec}$ distribution
for $\nu_\mu + {}^{12}{\rm C}_{g.s.} \to \mu^- + \pi^+ + {}^{12}{\rm C}_{g.s.}$
at $E_\nu = 1$ GeV.
 }
 \end{center}
\end{minipage}
\end{figure}

The SciBooNE collaboration has been pursuing a further analysis 
of the data on neutrino and anti-neutrino CC coherent pion production,
and some preliminary results have appeared
in Refs.\cite{hiraide_nuint09,anti_cc}.
These results contain detailed information 
on the differential observables for the pion and muon, 
and it seems informative to present our theoretical results
in a manner that allows ready comparison with these data.
To this end, we need to take into account
the muon momentum cut ($p_\mu >$ 350~MeV)
and the momentum transfer cut ($Q^2_{rec} <$ 0.1~GeV$^2$)
adopted in the SciBooNE experiment; 
$Q^2_{rec}$ has been defined in \Eq{q_rec}.
The theoretical results we present in the following
take account of these cuts unless otherwise stated.
We will present the results at $E_\nu$ = 1~GeV around which 
the event rate has a peak.
Although, for direct comparison, we need to convolute the observables with the
(anti-\ )neutrino flux used in the SciBooNE experiment,
the flux  has not been released yet.
We therefore present our results at a representative
value of $E_\nu$ = 1 GeV.
In Fig.~\ref{fig_theta_p.sci}, we show the $\cos\theta_\pi$-distribution for
the neutrino and anti-neutrino CC processes. 
In the recent data analysis by the SciBooNE collaboration, 
events are classified according to the pion emission angle ($\theta_\pi$).
Their preliminary results
exhibit a rather clear excess yield 
for $\theta_\pi<35^\circ$, which is thought to be 
ascribable to coherent pion production.
In our model, 85\% of the pions are emitted in $\theta_\pi<35^\circ$
for the neutrino CC process at $E_\nu =$ 1~GeV,
a feature that is in fair agreement
with the preliminary SciBooNE result .

\begin{figure}[t]
\begin{minipage}[t]{77mm}
 \begin{center}
 \includegraphics[width=77mm]{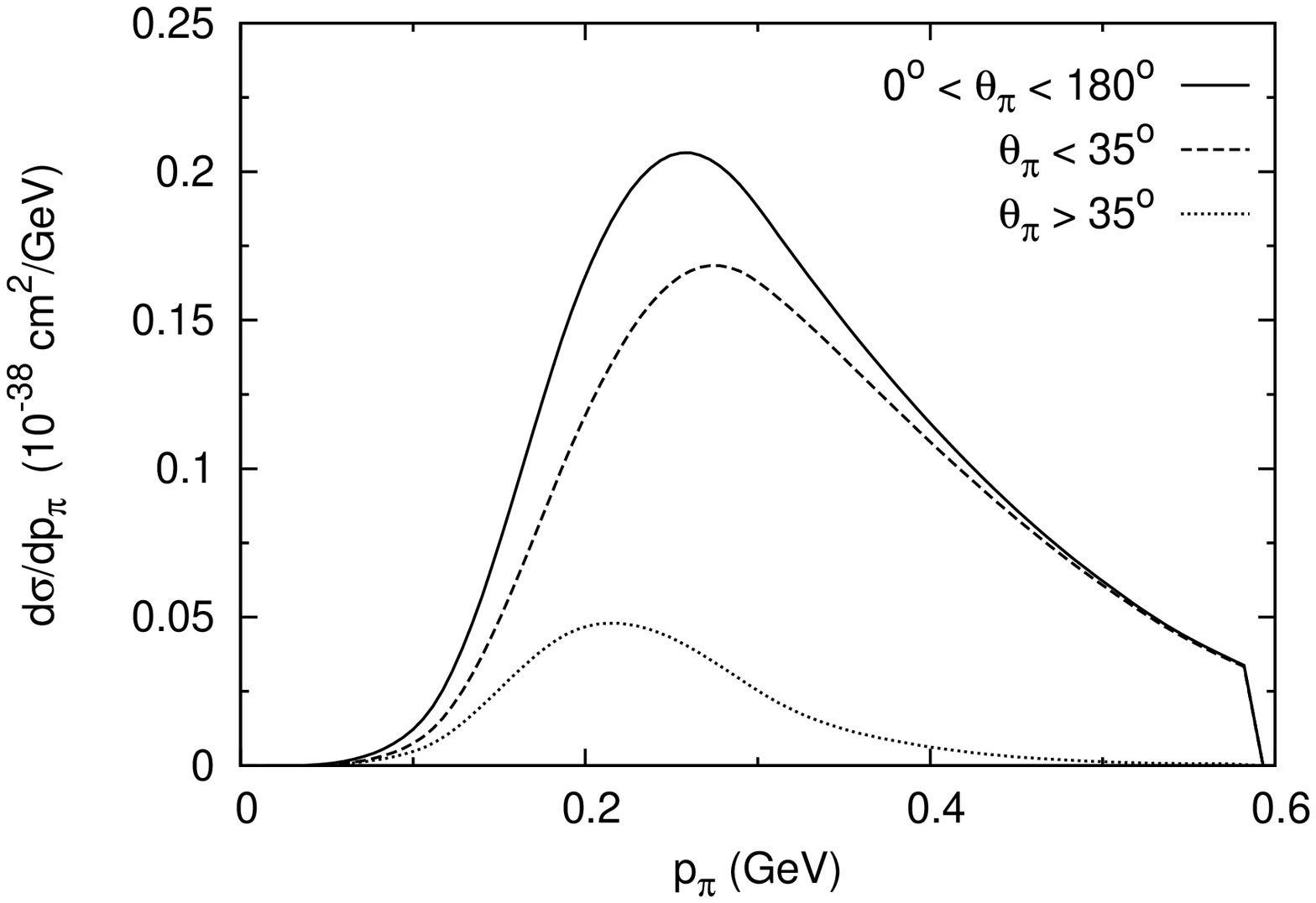}
 \caption{\label{fig_pmom.sci}
The pion momentum distribution
for $\nu_\mu + {}^{12}{\rm C}_{g.s.} \to \mu^- + \pi^+ + {}^{12}{\rm C}_{g.s.}$
at $E_\nu = 1$ GeV.
 }
 \end{center}
\end{minipage}
\hspace{5mm}
\begin{minipage}[t]{77mm}
 \begin{center}
 \includegraphics[width=77mm]{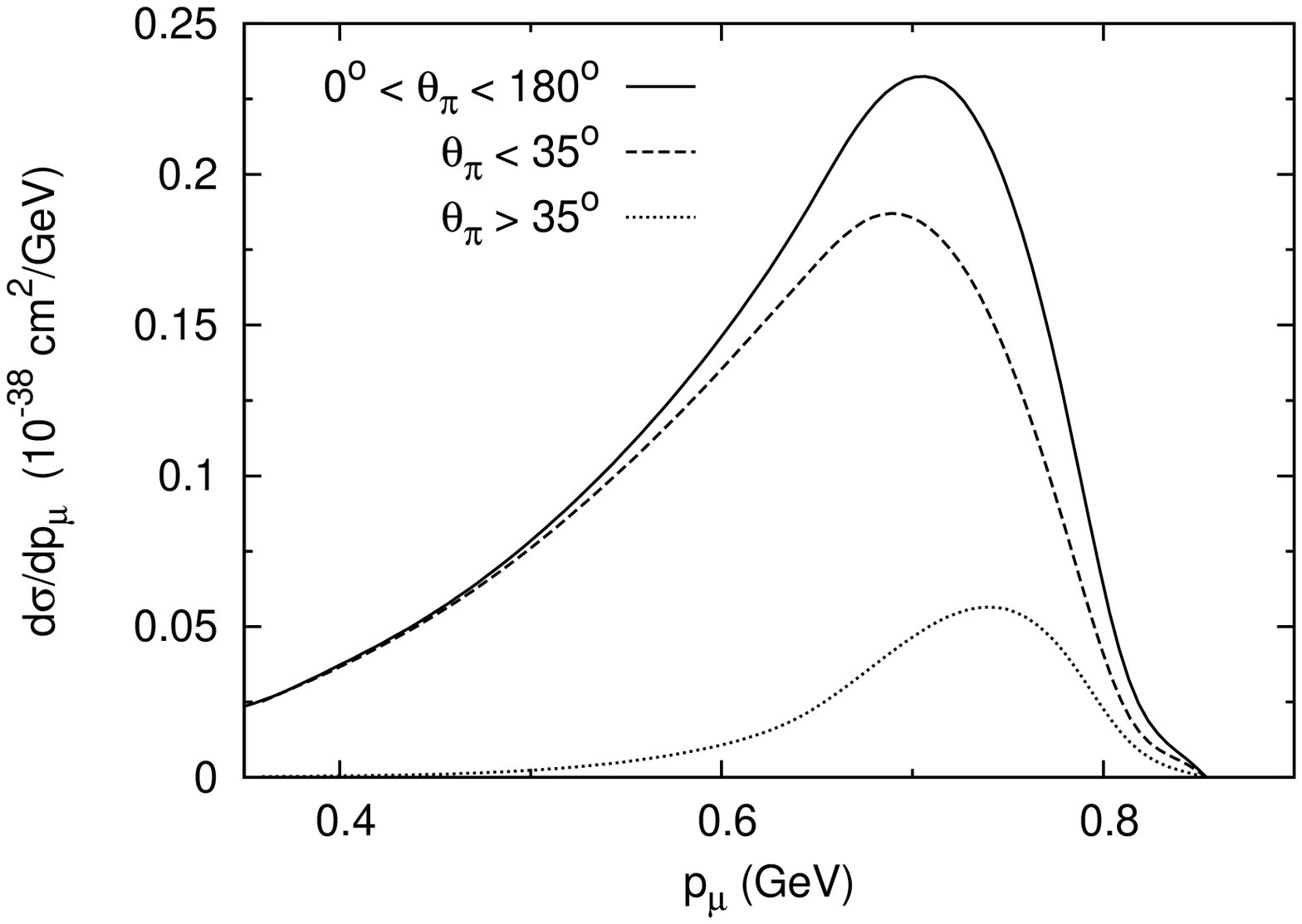}
 \caption{\label{fig_lmom.sci}
The muon momentum distribution
for $\nu_\mu + {}^{12}{\rm C}_{g.s.} \to \mu^- + \pi^+ + {}^{12}{\rm C}_{g.s.}$
at $E_\nu = 1$ GeV.
 }
 \end{center}
\end{minipage}
\end{figure}

\begin{figure}[t]
 \begin{center}
 \includegraphics[width=77mm]{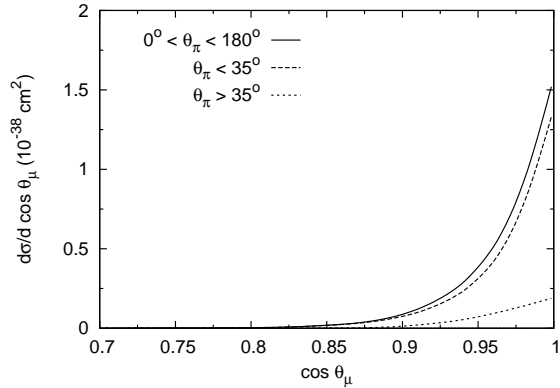}
 \caption{\label{fig_theta_l.sci}
The muon scattering angle distribution
for $\nu_\mu + {}^{12}{\rm C}_{g.s.} \to \mu^- + \pi^+ + {}^{12}{\rm C}_{g.s.}$
at $E_\nu = 1$ GeV.
 }
 \end{center}
\end{figure}

\begin{figure}[t]
\begin{minipage}[t]{77mm}
 \begin{center}
 \includegraphics[width=77mm]{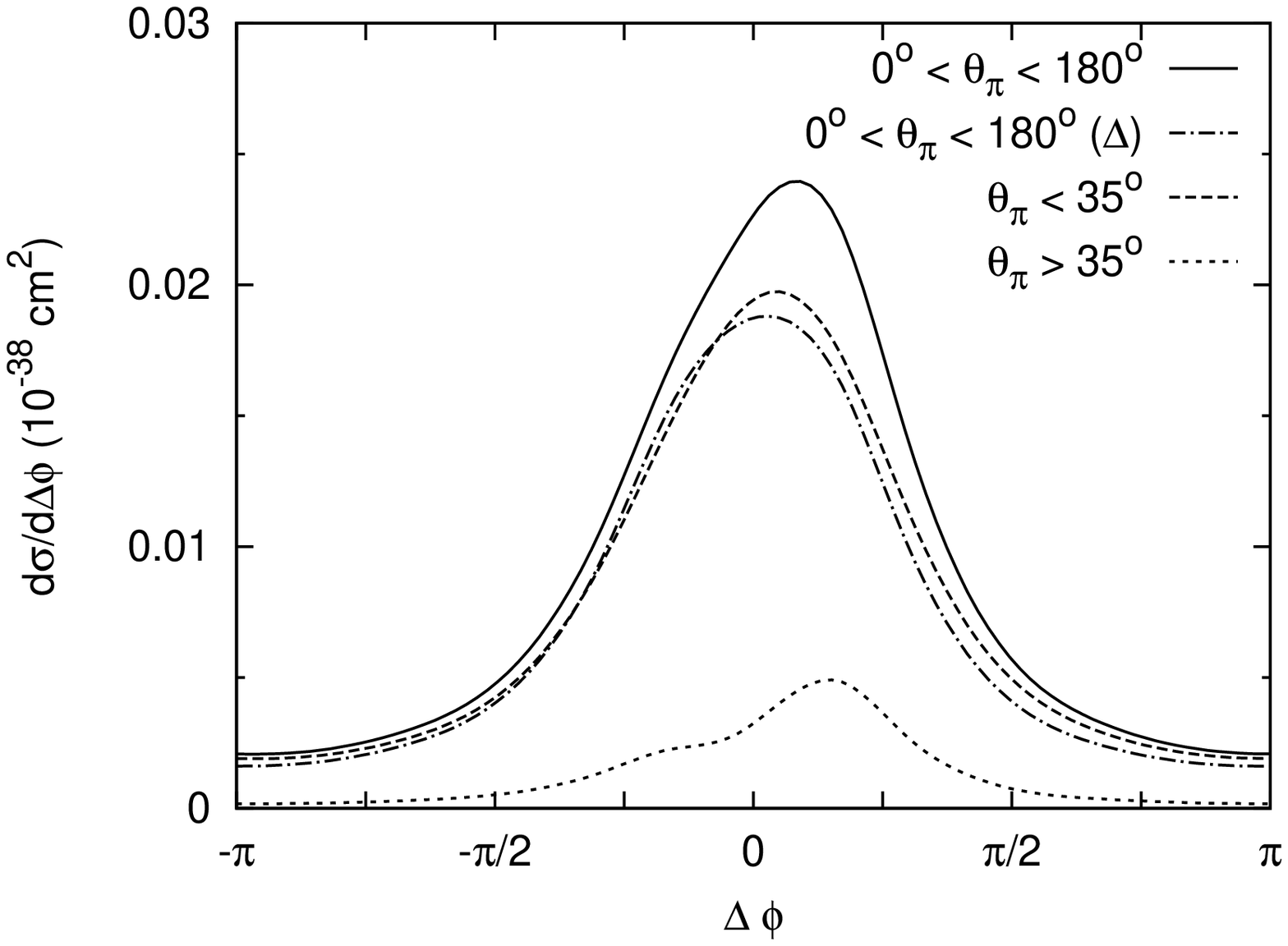}
 \caption{\label{fig_coplanar.sci}
The coplanar angle difference ($\Delta\phi$) distribution 
for $\nu_\mu + {}^{12}{\rm C}_{g.s.} \to \mu^- + \pi^+ + {}^{12}{\rm C}_{g.s.}$
at $E_\nu = 1$ GeV. The definition of $\Delta\phi$ is given in 
Fig.~\ref{fig_coplanar}.
 }
 \end{center}
\end{minipage}
\hspace{5mm}
\begin{minipage}[t]{77mm}
 \begin{center}
 \includegraphics[width=77mm]{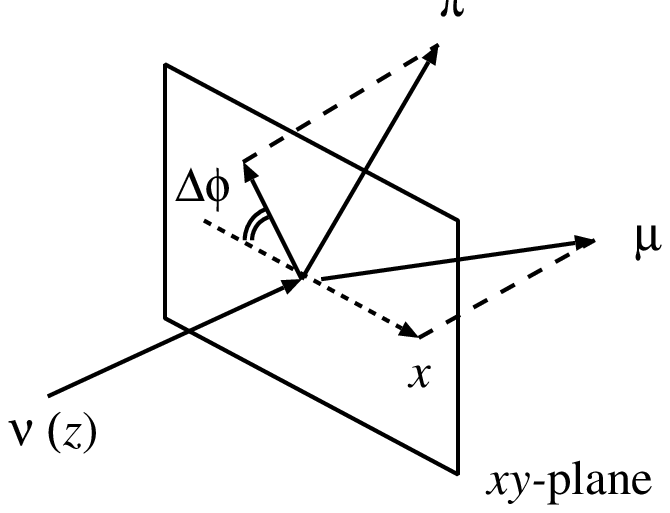}
 \caption{\label{fig_coplanar}
Graphical definition for the coplanar angle difference ($\Delta\phi$).
 }
 \end{center}
\end{minipage}
\end{figure}

Next we show in Fig.~\ref{fig_q2_rec.sci} (solid line)
the $Q^2_{\rm rec}$ distribution
for the neutrino reaction.\footnote{As discussed earlier,  
the neutrino and anti-neutrino cross sections differ only slightly.} 
Only the $p_\mu $ cut is applied here
for an obvious reason.
We can see that the contribution from above  
$Q^2_{\rm rec}$ = 0.1 GeV$^2$
(the value adopted for the $Q^2_{\rm rec}$ cut) 
constitutes only a small fraction 
of the entire contribution (3\% for the solid curve).
The decomposition of the total contribution (solid curve)
into two parts according to whether 
$\theta_\pi$ is smaller or larger than $35^\circ$ 
is shown by the dashed curve ($\theta_\pi \!< \!35^\circ$)
and the dotted curve ($\theta_\pi \!>\! 35^\circ$).
The pion and muon momentum distributions are shown in
Figs.~\ref{fig_pmom.sci} and \ref{fig_lmom.sci}.
The upper (lower) end of the pion (muon) momentum distribution is
sharply cut off because of the muon momentum cut ($p_\mu >$ 350~MeV).
The muon scattering angle distribution is shown in 
Fig.~\ref{fig_theta_l.sci}.
Figures \ref{fig_theta_p.sci}--\ref{fig_theta_l.sci} clearly show
the characteristics of coherent pion production, i.e., sharply forward
scattering (emission) of the muon (pion) with small momentum transfers.
Finally, we show in Fig.~\ref{fig_coplanar.sci} 
the spectrum with respect to the coplanar angle difference, $\Delta\phi$,
which is defined by
$\Delta\phi = \phi_\pi -\pi$,
where $\phi_\pi$ is the pion azimuthal angle in
the LAB frame. 
(See Fig.~\ref{fig_coplanar} for a graphical
representation of $\Delta\phi$.)
Fig.~\ref{fig_coplanar.sci} shows slight asymmetry 
in the $\Delta\phi$ distribution around $\Delta\phi=0$.
It is interesting to note that
this asymmetry is generated mostly by the contribution
from the non-resonant amplitudes. 
To demonstrate this point, we present in the same figure
the results obtained with the non-resonant amplitudes turned off,
(dash-dotted curve).
We also remark that the asymmetry 
arises mostly from the kinematical region satisfying 
$\theta_\pi > 35^\circ$ (see the dotted curve).
A similar asymmetry also arises for the anti-neutrino process.

The SciBooNE collaboration have recently presented their preliminary
results corresponding to
Figs.~\ref{fig_theta_p.sci}--\ref{fig_coplanar.sci} for both of the
neutrino and anti-neutrino CC coherent pion
production reactions~\cite{hiraide_nuint09,anti_cc}.
When the flux prediction for the SciBooNE experiment becomes available,
we will be able to convolute the results of our calculation
with the flux and make direct comparison with the data.

\begin{figure}[t]
\begin{minipage}[t]{77mm}
 \begin{center}
 \includegraphics[width=77mm]{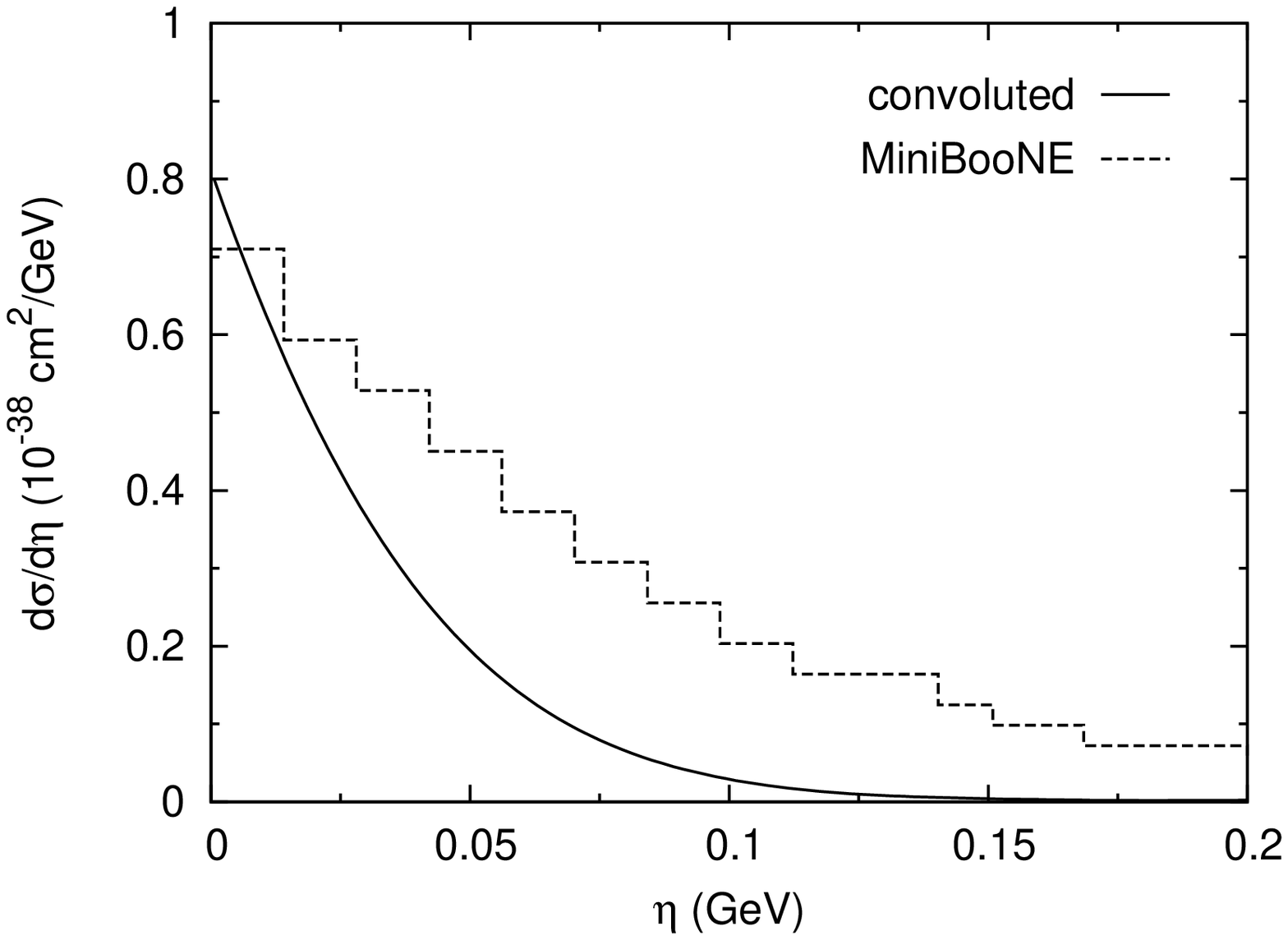}
 \caption{\label{fig_eta_average}
The flux-convoluted $\eta$-distribution 
for $\nu + {}^{12}{\rm C}_{g.s.} \to \nu + \pi^0 + {}^{12}{\rm C}_{g.s.}$ obtained in our full calculation.
The neutrino flux is taken from MiniBooNE~\cite{miniboone_flux}.
Also shown is the Monte Carlo result 
from MiniBooNE~\cite{miniboone} rescaled
to match our result at $\eta$ = 0.005 GeV.
 }
\end{center}
\end{minipage}
\hspace{5mm}
\begin{minipage}[t]{77mm}
 \begin{center}
 \includegraphics[width=77mm]{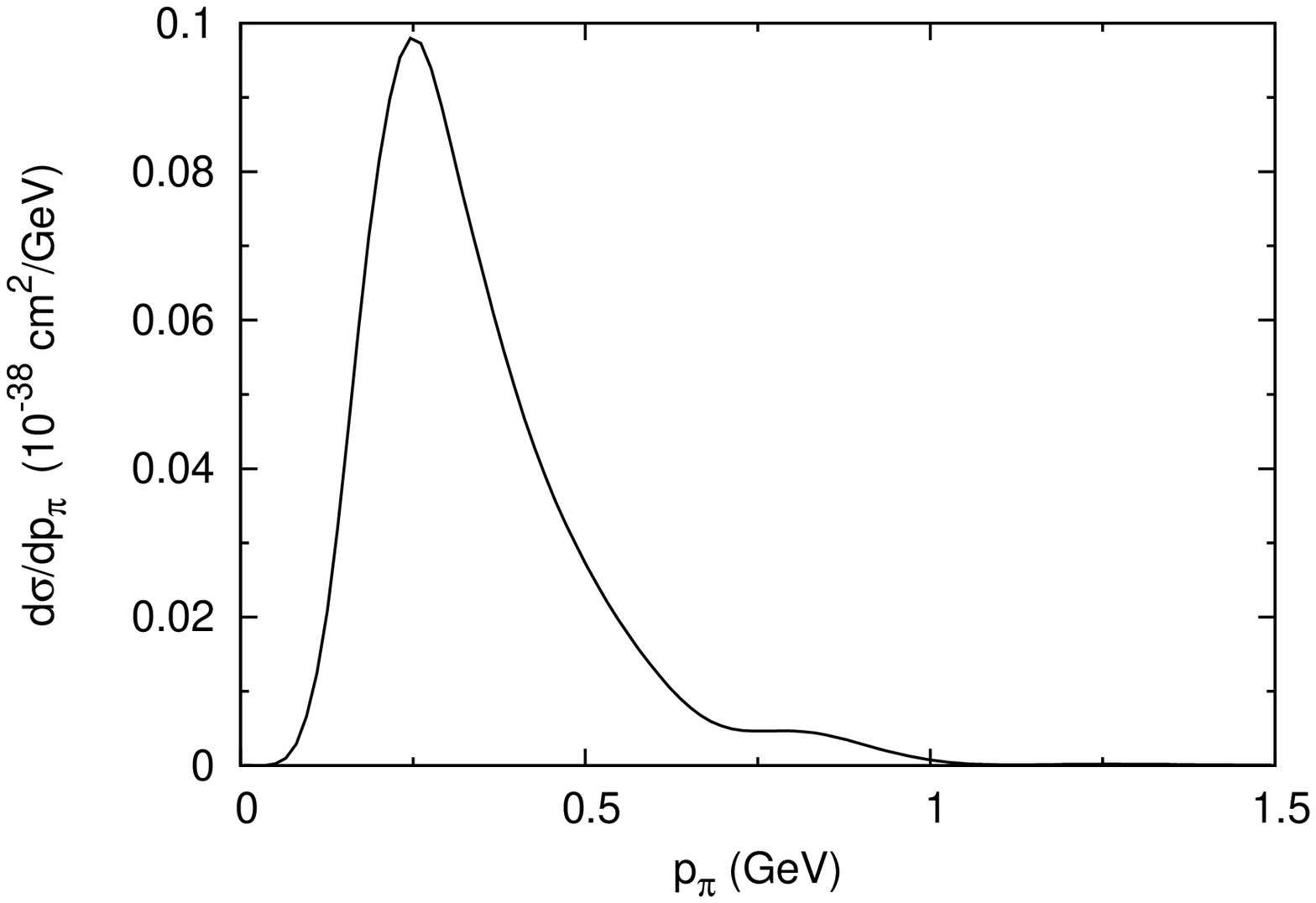}
 \caption{\label{fig_pi0_miniboone}
The flux-convoluted $\pi^0$ momentum distribution 
for $\nu + {}^{12}{\rm C}_{g.s.} \to \nu + \pi^0 + {}^{12}{\rm C}_{g.s.}$.
The neutrino flux is taken from MiniBooNE~\cite{miniboone_flux}.
 }
 \end{center}
\end{minipage}
\end{figure}

Meanwhile, the MiniBooNE collaboration has been investigating
the NC process in \\(anti-)neutrino-nucleus scattering,
and some results for the neutrino process have been
published~\cite{miniboone}, and more results are expected to be
released.
Since the neutrino flux information for the MiniBooNE 
experiment is available~\cite{miniboone_flux}, we can give
the theoretical values of relevant observables 
convoluted with the flux.
At present, data are publicly available 
only for the $\eta$-distribution
[\,$\eta \equiv E_\pi (1-\cos\theta_\pi)$],
and we compare our calculation for this quantity with the data.
In the analysis of the MiniBooNE NC data,
the $\eta$-distribution was used 
to distinguish coherent pion production from other processes
contributing to the $\pi^0$-production events.
To be more specific, MiniBooNE used the ``shape`` 
of the $\eta$-distribution obtained from the RS model~\cite{RS}
with the momentum reweighting function applied.
It has been found, however, 
that a microscopic calculation in Ref.~\cite{amaro}
gives an $\eta$-distribution appreciably different 
from that obtained in the RS model,
and the authors of Ref.~\cite{amaro}
have pointed out that the MiniBooNE
might have substantially overestimated the NC events.
Figure~\ref{fig_eta_average} shows the ``average'' $\eta$-distribution
obtained by convoluting the $\eta$-distribution 
given by our present calculation 
with the MiniBooNE neutrino flux~\cite{miniboone_flux}. 
For comparison, the figure also shows
the MiniBooNE Monte Carlo results 
({\it cf.} Fig.~3b of Ref.~\cite{miniboone}),
arbitrarily rescaled to match the theoretical curve
at $\eta$ = 0.005 GeV.
We remark that the $\eta$-distribution we have obtained 
is fairly close to that given in Ref.~\cite{amaro},
because the non-resonant amplitudes do not change the shape of
the $\eta$-distribution significantly.
Therefore, we arrive at the same conclusion as in Ref.~\cite{amaro}
that it is possible that MiniBooNE
substantially overestimated the NC events.

\begin{figure}[t]
\begin{minipage}[t]{77mm}
 \begin{center}
 \includegraphics[width=77mm]{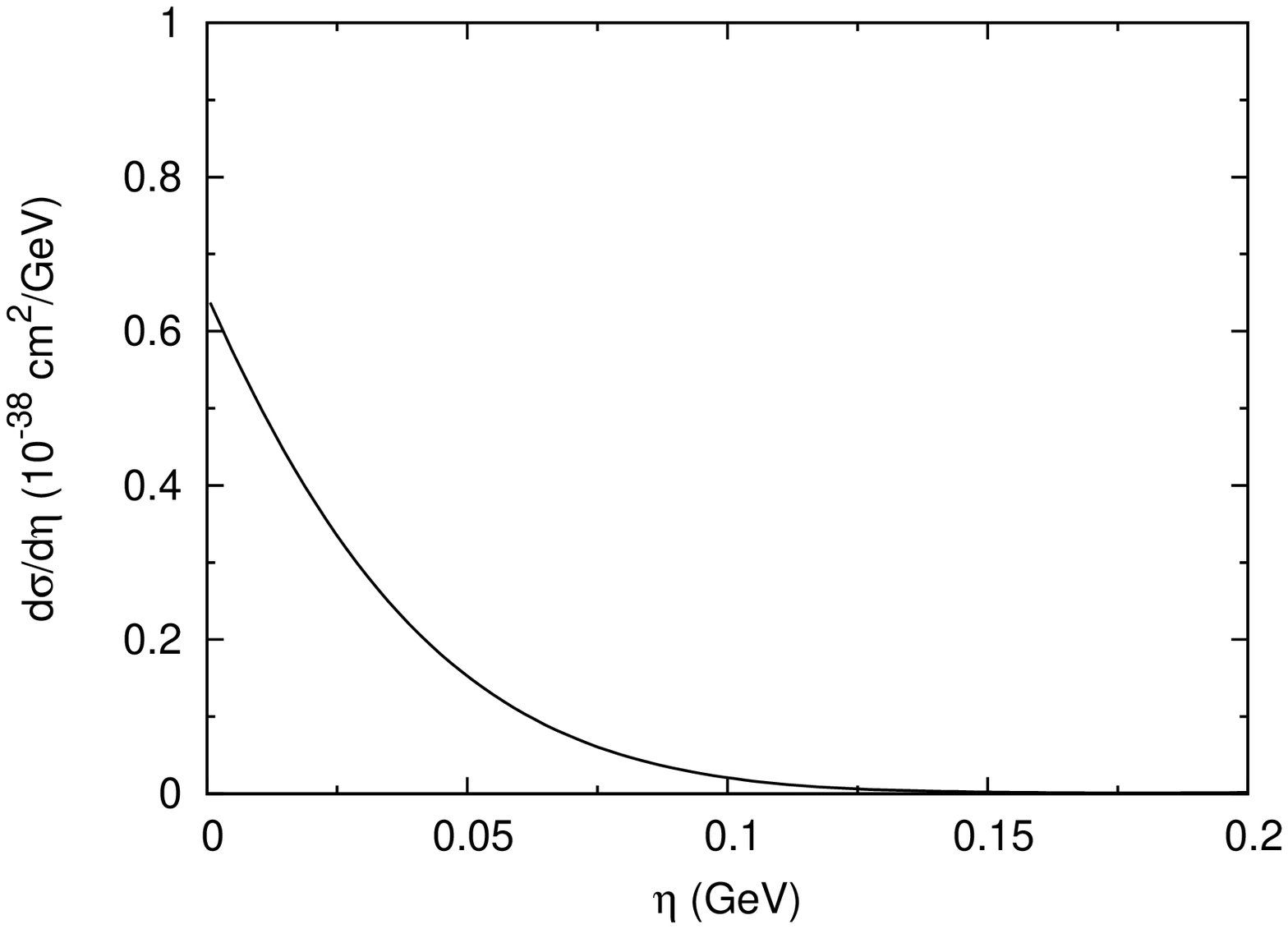}
 \caption{\label{fig_bnu_eta_average}
The flux-convoluted $\eta$-distribution 
for $\bar{\nu} + {}^{12}{\rm C}_{g.s.} \to \bar{\nu} + \pi^0 + {}^{12}{\rm C}_{g.s.}$.
The anti-neutrino flux is taken from MiniBooNE~\cite{miniboone_flux}.
 }
\end{center}
\end{minipage}
\hspace{5mm}
\begin{minipage}[t]{77mm}
 \begin{center}
 \includegraphics[width=77mm]{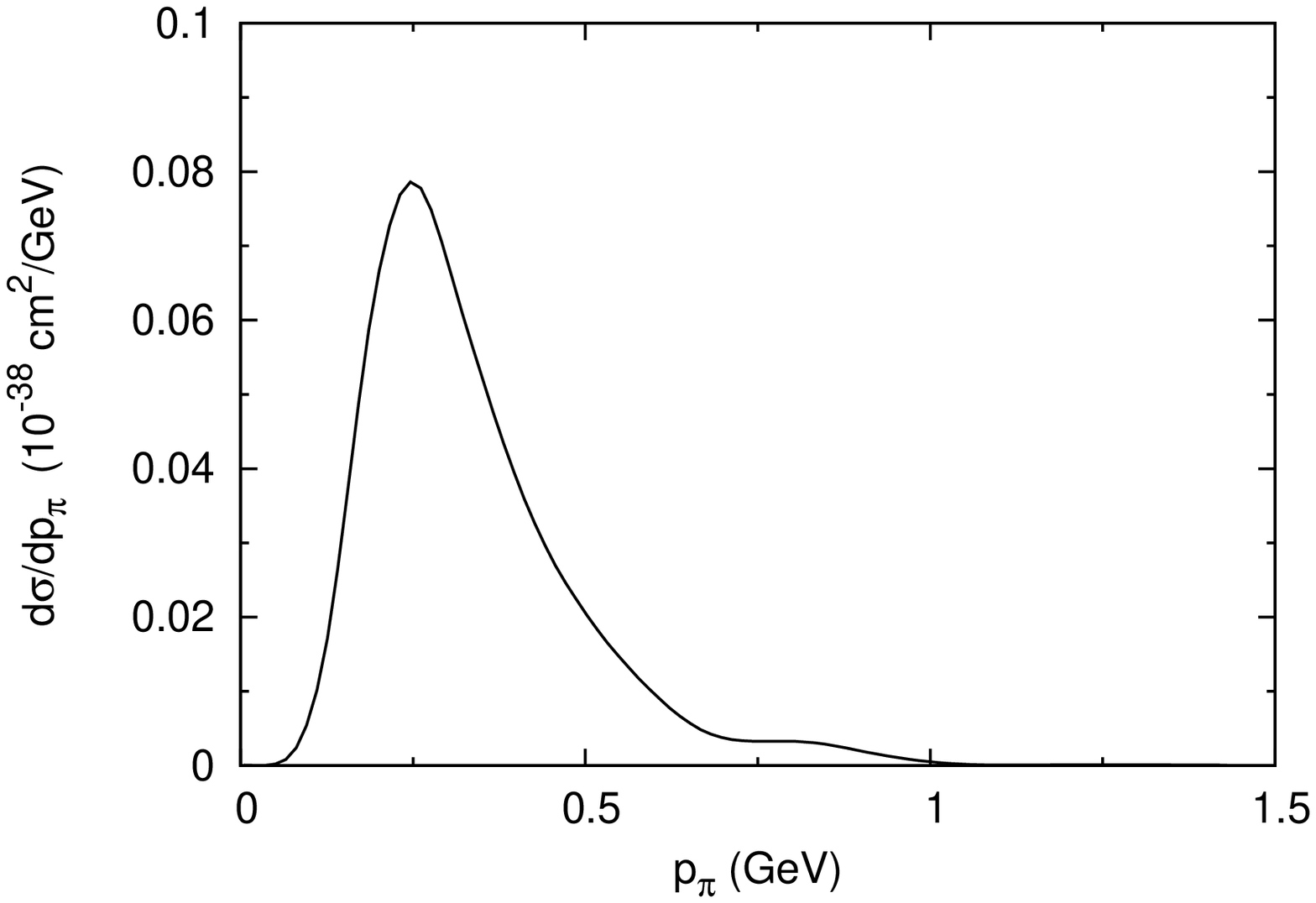}
 \caption{\label{fig_bnu_pi0_miniboone}
The flux-convoluted $\pi^0$ momentum distribution 
for $\bar{\nu} + {}^{12}{\rm C}_{g.s.} \to \bar{\nu} + \pi^0 + {}^{12}{\rm C}_{g.s.}$.
The anti-neutrino flux is taken from MiniBooNE~\cite{miniboone_flux}.
 }
 \end{center}
\end{minipage}
\end{figure}

To facilitate a comparison of our calculation with
data that are expected to be become available 
soon from MiniBooNE, 
we present theoretical predictions for some 
more quantities that are likely to be relevant.
Figure~\ref{fig_pi0_miniboone} shows 
the flux-convoluted $\pi^0$ momentum distribution 
predicted by our calculation.
As far as observables for the anti-neutrino process are concerned,
the flux-convoluted $\eta$-distribution resulting from our calculation
is given in Fig.~\ref{fig_bnu_eta_average},
and the flux-convoluted $\pi^0$ momentum distribution 
obtained in our model is shown in Fig.~\ref{fig_bnu_pi0_miniboone}.


\subsection{Comparison of Microscopic Models}
\label{sec_comp}

As mentioned, there are mainly two different theoretical
approaches to coherent pion production in neutrino-nucleus
scattering; a PCAC-based model and a microscopic model.
The relation between the RS model (a PCAC-based model) and 
a microscopic model has been discussed in great detail in Ref.~\cite{amaro}, 
and comparison of those two models, 
including some improvement of the RS model, 
has been made in Refs.~\cite{amaro,hernandez}.
The authors of Refs.~\cite{amaro,hernandez}
have emphasized that it can be problematic to
use the RS model for $E_\nu$\ltap 2 GeV.
To shed some more light on this issue,
we consider it useful to make comparison 
of different microscopic models.
In particular, we focus here on comparison between our model
and the model of Amaro et al.~\cite{amaro},
which is the most sophisticated among the existing microscopic models 
for neutrino-induced coherent pion production.
\footnote{
A rather extensive comparison of numerical results from various
calculations for the neutrino-induced coherent pion production,
including those of Amaro et al.~\cite{amaro}, 
recent PCAC-based models~\cite{paschos,bs_pcac} and
ours, has been presented at NuInt09 by Boyd et al.~\cite{nuint09}.
}
The other microscopic calculations in the literature
lack one or more aspects that are obviously important,
such as the distortion of the final pion and the non-resonant mechanism 
for the weak currents.

Here, we particularly focus on the elementary amplitudes 
for pion production off the nucleon.
Our approach employs the SL model while Amaro et al.~\cite{amaro} used 
a model developed in Ref.~\cite{hnv} (to be referred to as HNV).
Both SL and HNV include the resonant and non-resonant amplitudes.
A point to be noted, however, 
is that, although both models reproduce reasonably well
the data for the $\nu_\mu + N  \to \mu^- + \pi^+ + N$ reactions 
after an appropriate adjustment of the axial-$N\Delta$ coupling,
the two models involve rather different reaction mechanisms.
In the SL model, we derive a set of tree diagrams from a given
Lagrangian with the use of a unitary transformation, and then 
we embed these tree diagrams in the Lippmann-Schwinger equation,
which is solved exactly to yield a non-perturbative pion
production amplitude that satisfies $\pi$-$N$ two-body unitarity.
In HNV, on the other hand, a set of tree diagrams
are calculated from a chiral Lagrangian.
Then the sum of the contributions of these tree diagrams
is identified with the pion production amplitude.
At the tree level, the SL and the HNV models have essentially the same
non-resonant mechanisms; a contact vertex in HNV may be interpreted as
the vector meson exchange mechanism in SL.
However the role of the non-resonant amplitude
appears differently in the two models.
In the SL model, non-resonant amplitude contributes
constructively (destructively) to the resonant amplitudes 
below (above) the resonance energy.
For $\nu_\mu + p  \to \mu^- + \pi^+ + p$,
the interference of the non-resonant amplitude 
with the resonant amplitude changes in the SL model
the total cross sections by a factor of 1.5, 1.02, 0.96 at
$E_\nu =$ 0.5, 1, 1.5 GeV
\footnote{See footnote \ref{footnote:non-res}.
},
while the interference in the HNV always enhances
the total cross sections; e.g., enhancement of a factor of 1.1 
at $E_\nu =$ 1.5~GeV.
The difference of the non-resonant mechanism appears also in the 
coherent pion production on ${}^{12}$C,
where only the spin and isospin non-flip amplitude contributes.
Whereas the non-resonant amplitude plays an important role in our
model (as seen in Figs.~\ref{fig_tpi.1gev} and \ref{fig_tpi.0p5gev}),
it plays essentially no role in the HNV model.
In the neutrino CC coherent pion production, the full (tree)
non-resonant amplitude
increases the total cross section by 36\% (19\%) at $E_\nu$ = 0.5~GeV
and 18\% (0.4\%) at $E_\nu$ = 1~GeV in our model.
Thus the non-resonant mechanism in the spin-isospin non-flip amplitude
is enhanced by the rescattering process.
In the SL model, the non-resonant and resonant $\pi N$ dynamics in the
 $\Delta$ resonance region has been  tested using the extensive 
data of $(\gamma,\pi)$ and $(e,e^\prime\pi)$ reactions.
Although the SL model, which provides a unified description of the 
electroweak pion production reactions,  describes very well the available data
of the $(\nu,\ell\pi)$ processes,
the current data do not yet allow to test the details of the  reaction mechanism.

Furthermore, utilizing the consistency of 
$(\nu,\ell\pi)$, $(e,e^\prime\pi)$ and $(\pi,\pi)$ reactions
in the SL model,
we have developed a model which 
treats photo- and neutrino-induced coherent pion production
processes in a unified manner.
Thus we were able to calibrate the reliability of our approach
with data for the photo-processes, which is an aspect specific to
our approach.

\section{Conclusions}
\label{sec_conclusion}

We have developed a microscopic dynamical model
for describing neutrino-induced coherent pion production on nuclei.
Because experimental data for neutrino (both elementary and nuclear) processes
are rather limited, it is not straightforward to assess the reliability of
theoretical calculations.
A reasonable strategy to take
seems to develop a model which describes strong and electroweak
processes in a unified way, and then to test the model extensively 
by comparing with a large collection of data for the strong-interaction
and photo-induced processes and with limited available data for weak processes.
We have carried out this program here
for the case of  the neutrino-induced coherent pion production process.
By virtue of the mentioned strategy, our model is probably 
the most 
extensively tested one among the existing models for this process.
To achieve the stated goal, we need a theoretical framework
that provides a unified description for the elementary
$(\pi,\pi')$, $(\gamma,\pi)$ and $(\nu,\ell\pi)$ processes 
on a single nucleon.
We have adopted the SL model, which is known to give
satisfactory descriptions of these elementary amplitudes.
We then have combined the SL model with
the $\Delta$-hole model to construct a theoretical framework 
that can describe in a unified way 
pion-nucleus scattering and electroweak coherent 
pion production.
The unified nature of this approach allows us to fix free parameters
in the model using the data for pion-nucleus scattering, 
which in turn enables us to make parameter-free predictions 
on electroweak coherent pion production off a nucleus.
Another benefit of the present unified approach is 
that we can assess the reliability of our model
by comparing the results for coherent pion photo-production
with data.
Our model is found to describe reasonably well
both pion-nucleus scattering and coherent photo-processes, 
which establishes a basis for applying the same model
to the neutrino-induced processes.

Comparing our numerical results with the recent data
on neutrino-induced coherent pion production,
we have found that the result for the CC process is consistent 
with the upper limit from
K2K\cite{hasegawa}, and that the result for the NC process is somewhat 
smaller than the preliminary experimental value from MiniBooNE\cite{raaf}.
However, as discussed in the literature, MiniBooNE's analysis may have
overestimated the cross section due to the use of the RS model in their
analysis.
We have examined to what extent the various aspects 
of physics involved in our model individually affect 
the cross sections. 
We have shown that the medium effect on the $\Delta$ 
(the spreading potential effect in particular)
and the FSI change the cross sections significantly.
It is to be noted, however, that
these rather drastic changes in the cross sections due
to the medium effects are well under control because:
(i) the spreading potential
and the pion distorted wave function
have been fitted to and tested by
the empirical total and elastic cross sections 
for pion-nucleus scattering in and around the $\Delta$ region;
(ii) the medium effects of a similar magnitude for the
photo-process have been shown to 
bring our calculation into good agreement with the data.

An interesting feature of our model is that the unitarized non-resonant
amplitudes give a significant contribution to the cross sections. 
This is in sharp contrast with the results of the previous calculations;
for instance, the calculations in Refs.~\cite{amaro,valencia2},
which considered a tree-level non-resonant mechanism,
found almost no contribution from it.
It is worth emphasizing that this noticeable difference
should not be taken as a measure of uncontrollable model dependence
because (as we confirmed) the difference arises 
largely from unitarization of the non-resonant amplitude,
which clearly needs to be implemented.

We have reexamined the non-local effect in $\Delta$-propagation
in nuclei. 
It was emphasized in Ref.~\cite{non-local}
that this non-local effect, despite its large size,
was not considered explicitly
in any of the existing models
for neutrino-induced coherent pion production
(whether based on a microscopic model or the RS model).
The authors of Ref.~\cite{non-local} made this remark based 
on their calculation that only included the $\Delta$ mechanism.
Our present calculation, which additionally incorporates
the spreading potential and FSI,
also indicates that the non-locality gives a large effect.
Thus, regardless of the level of sophistication in the treatment
of medium effects,
one should always include the non-locality effect explicitly.

Because it is expected that the SciBooNE and the MiniBooNE
collaborations will
report more detailed data on (anti)neutrino-induced coherent CC and
NC pion productions, we have presented numerical results relevant to these
experiments. 

Finally, we made a comparison of
the elementary amplitude (HNV\cite{hnv}) used by Amaro et al.\cite{amaro} 
and ours (SL~\cite{SL,SUL}) to clarify similarities and differences
between them.
The noteworthy points are:
(i) At tree-level, both SL and HNV have essentially the same non-resonant
mechanism;
(ii) In the SL model, a unitary pion-production amplitude is obtained by
solving the Lippmann-Schwinger equation 
in which the tree-diagrams are embedded,
whereas, in the HNV model, the sum of the tree-diagrams are identified with
the pion-production amplitude;
(iii) The non-resonant amplitudes of SL and HNV work differently both for the
elementary processes (e.g., $\nu_\mu+p\to \mu^-+\pi^++p$), and for
coherent pion production;
(iv) In SL, the rescattering contribution contained in the non-resonant amplitude
considerably enhances the cross section for coherent pion production.

\begin{acknowledgments}
S. X. N. acknowledges informative discussions with Hidekazu Tanaka and
 Hirohisa Tanaka about the SciBooNE and MiniBooNE experiments.
S. X. N. also thanks Akira Konaka and Issei Kato for stimulating
 discussions.
This work is supported by
the Natural Sciences and Engineering Research Council of Canada
and Universidade de S\~ao Paulo (SXN),
by the U.S. Department of Energy, Office of Nuclear Physics,
under contract DE-AC02-06CH11357 (TSHL),
by the Japan Society for the Promotion of Science,
Grant-in-Aid for Scientific Research(C) 20540270 (TS),
and by the U.S. National Science Foundation
under contract PHY-0758114 (KK).
\end{acknowledgments}

\appendix
\section{multipole amplitudes}
\label{app_mutlipole}

The amplitudes $F_i^V, F_i^A$ in \Eqs{fvec}{faxi} are expressed
in terms of multipole amplitudes $E_{l\pm}^{V,A},M_{l\pm}^{V,A},
S_{l\pm}^{V,A}$ and $L_{l\pm}^A$ as
\begin{eqnarray}
F_1^V   =   \sum_l[
  P_{l+1}' E_{l+}^V  + P_{l-1}'     E_{l-}^V 
+ lP_{l+1}'M_{l+}^V + (l+1)P_{l-1}' M_{l-}^V]\,,\\
F_2^V   =   \sum_l[
                    (l+1)P_l'M_{l+}^V + lP_l' M_{l-}^V]\,,\\
F_3^V   =   \sum_l[
  P_{l+1}'' E_{l+}^V  + P_{l-1}''   E_{l-}^V 
- P_{l+1}'' M_{l+}^V  + P_{l-1}''   M_{l-}^V]\,,\\
F_4^V   =   \sum_l[
- P_{l}'' E_{l+}^V  - P_{l}''   E_{l-}^V 
+ P_{l}'' M_{l+}^V  - P_{l}''   M_{l-}^V]\,,\\
F_5^V   =   \sum_l[
  (l+1) P_{l+1}' L_{l+}^V  -  l  P_{l-1}' L_{l-}^V]\,, \\
F_6^V   =   \sum_l[
 -(l+1) P_{l}' L_{l+}^V  +  l  P_{l}' L_{l-}^V]\,, \\
F_7^V   =   \sum_l[
 -(l+1) P_{l}' S_{l+}^V  +  l  P_{l}' S_{l-}^V]\,, \\
F_8^V   =   \sum_l[
  (l+1) P_{l+1}' S_{l+}^V  -  l  P_{l-1}' S_{l-}^V]\,,
\end{eqnarray}
and
\begin{eqnarray}
F_1^A   =   \sum_l[
  P_{l}' E_{l+}^A  + P_{l}'     E_{l-}^A 
+ (l+2)P_{l}'M_{l+}^A + (l-1)P_{l}' M_{l-}^A]\,,\\
F_2^A   =   \sum_l[
  (l+1)P_{l+1}'M_{l+}^A + lP_{l-1}' M_{l-}^A]\,,\\
F_3^A   =   \sum_l[
  P_{l}'' E_{l+}^A  + P_{l}''   E_{l-}^A 
+ P_{l}'' M_{l+}^A  - P_{l}''   M_{l-}^A]\,,\\
F_4^A   =   \sum_l[
- P_{l+1}'' E_{l+}^A  - P_{l-1}''   E_{l-}^A 
- P_{l+1}'' M_{l+}^A  + P_{l-1}''   M_{l-}^A]\,,\\
F_5^A   =   \sum_l[
 -(l+1) P_{l}' L_{l+}^A  +  l  P_{l}' L_{l-}^A] \,,\\
F_6^A   =   \sum_l[
  (l+1) P_{l+1}' L_{l+}^A  -  l  P_{l-1}' L_{l-}^A]\,, \\
F_7^A   =   \sum_l[
  (l+1) P_{l+1}' S_{l+}^A  -  l  P_{l-1}' S_{l-}^A] \,,\\
F_8^A   =   \sum_l[
 -(l+1) P_{l}' S_{l+}^A  +  l  P_{l}' S_{l-}^A] .
\end{eqnarray}
$P_L(x)$ is the Legendre function and $x=\hat{k}\cdot\hat{q}$;
$\bm{k}$ and $\bm{q}$ are the pion momentum and the momentum transfer to the
nucleon, respectively.

The multipole amplitudes from isovector currents are further decomposed
according to the total isospin ($T$) in the final $\pi N$ state as
\begin{eqnarray}
\eqn{amp_iso}
X_{l\pm}^{V,A}  = \sum_{T=1/2,3/2} X_{l\pm}^{(T)V,A}\Lambda^{T}_{ij}\ ,
\end{eqnarray}
with $X$ being $E$, $M$, $L$ or $S$.
We have introduced the projection operator $\Lambda^{T}_{ij}$
defined by
\begin{eqnarray}
\Lambda^{3/2}_{ij} = {2 \delta_{i,j}-i\epsilon_{ijk}\tau_k \over 3}\\
\Lambda^{1/2}_{ij} = {\delta_{i,j}+i\epsilon_{ijk}\tau_k \over 3} \ ,
\end{eqnarray}
where the indexes $i$ and $j$ refer to the final pion isospin state and the
component of the isovector current, respectively.
For electromagnetic or NC processes, 
$M_{l\pm}^{(0)V}\tau_i$, which is due to an isoscalar current,
is also added to \Eq{amp_iso}.

In the main text we sometimes use the notation
$X_{l\pm}^{V(A),\zeta}$, where $\zeta$ collectively denotes the pion
charge and the nucleon isospin state;
$X_{l\pm}^{V(A),\zeta}$ is a matrix element (in isospin space) of
\Eq{amp_iso}.
Since we are only concerned with coherent pion production,
the specification of the pion charge determines $i$ and $j$ in \Eq{amp_iso}.
We can find the matrix element (in isospin space) of \Eq{amp_iso} by
specifying the nucleon isospin state.

\section{Lorentz transformation from ACM to 2CM}
\label{app_2cm}

In coherent pion production in neutrino-nucleus scattering
($\nu_\ell + t \to \ell^- + \pi^+ + t$),
the elementary process is
$W^+ (q_A) + N (p_N) \to \pi^+(k_A) + N(p_N^\prime)$, where the four-momenta
in the pion-nucleus center-of-mass frame (ACM) are given in the parentheses.
We suppose here that the pion momentum is on-shell.
In a prescription we employ, the nucleon momenta are fixed as
\begin{eqnarray}
\eqn{p_fix2}
\bm{p}_N = - {\bm{q}_A\over A} - {A-1\over 2A}(\bm{q}_A-\bm{k}_A)\ ,
\qquad
\bm{p}_N^\prime = - {\bm{k}_A\over A} + {A-1\over 2A}(\bm{q}_A-\bm{k}_A)
\ ,
\end{eqnarray}
and the invariant mass ($W$) of the pion and nucleon is
\begin{eqnarray}
\eqn{w_app}
W = \sqrt{(p_N^0+q_A^0)^2 - (\bm{p}_N+\bm{q}_A)^2}\ ,
\end{eqnarray}
where $p^0_N$ is the nucleon energy on the mass-shell.
We note that $W$ depends on $x_A (\equiv \hat{k}_A\cdot\hat{q}_A)$
as well as $|\bm{q}_A|$ and $|\bm{k}_A|$. 
For convenience,
we write $W(|\bm{q}_A|,|\bm{k}_A|,x_A)$.

We perform the standard Lorentz transformation from ACM to 
the $\pi N$ CM frame (2CM).
An arbitrary four-momentum in 2CM ($p_2$) is written with the corresponding 
four-momentum in ACM ($p_A$) as
\begin{eqnarray}
\eqn{lorentz}
\bm{p}_2 &=& \bm{p}_A - {p_A^0\over W} \bm{P} + {P^0-W\over W} 
(\bm{p}_A\cdot\hat{P})\hat{P} \ , \\\nonumber
p^0_2 &=& {P^0 p_A^0 - \bm{p}_A\cdot\bm{P}\over W}  \ ,
\end{eqnarray}
with $P = p_N + q_A$.

We now consider a case in which the pion momentum is off-shell ($k_A^\prime$). 
We encounter this situation when we consider the final-state interaction
in the coherent process.
As before, the nucleon momenta are fixed using \Eq{p_fix2} with $k_A$
replaced by $k_A^\prime$.
However, we do not use the nucleon energy on the mass-shell.
Instead, we take $p_N^0$ so that 
\begin{eqnarray}
W(|\bm{q}_A|,|\bm{k}_A^\prime|,x^\prime_A)=W(|\bm{q}_A|,|\bm{k}_A|,x_A) 
\qquad {\rm for}\ x^\prime_A=x_A\ , 
\end{eqnarray}
where $W$ is obtained with \Eq{w_app}.
With the nucleon four-momentum ($p_N$) obtained in this way, we can
perform the Lorentz transformation as \Eq{lorentz}.
This prescription greatly reduces the amount of labor
involved in our numerical calculation,
because the SL amplitudes need to be calculated at each value of $W$.
With the variables obtained above, we can calculate $\Gamma_{2AL}$
used in \Eqs{i_mtx}{j_mtx}:
\begin{eqnarray}
\eqn{gam_4}
\Gamma_{2AL} = \sqrt{\omega^\prime_{\pi,2} p^{\prime\,0}_{N,2}p^{0}_{N,2}
\over \omega^\prime_{\pi,A}p^{\prime\,0}_{N,L}p^{0}_{N,L}} \ ,
\end{eqnarray}
with $\omega^\prime_{\pi,A}=\sqrt{\bm{k}_A^\prime+m_\pi^2}$.

Finally, we discuss the factor $\Gamma^\chi$,used in \Eqs{i_mtx}{j_mtx},
which originates from
the pion wave function due to the Lorentz transformation.
Among the final-state interactions,
the simplest process 
is the scattering of the pion off a single nucleon 
$\pi(k_A^\prime) + N(p_N^{\prime\prime}) \to \pi(k_A) + N(p_N^f)$,
where the variables in ACM are shown in the parentheses;
only $k_A$ is on-shell.
Similarly to \Eq{p_fix2}, we fix the nucleon momenta as
\begin{eqnarray}
\bm{p}_N^{\prime\prime} = - {\bm{k}^\prime_A\over A} - {A-1\over 2A}(\bm{k}^\prime_A-\bm{k}_A)\ ,
\qquad
\bm{p}_N^f = - {\bm{k}_A\over A} + {A-1\over
2A}(\bm{k}^\prime_A-\bm{k}_A) \ .
\end{eqnarray}
We assume here that
the energies of all the nucleons are on the mass-shell.
For the Lorentz transformation from ACM to LAB specified this way,
we can calculate the Lorentz factor as
\begin{eqnarray}
\eqn{gam_5}
\Gamma^\chi = \sqrt{\omega_{\pi,A}E^{\prime\prime}_{N,A}E^{f}_{N,A}
\over \omega_{\pi,L}E^{\prime\prime}_{N,L}E^{f}_{N,L}} 
\simeq \sqrt{\omega_{\pi,A}\over \omega_{\pi,L}} \ ,
\end{eqnarray}
Although the actual final-state interaction includes 
multiple scattering processes,
it is beyond our framework to calculate $\Gamma^\chi$ with 
multiple scattering taken into account.
We therefore use $\Gamma^\chi$ calculated for the elementary process in
\Eqs{i_mtx}{j_mtx}.
Actually,  the Lorentz factor for the plane wave term in \Eq{pi_wave}
% requires 
is given by the the rightmost expression in \Eq{gam_5}.
Because the approximate equality in \Eq{gam_5} is quite accurate for
$\bm{k}^\prime_A=\bm{k}_A$, we use the middle expression
in \Eq{gam_5} to evaluate the matrix elements in \Eqs{i_mtx}{j_mtx}.

\section{Expressions for some components in the $\Delta$ propagator}
\label{app_misc}

\subsection{Pauli correction to the $\Delta$ self energy}

We follow Ref.~\cite{pauli} to calculate the Pauli correction to the
$\Delta$ self energy ($\Sigma_{\rm Pauli}$). The $\pi N\Delta$ coupling
is from the SL model.
\begin{eqnarray}
\Sigma_{\rm Pauli} &=& {m_N\over W}
\left[2\theta(k_F-\beta)\int_0^{k_F-\beta} dq q^2
{\omega_\pi(q) F^{\rm bare}_{\pi N\Delta}(q) F_{\pi N\Delta}(q)\over
K^2 - q^2 + i\epsilon} \right.\\\nonumber
&+&\left.\int_{|k_F-\beta|}^{k_F+\beta} dq q^2 
\left(1 - {q^2+\beta^2-k_F^2\over 2q\beta}\right)
{\omega_\pi(q) F^{\rm bare}_{\pi N\Delta}(q) F_{\pi N\Delta}(q)\over
K^2 - q^2 + i\epsilon}
\right] \ ,
\end{eqnarray}
where $\theta(x)$ is the step function,
$k_F$ is the Fermi momentum [\Eq{fermi_mom}],
$W$ is the $\pi N$ invariant mass [\Eq{inv_mass}],
$\omega_\pi(q)=\sqrt{q^2+m_\pi^2}$, and
\begin{eqnarray}
K^2 &=& {m_N\over W} \left[(W-m_N)^2 - m_\pi^2\right] \ .
\end{eqnarray}
Furthermore, for electroweak pion production amplitude [\Eq{i_mtx}],
\begin{eqnarray}
\bm{\beta} &=& {m_N\over W} (\bm{p}_N + \bm{q}_A) \ ,
\end{eqnarray}
where $\bm{p}_N$ is fixed using \Eq{p_fix}, and
$\bm{q}_A$ is the momentum transfer to a nucleus in ACM;
for the optical potential [\Eq{v_res}], $\bm{q}_A$ is replaced with
$\bm{k}_A$ (the incoming pion momentum).
We use the on-shell pion momentum to fix $\bm{p}_N$.
The dressed $\pi N\Delta$ vertex ($F_{\pi N\Delta}$) is 
taken from \Eq{t-res},
and the bare $\pi N\Delta$ vertex denoted by
$F^{\rm bare}_{\pi N\Delta}$ is given as \cite{SL}
\begin{eqnarray}
F^{\rm bare}_{\pi N\Delta}(q) = -i {f_{\pi N\Delta}\over m_\pi}
\sqrt{E_N(q)+m_N\over 24\pi^2E_N(q)\omega_\pi(q)}
\left(\Lambda^2_{\pi N\Delta}\over \Lambda^2_{\pi N\Delta} + q^2
\right)^2 \!q \ .
\end{eqnarray}

\subsection{$\Delta$ spreading potential}

We consider the following spreading potential consisting of the central
and the LS parts:
\begin{eqnarray}
\eqn{spr}
\Sigma_{\rm spr} &=& V_C {\rho_t(r)\over \rho_t(0)} 
+ V_{LS} f_{LS}(r) 2 \bm{L}_\Delta\cdot\bm{\Sigma}_\Delta \ , \\
f_{LS}(r) &=& \mu r^2 e^{-\mu r^2} \ ,
\end{eqnarray}
with $\mu=0.3$ fm$^{-2}$.
We have two complex coupling constants $V_C$ and $V_{LS}$ which are
fitted to pion-nucleus scattering data.
The radial dependence of the LS spreading potential is taken from
Ref.~\cite{LS_spr}.
We implement the spreading potential [\Eq{spr}] in the
$\Delta$-propagator after evaluating the doorway state expectation value
of the LS term.
Thus, the LS term provides an L-dependent shift of the resonance mass and
width as\cite{LS_spr}
\begin{eqnarray}
\Sigma_{LS}^L = -5 V_{LS}{\bra{\phi_L}\rho_t f_{LS}k^2 -
 \left(\rho_tf_{LS}\right)^\prime {d\over dr}
+ {L(L+1)\over 2r}\left(\rho_tf_{LS}\right)^\prime\ket{\phi_L}
\over
\bra{\phi_L}\rho_t k^2 -
(\rho_t)^\prime {d\over dr}
\ket{\phi_L}} \ ,
\end{eqnarray}
with the plane wave pion function $\phi_L(r)=j_L(kr)$.

\subsection{$\Delta$ (nucleon) potential}
\begin{eqnarray}
\eqn{nucl_potential}
 V_{\Delta}(r)= V(r) = (-55\, {\rm MeV}) \left( {\rho_t(r)\over\rho_t(0)}\right) \ .
\end{eqnarray}
\newpage
\subsection{$\Delta$ Coulomb potential}
\begin{eqnarray}
&&\hspace{7mm} (r \ge R_e)
\hspace{30mm} (r < R_e) \\[5mm]\nonumber
 V^C_{\Delta}(r)&=& 
\left\{ \begin{array}{ccl}
{\displaystyle 2 (Z-1) \alpha \over\displaystyle r} \ , \qquad
 & - {\displaystyle (Z-1)\alpha r^2 \over\displaystyle R_e^3}
 + {\displaystyle 3 (Z-1)\alpha \over\displaystyle R_e}
\ , \qquad
&(\pi^++p\to\Delta^{++})\\[5mm]
{\displaystyle Z \alpha \over\displaystyle r} \ , \qquad
 &- {\displaystyle Z\alpha r^2 \over\displaystyle 2R_e^3}
 + {\displaystyle 3 Z\alpha \over\displaystyle 2R_e}
\ , \qquad
&(\pi^++n\to\Delta^{+})\\[5mm]
0 \ , \qquad
 &
0 \ , \qquad
&(\pi^-+p\to\Delta^{0})\\[5mm]
-{\displaystyle Z \alpha \over\displaystyle r} \ , \qquad
 & {\displaystyle Z\alpha r^2 \over\displaystyle 2R_e^3}
 - {\displaystyle 3 Z\alpha \over\displaystyle 2R_e}
\ , \qquad
&(\pi^-+n\to\Delta^{-})
\end{array}
\right.
\end{eqnarray}
In the above $Z$ is the atomic number. 
The equivalent square well radius, 
denoted by $R_e$, 
is related to the mean square radius
 ($\langle r^2 \rangle$) of a nucleus by
\begin{eqnarray}
R_e = \sqrt{{5\over 3}\langle r^2 \rangle} \ .
\end{eqnarray}

\end{document}